\newcommand{\BE}{\begin{eqnarray}}
\newcommand{\EE}{\end{eqnarray}}
\newcommand{\be}{\begin{eqnarray}}
\newcommand{\ee}{\end{eqnarray}}
\newcommand{\BEN}{\begin{eqnarray*}}
\newcommand{\EEN}{\end{eqnarray*}}
\newcommand{\ben}{\begin{eqnarray*}}
\newcommand{\een}{\end{eqnarray*}}
\newcommand{\BA}{\begin{array}}
\newcommand{\EA}{\end{array}}
\newcommand{\ba}{\begin{array}}
\newcommand{\ea}{\end{array}}
\newcommand{\BLN}{\begin{align*}}
\newcommand{\ELN}{\end{align*}}

\newcommand{\f}{\frac}

\newcommand{\bl}{\begin{lstlisting}}
\newcommand{\el}{\end{lstlisting}}

\documentclass[11pt,A4,onecolumn]{article}

\usepackage{hyperref}
\usepackage{url}
\usepackage{amsmath,amssymb,mathrsfs}
\usepackage{setspace}
\usepackage{tabularx}
\usepackage[format=plain,font=footnotesize]{caption}
\usepackage{array,multirow,graphicx}
\usepackage{subcaption}
\usepackage{tikz}
\usetikzlibrary{math}
\usetikzlibrary{arrows, decorations.markings}
\usepackage{hhline}
\usepackage{mathtools}
\usepackage{pdfpages}

\usepackage{booktabs}
\usepackage{soul} 
\usepackage{pifont}
\newcommand{\cmark}{\ding{51}}%
\newcommand{\xmark}{\ding{55}}%
\usepackage{cleveref}
\crefformat{equation}{Eq.~(#2#1#3)}
\Crefformat{equation}{Equation~(#2#1#3)}
\crefformat{lemma}{Lemma~#2#1#3}
\Crefformat{lemma}{Lemma~#2#1#3}
\crefformat{proposition}{Prop.~#2#1#3}
\Crefformat{proposition}{Proposition~#2#1#3}
\crefformat{theorem}{Theorem~#2#1#3}
\Crefformat{theorem}{Theorem~#2#1#3}
\crefformat{figure}{Fig.~#2#1#3}
\Crefformat{figure}{Fig.~#2#1#3}

\title{Identification of gatekeeper diseases on the way to cardiovascular mortality}

\author{Nils Haug$^\text{a,b}$, Stefan Thurner$^\text{a,b,c,d}$, Alexandra Kautzky-Willer$^\text{e}$, \\ Michael Gyimesi$^\text{f}$ and Peter Klimek$^{\text{a,b}}$\footnotemark[1] \vspace{5mm}\\
\footnotesize
\begin{tabular}{c}
$^\text{a}$Section for Science of Complex Systems, CeMSIIS,  Medical University of Vienna,\\
Spitalgasse 23, A-1090 Wien, Austria\\
$^\text{b}$Complexity Science Hub Vienna, Josefst\"adter Stra\ss e 39, A-1080 Wien, Austria\\
$^\text{c}$IASA, Schlossplatz 1, A-2361 Laxenburg, Austria\\
$^\text{d}$Santa Fe Institute, 1399 Hyde Park Road, Santa Fe, NM 85701, USA\\
$^\text{e}$Gender Medicine Unit, Division of Endocrinology and Metabolism, \\
Department of Internal Medicine III, Medical University of Vienna,\\
Spitalgasse 23, A-1090 Wien, Austria\\
$^\text{f}$Gesundheit \"Osterreich GmbH, Stubenring 6, A-1010 Wien, Austria\\
$*$peter.klimek@meduniwien.ac.at
\end{tabular}
}

\begin{document}

\maketitle

\begin{abstract}
Multimorbidity, the co-occurrence of two or more chronic diseases such as diabetes, obesity or cardiovascular diseases in one patient, is a frequent phenomenon. 
To make care more efficient, it is of relevance to understand how different diseases condition each other over the life time of a patient. 
However, most of our current knowledge on such patient careers is either confined to narrow time spans or specific (sets of) diseases.   
Here, we present a population-wide analysis of long-term patient trajectories by clustering them according to their disease history observed over 17 years. 
When patients acquire new diseases, their cluster assignment might change. A health trajectory can then be described by a temporal sequence of disease clusters. From the transitions between clusters we construct an age-dependent multilayer network of disease clusters.
Random walks on this multilayer network provide a more precise model for the time evolution of multimorbid health states when compared to models that cluster patients based on single diseases. Our results can be used to identify decisive events that potentially determine the future disease trajectory of a patient. We find that for elderly patients the cluster network consists of regions of low, medium and high in-hospital mortality. Diagnoses of diabetes and hypertension are found to strongly increase the likelihood for patients to subsequently move into the high-mortality region later in life.
\end{abstract}

Noncommunicable (or chronic) diseases (NCDs) such as dorsalgia, hypertension, respiratory diseases or diabetes affect a large fraction of the world's population and decrease the quality of life of people burdened by them \cite{WHO2018}. 
For example, in 2004, almost $50\%$ of the population of the United States was suffering from at least one chronic condition \cite{RWJ2004} and in 2014, approximately 25$\%$ of the adults had been diagnosed with more than one NCD \cite{Ward2016}, with both numbers strongly increasing with age. 
Besides having a negative impact on people's wellbeing, NCDs also pose a substantial burden on the healthcare systems of countries \cite{OECD2018,Berwick2012}. Apart from combatting common risk factors such as unhealthy lifestyle and diet, which many NCDs are attributed to \cite{Goodarz2009}, early detection and management play a key role in decreasing the burden of chronic diseases \cite{ElNahas2005}. Personalized (or stratified) medicine promises to make care more efficient by tayloring treatments individually to patients while taking their individual risk profile for potential future diseases into account \cite{PMC2014}. 
A necessary precondition to enter this new era of medicine is a deepened understanding of how the long-term developments of multiple diseases condition each other. 

The field of network medicine holds as central tenet that diseases cannot be studied independently from each other but arise from complex interactions between molecular units in the human body in terms of discretized structures such as protein--protein, metabolic, regulatory and RNA networks \cite{Barabasi2012}. 
For example, the fact that many diseases tend to be comorbid, i.e. that they tend to co-occur in patients, 
can be understood based on disease-causing genes or pathways shared by the comorbid diseases \cite{Goh2007, Rzhetsky2007,Park2009}.
However, not all diseases linked by shared genes exhibit significant comorbidities, as different mutations of the same gene can have different consequences. 
Conversely, not all comorbidities can be explained from molecular data. One reason is that comorbidities can also be the result of the exposure to the same environmental risk factors, for example, due to a particular lifestyle \cite{Klimek2016}. 
In addition, present data on molecular interactions is far from being complete \cite{Menche2015}. 

To obtain a more empirically-based understanding of disease comorbidity, observational healthcare data regarding hospital admissions, pharmaceutical prescriptions and doctor visits has been leveraged for the purpose of disease prediction and to reveal previously unknown molecular connections between diseases \cite{Jensen2012,Schneeweiss14}.
Using so-called phenotypic comorbidity networks \cite{Hidalgo09,Chmiel14,Jeong2017,Fotouhi2018}---networks where nodes represent diseases and two nodes are linked if the two corresponding diseases are comorbid--- it was shown that diseases which are comorbid to many other diseases have a higher mortality and that patients tend to develop diseases in close network proximity to the ones they already have.
The latter fact opens up the possibility to predict future diseases of patients based on their medical history. 
This has been accomplished with approaches that include collaborative filtering \cite{Davis2008,Steinhaeuser2009}, frequent itemsets \cite{Folino2010,Folino2015}, learning of transition probabilities between states represented by binary vectors \cite{Arandjelovic2015,Vasileva2017}, 
deep learning \cite{Pham2016,Miotto2016} and point processes \cite{Weiss2013}.

Next to predicting future disease, longitudinal data on hospital admissions allows the identification of previously unknown patterns in the temporal sequence of diseases (health trajectories) patients are diagnosed with. 
By analyzing the temporal order of diagnosis pairs co-occuring in a significant number of patients with a particular directionality in time, it was possible to identify such trajectories with a length of up to four diseases \cite{Jensen14}. 
Alternative methodological approaches to identify typical health trajectories include dynamic time warping \cite{Giannoula18}, a method originally developed for speech recognition, or the detection of putative causal relations between pairs of diseases using information-theoretic approaches \cite{Kannan2016}. 
The major challenge all these works are confronted with is the high number of different diagnosis codes, leading to a combinatorial explosion of the number of possible disease combinations, many of which occur only in a single patient. 

\begin{figure*}
\centering
\includegraphics[width=11.4cm]{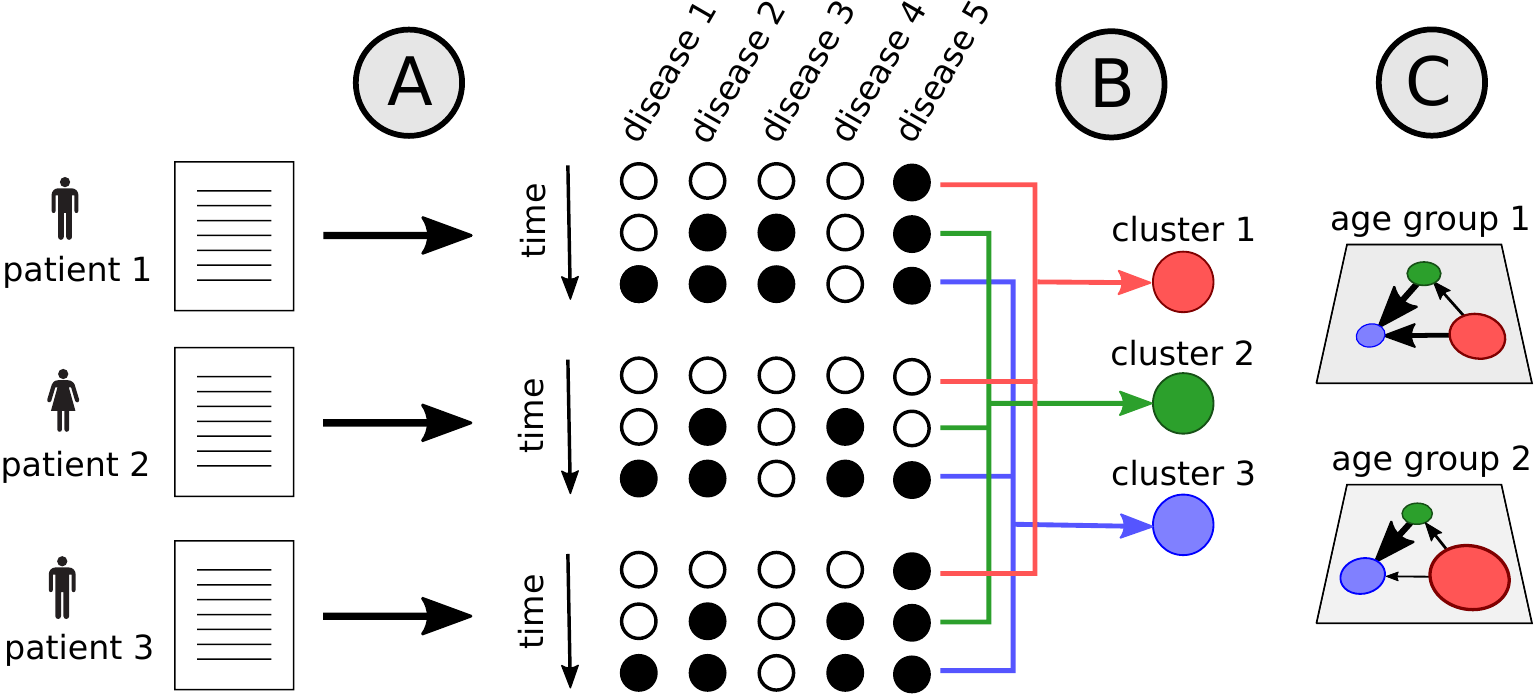}
\caption{Illustration of the workflow from the input data to a multiplex network model of disease progression. (A) From the input data, we extract the observed health states of patients in the cohort at the end of each year of the observation period, represented by binary vectors. (B) All sampled health states are assigned to clusters using the DIVCLUS-T algorithm. For each patient we thereby obtain a sequence of clusters he or she is assigned to over the years (the health trajectory). (C) From the health trajectories of patients, we construct a multiplex network, where nodes represent clusters, layers represent age and sex groups and directed links between the nodes are weighted according to the rates at which patients change between the corresponding clusters.}
\label{fig_workflow}
\end{figure*}

Here we present an alternative approach to characterize the population of an entire country in terms of its long-term health trajectories.
An illustration of our work flow is shown in \cref{fig_workflow}.  First, \cref{fig_workflow}A, we represent the health state of a patient at a given point in time by a binary vector encoding the set of diseases he or she has been diagnosed with so far. 
Based on a longitudinal data set covering all intramural stays in Austria from 1997 until 2014, see Methods, we consider the yearly disease vectors of all patients without any hospital admission between 1997 and 2002.
The set of all health states (disease vectors) is then partitioned using a divisive clustering algorithm called DIVCLUS-T \cite{Chavent07}, \cref{fig_workflow}B. 
Each cluster is defined by a set of diseases which each patient in that cluster has been diagnosed with (inclusion criteria) and a set of diseases each patient in that cluster has {\it not} been diagnosed with (exclusion criteria). 
As time proceeds, patients acquire new diseases and consequently can change the cluster they belong to, see \cref{fig_workflow}B. 
The sequence of clusters a patient belongs to throughout the observation period describes his or her health trajectory. 
We construct a multiplex network where nodes represent clusters, layers correspond to different patient sex and age groups.
Directed links in the multiplex layers are weighted according to the rate at which patients of the corresponding sex and age group change between the two clusters corresponding to the nodes, see \cref{fig_workflow}C. 
Our approach therefore provides a statistical description of the disease progression of patients which takes into account their sex, age and entire observed history of hospital diagnoses.

\section{Results}
\label{section_results}

Our study cohort consists of the $M = 5,112,811$ patients who did not receive a diagnosis with ICD-10 code from A00 to N99 in 1997--2002 and received at least one diagnosis with ICD-10 code from A00 to Z99 in 2003--2014. 
Figure S2 in the Supplementary Information (SI) shows the distribution of the dates of birth of patients of this cohort; 53$\%$ of them are female.

\subsection{Disease clusters}
\label{ssct_dis_clust}

For each patient we observe thirteen disease vectors (from the end of each year in the observation period 2002--2014) which are grouped into 132 disease clusters by DIVCLUS-T, see Methods.
Tables detailing the inclusion and exclusion criteria of all clusters can be found in SI, section 8. There and in the main text, the symbols \xmark~and~\cmark~next to a diagnosis block indicate that this block is an exclusion, respectively an inclusion criterion for that cluster. Since all patients are free from any (known) prior diagnoses, all patient trajectories start in cluster 0 (the ``healthy cluster''), where all diagnoses are exclusion criteria.

\begin{figure*}
\centering
\begin{subfigure}[h]{5.8cm}
\includegraphics[angle=0,width=\linewidth]{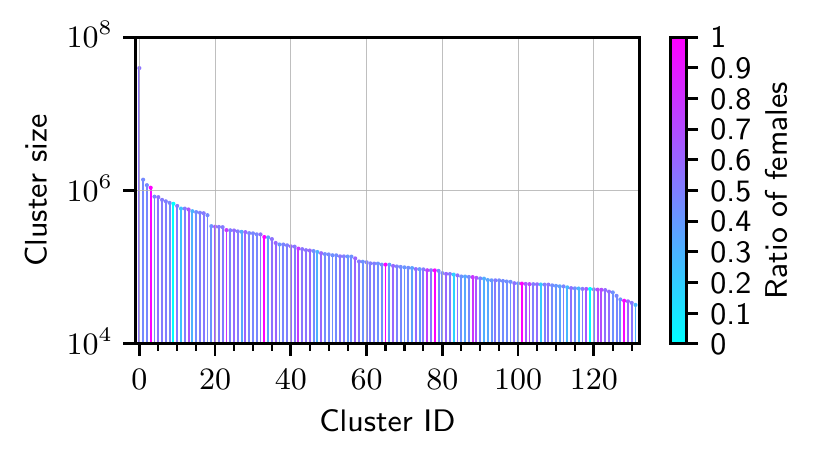}
\caption{}
\label{fig_cluster_sizes_sex_ratio}
\end{subfigure}
\hspace{10mm}
\begin{subfigure}[h]{5.8cm}
\includegraphics[angle=0,width=\linewidth]{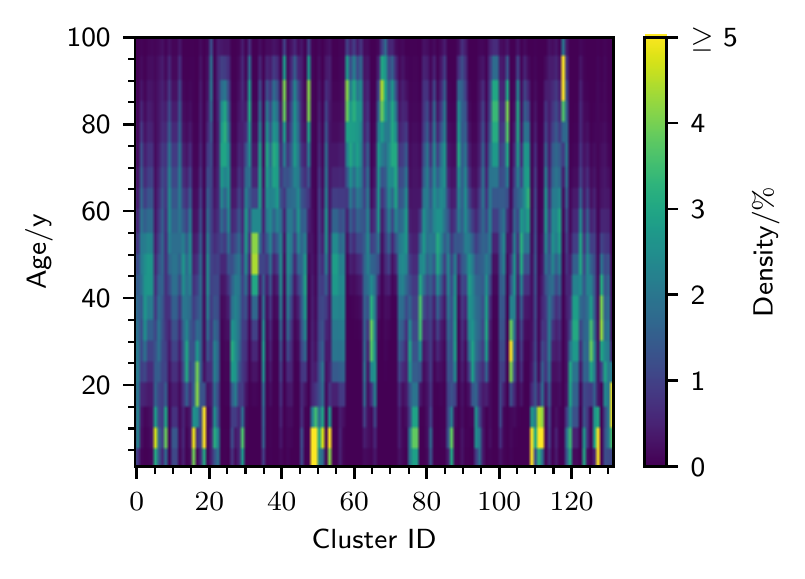}
\caption{}
\label{fig_2d_hist_ages_in_clusters}
\end{subfigure}
\begin{subfigure}[h]{5.8cm}
\includegraphics[angle=0,width=\linewidth]{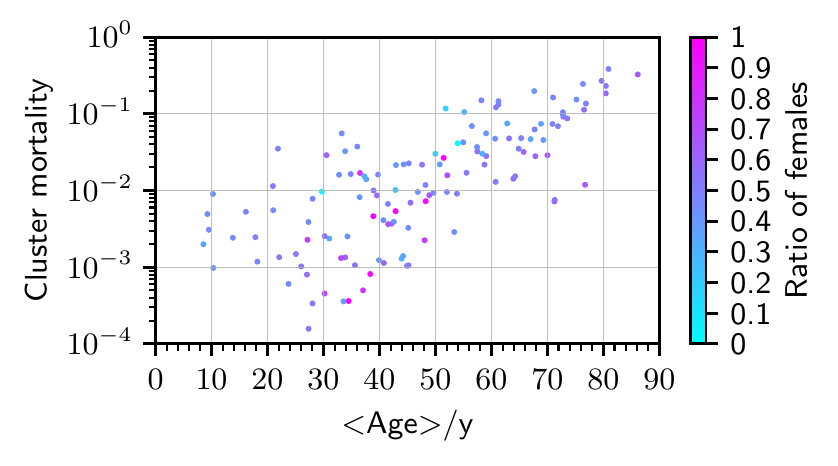}
\caption{}
\label{fig_mortality_in_clusters}
\end{subfigure}
\hspace{15mm}
\begin{subfigure}[h]{4.3cm}
\centering \includegraphics[angle=0,width=\linewidth]{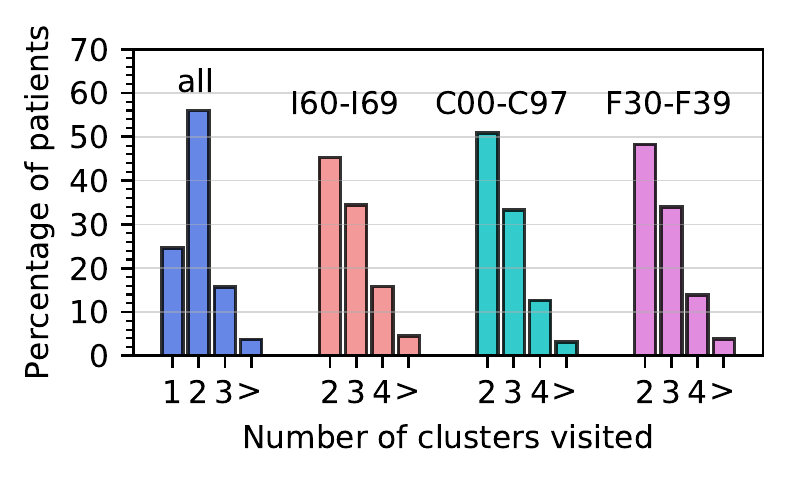}
\caption{}
\label{fig_hists_no_of_clusters_visited}
\end{subfigure}
\begin{subfigure}[h]{4.3cm}
\centering \includegraphics[angle=0,width=\linewidth]{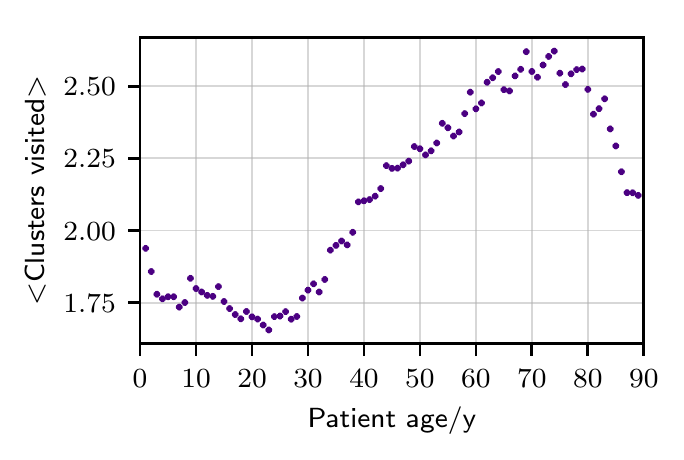}
\caption{}
\label{fig_mean_no_of_clusters_by_age}
\end{subfigure}
\hspace{2.5mm}
\begin{subfigure}[h]{9cm}
\footnotesize
\begin{tabular}{lc}
\multicolumn{2}{c}{\textbf{Cluster 0}}\\
\toprule
all & \xmark\\ \bottomrule
\end{tabular}
$\underset{\text{\sf 129,961}}{\longrightarrow}$
\footnotesize
\begin{tabular}{ll}
\multicolumn{2}{c}{\textbf{Cluster 2}}\\
\toprule
E70-E90 & \xmark\\ 
I10-I15 & \xmark\\
I80-I89 & \xmark\\
J30-J39 & \xmark\\
M00-M25 & \cmark\\
M40-M54 & \xmark\\
M60-M79 & \xmark\\
M80-M94 & \xmark\\ \bottomrule
\end{tabular} $\underset{\text{\sf 5,983}}{\longrightarrow}$
\begin{tabular}{ll}
\multicolumn{2}{c}{\textbf{Cluster 31}}\\
\toprule
I10-I15 & \xmark\\
J30-J39 & \xmark\\
M00-M25 & \cmark\\
M60-M79 & \cmark\\
M80-M94 & \xmark\\ \bottomrule
\end{tabular}
\newline 
\caption{}
\label{fig_most_freq_red_traj}
\end{subfigure}
\caption{(A) Number of observations assigned to the different clusters, color coded by the ratio of females in the cluster. (B) Two-dimensional histogram of the age distribution in the different clusters. (C) Scatter plot of mortality versus mean age in the different clusters with color indicating their sex ratios. (D) Relative distribution of the number of clusters visited during the observation period for all people in the cohort (blue), patients diagnosed with cerebrovascular diseases (red, I60--I69), malignant neoplasms (green, C00--C97) and mood [affective] disorders (magenta, F30--F39) during the observation period. (E) Mean number of clusters visited by patients during the observation period depending on their age at the beginning of the observation period. (F) Illustration of the reduced trajectory (0,2,31), which is the most frequent reduced trajectory of length 3 and followed by 5,983 patients of the cohort. The numbers underneath the arrows give the total number of patients who step from cluster 0 to cluster 2 and from cluster 0 to cluster 2 to cluster 31, respectively.}
\label{fig_cluster_descriptives}
\end{figure*}

In \cref{fig_cluster_sizes_sex_ratio} we show the sizes, meaning the total number of observations, assigned to each cluster. 
Note that the cluster IDs are given by their rank after being sorted by size. 
Color in \cref{fig_cluster_sizes_sex_ratio} shows the sex distribution within the clusters. 
There are several clusters only containing patients of one sex; this is explained by sex specific inclusion criteria such as `Diseases of male genital organs' (N40--N51). There are, however, also clusters with a high sex imbalance that is less easily explained. For example, 82$\%$ of the patients in cluster 83 are male. For this cluster, the inclusion criteria are `Mental and behavioral disorders due to psychoactive substance use' (F10--F19) and `Diseases of liver' (K70--K77).

A two-dimensional histogram of the age distributions in the different clusters is shown in \cref{fig_2d_hist_ages_in_clusters}. Clusters differ strongly in their age composition, with a high number of clusters where the distribution is centered around ages 60--80. In \cref{fig_mortality_in_clusters} we compare the mean age of patients in a cluster with the probability that a patient who enters this cluster stays there and dies in-hospital within the observation period (hospital release type death). We refer to this probability as the cluster mortality. Mortality tends to increase with the mean age of patients in a cluster, however, for a given mean age, the spread of the cluster mortality can extend over more than one order of magnitude. The maximum mortality value of 38$\%$ is measured for cluster 68, which has no exclusion criteria and includes hypertensive diseases (I10--I15), other forms of heart disease (I30--I52), diseases of arteries, arterioles and capillaries (I70-I79) \textit{and} a diagnosis of renal failure (N17--N19). Moreover, 67$\%$ of the health states assigned to that cluster include metabolic disorders (E70-E90) and 49$\%$ include diabetes mellitus (E10-E14).

\subsection{Health trajectories}

As patients acquire new diseases from one year to the next, they can transition between clusters, see Methods.
The time sequence of clusters that a patient has visited defines his or her trajectory.
 We also define the reduced health trajectory of a patient as the health trajectory with all repetitions removed. For example, if patient $i$ has the health trajectory $(1,1,5,5,5,3)$, then his or her reduced health trajectory is $(1,5,3)$.
The blue bars in \cref{fig_hists_no_of_clusters_visited} show the distribution of the mean length of the reduced health trajectory of patients during the observation period. 
Almost 25$\%$ of the patients stay in cluster 0 during the entire time span (only received codes from the ICD chapter O--Z); 2$\%$ visit four or more clusters. \Cref{fig_mean_no_of_clusters_by_age} shows how the length of the reduced health trajectory of patients varies with age. Adolescent patients visit the least number of clusters, and patients around the age of 70 visit the highest number of clusters. The most frequent reduced trajectory of length 3 is (0,2,31), and is followed by 5,983 individuals; \cref{fig_most_freq_red_traj}. Patients with this reduced trajectory first acquire a diagnosis from the block `Arthropathies' (M00--M25) and subsequently one from the block `Soft tissue disorders' (M60--M79). 

We now describe results for typical health trajectories involving the diagnosis blocks of `Cerebrovascular diseases', `Malignant neoplasms' and `Mood [affective] disorders'.

{\bf Cerebrovascular diseases (I60--I69).} 
A total of 199,681 patients (3.9$\%$) were diagnosed with cerebrovascular diseases. 
The distribution of the length of their reduced trajectories is clearly shifted towards higher values compared to the total cohort; see \cref{fig_hists_no_of_clusters_visited}.
The most frequent reduced trajectory of length three, followed by 1,447 patients, is (0,32,48),
meaning that patients from the ``healthy cluster'' 0 (all diseases are exclusion criteria) move to cluster 32 with hypertensive diseases (I10--I15) and heart diseases (I30--I52) to cluster 48 
where cerebrovascular diseases (I60--I69) changed from an exclusion (cluster 32) to an inclusion criterion (cluster 48).

{\bf Malignant neoplasms (C00--C97).}
A total of 312,787 patients (6.1$\%$) received a diagnosis from the block `Malignant neoplasms'. 
The distribution of the number of visited clusters visited by these patients is shown in \cref{fig_hists_no_of_clusters_visited} and shifted toward higher values with respect to all other patients. 
More than $50\%$ of them only visit two different clusters, and $44\%$ (69,552) of those patients have the reduced trajectory (0,12), i.e., from the healthy cluster to a cluster where cancer (C00--C97) is the only inclusion criterion. 
The most frequent reduced trajectory of length 3, followed by 3,093 cancer patients, is (0,12,38) where patients acquire hypertensive diseases (I10--I15) after cancer.

{\bf Mood [affective] disorders (F30--F39).}
A total of 210,589 patients (4.1$\%$) of the study cohort are diagnosed with mood or affective disorders during the observation period. The distribution of their reduced trajectory lengths is shown in \cref{fig_hists_no_of_clusters_visited} and again shifted toward higher lengths compared to all patients. 
Among these patients, 9.5$\%$ end their trajectory in cluster 55, where additional to `Mood [affective] disorders', patients are diagnosed with a diagnosis from the block `Mental and behavioral disorders due to psychoactive substance use' (F10--F19). 
The most common reduced trajectory of length 3 in this subcohort, followed by 1,936 patients, is (0,14,55). These patients first acquire a diagnosis from the block F10--F19, and subsequently one from the block F30--F39.  

\begin{figure*}
\centering
\includegraphics[width=14cm,angle=0,trim={0cm 0cm 0 0cm},clip=false]{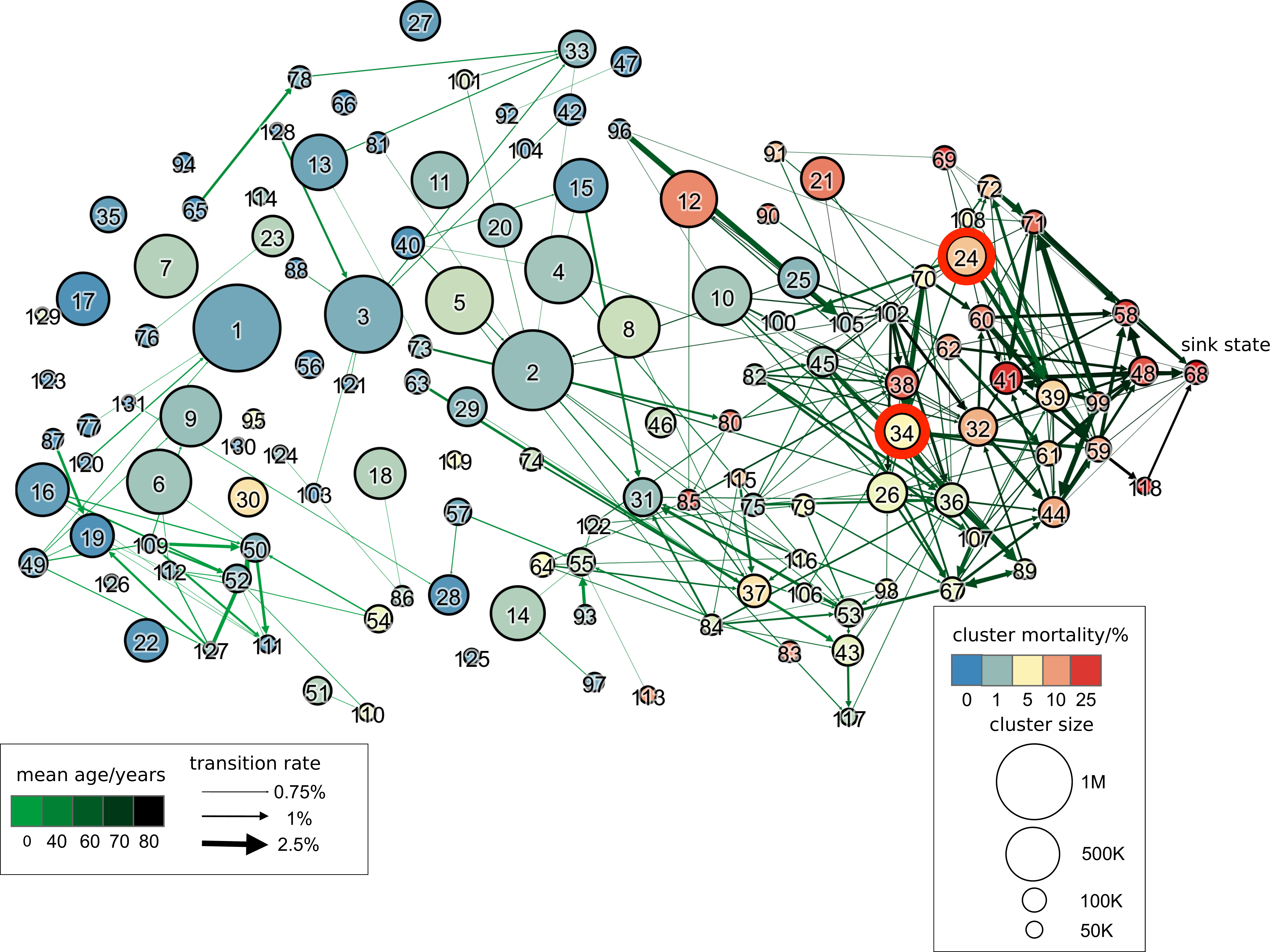}
\caption{Plot of the collapsed dynamic network $\mathcal{M}$, where nodes represent clusters and directed edges between the nodes are weighted according to the transition probability between the two nodes. The edges are colored depending on the mean age of the patients who make the corresponding step. Light green edges indicate a low mean age and dark green edges indicate a high mean age of the patients making the corresponding step. The size of the nodes is scaled according to the size of the corresponding cluster and the node color indicates the cluster mortality, with blue indicating low mortality and red indicating high mortality. The node representing cluster 0, which is connected to all other nodes, has been removed in this visualization. The layout has been generated by a force-directed algorithm after which nodes have been displaced from left to right by a distance proportional to the average age associated with its links, i.e., patient age increases from left to right. To increase the readability, we only display the 255 strongest links of the network. For elderly patients we observe network regions of low, medium and high mortality. Clusters 24 and 34, which are identified as gatekeepers to the sink state, are marked with red circles; the sink state cluster 68 is annotated.}
\label{fig_cluster_network}
\end{figure*}

\subsection{Disease cluster multiplex network}

Separately for males and females, we construct a directed multiplex dynamic network, $\mathcal{M}$, where nodes represent patient clusters and layers (time slices) correspond to age and sex groups. 
A directed link from node $i$ to node $j$ within time slice $a$ in the network for patients of sex $g$ has a weight proportional to the number of observed transitions from cluster $i$ to $j$ of patients with corresponding age and sex.
\Cref{fig_cluster_network} shows a visualization of the dynamic network of clusters of patient health states, filtered and collapsed into one layer. 
Nodes correspond to clusters and their size is the number of observations in that cluster; node color corresponds to cluster mortality on a scale from 0\% (blue) to 38\% (red).
Links indicate a significant number of transitions between two clusters of patients with an average age that is encoded into the link color from 0y (light green) to 80y (black).
The layout in \cref{fig_cluster_network} has been generated in a force-directed way, after which the nodes were displaced from left to right by a distance proportional to the average age of patients in the cluster.
As patients age (move from left to right in the network), they either reach a region of low or high mortality.
The low mortality clusters for the elderly (clusters 25, 102, 105) are characterized by having various diseases of the eye as inclusion criteria and most other diseases as exclusion criteria.
Contrast this to the high mortality region for the elderly that is often accessed via cluster 24 (including hypertensive, I10--I15, and diabetes, E10-E14) or cluster 26 (with hypertensive diseases and metabolic disorders, E70--E90).
Cluster 68 is a sink state of our dynamic multiplex network, i.e. has zero out-strength, and lies within the high-mortality region. Note that the fact that there exists such a sink state is a consequence of our choice of clustering algorithm.
There are clusters for younger patients with links pointing toward the low and high mortality regions, such as cluster 2 (including arthropathies, M00--M25) or cluster 29 (diseases of the oesophagus, stomach, and duodenum, K20--K31).
While cluster 68 has a balanced sex ratio (49\% females), many clusters with direct links to 68 have an overrepresentation of females (65\% in cluster 118, 52\% in 58, 59\% in 48).
Additionally, we find a region of medium mortality for the elderly made up of clusters that include arthropathies and osteopathies (M80--M94), cluster 43, or additionally soft tissue disorders (M60--M79), cluster 117.

\subsection{Gatekeeper clusters}

We identify ``gatekeeper clusters''  strongly increasing the risk of cardiovascular mortality by using a stochastic model of disease progression. Precisely, we model health trajectories as random walks on the disease cluster multiplex network $\mathcal{M}$, see Methods.

For a given cluster $k$, we term as gatekeepers to $k$ those clusters which tend to be visited by many patients early in their medical career and where patients visiting that cluster have, according to our random walk model, an increased probability to visit cluster $k$ during their lifetime. For the high-mortality cluster 68 (the sink state), we identify cluster 24 as a gatekeeper, highlighted red in \cref{fig_cluster_network}; see SI for details. Patients in that cluster necessarily have been diagnosed with diabetes mellitus (E10-E14) and hypertensive diseases (I10-I15), but have not yet received diagnoses of obesity and other hyperalimentation (E65-E68), metabolic disorders (E70-E90) and other forms of heart disease (I30-I52). The probability that a patient of  the cohort who enters into cluster 24 will reach cluster 68 during his or her lifetime according to our model is 0.23, and is hence increased by approx. $40\%$ compared to the baseline probability of 0.16 for all patients of the cohort. 
Another closely related gatekeepers is clusters 34 (metabolic, hypertensive, and ischaemic heart diseases but no diabetes and no other forms of heart diseases), for which the probability to reach cluster 68 increases by approx. 50--60\%. Clusters 34 is also highlighted red in \cref{fig_cluster_network}.

\section{Discussion}

We presented a novel method to describe patient trajectories as observed over 17y from a population-wide data set on hospital diagnoses. 
The method is based on partitioning observed medical histories of patients into a small number of clusters. 
A statistical analysis of the transition rates between clusters enabled us to create a model which can be used to simulate synthetic patient trajectories as random walks on a multiplex network. 

We show that elderly patients can be found in regions of low, medium and high mortality. The high mortality region is characterized by the co-occurrence of multiple chronic conditions ranging from diabetes, cardiovascular diseases, coronary heart disease, to renal failure; most of these diseases are exclusion criteria in the low mortality region for the elderly, where they only suffer from diseases such as senile cataract.
Although morbidity and mortality attributable to cardiovascular diseases decreased in the last decade (with greater improvement in men), they remain the predominant causes of mortality in both sexes. On the other hand NCDs like obesity, metabolic disorders, diabetes and renal diseases increased in men and women, partly due to population aging.
 The comorbid prevalence of diabetes and renal disease and failure markedly increases cardiovascular morbidity and mortality.  Patients with renal diseases hardly achieve the targets for blood pressure and glycemic control. However, the use of new classes of antidiabetic drugs may change the paradigm of ineluctable cardiorenal risk in the future \cite{Giugliano2019}.

Some patients enter the region of high mortality and high disease burden at an older age via the low mortality region in \cref{fig_cluster_network}.
For others, however, the course toward high mortality is already set at much younger ages.
For instance, cluster 46 consists of patients with an  average age of 44y, metabolic disorders (E70--E90) and about thirty other diagnosis blocks as exclusion criteria.
From this cluster there is a link to cluster 34 where patients additionally receive diagnoses of hypertensive and ischaemic heart diseases, after which disease and mortality burden are bound to increase further. This is in line with findings that patients with metabolic syndrome have an almost threefold increased mortality because of cardiovascular or coronary heart disease in an 11.4y follow up \cite{Lakka2002}. 
In this sense, patients with metabolic disorders appear to be locked in on a path toward high mortality already at much younger age (say, between 40y and 50y).

We  identified several gatekeeper clusters that can be loosely seen as ``points of no return'' toward the high mortality network region.
By identifying how patients transition into these gatekeeper clusters, we find that diabetes mellitus and hypertension often stand at the beginning of the medical career of patients.
These diseases substantially increase their probability to subsequently move into the high-mortality region of the cluster network. Diabetes and hypertension are the most prominent risk factors for development of both renal failure and cardiovascular disease.   Unfortunately, they are often undetected and only diagnosed because of acute complications caused by the underlying disorders. For example, in Denmark 35\% of the patients with newly diagnosed diabetes featured diabetic complications around diagnosis \cite{Gedebjerg2018}. As patients usually suffer from diabetes for many years before the diagnosis, screening and prevention programmes are necessary at least for subjects at high risk. It is highly recommended that patients with the clinical features hypertension, vascular diseases, dyslipidemia, prediabetes or abdominal obesity, often clustering as “metabolic syndrome”, should as early as possible and all (other) subjects older than 45 years undergo diabetes screening at regular intervals \cite{ADA2019}. Such procedure could help to identify more patients at risk earlier and to implement secondary prevention of long-term complications and assure guidelines-based therapy.

Females are overrepresented in several clusters with direct links to the high-mortality cluster 68, which in itself has a balanced sex ratio. 
Both diabetes and renal failure appear to attenuate the generally positive effect of female sex on life expectancy: these disorders seem to increase risk in women to a greater extent, thereby equalizing mortality risks \cite{Kautzky-Willer2016,Regensteiner2015,Hecking2014}. This may be ascribed to sex-dimorphic changes in risk factor burden and enviromental factors. Among patients with type 2 diabetes, which comprises more than 90\% of all diabetes cases, women with type 2 diabetes were shown to be more likely obese, hypertensive and to have hypercholesterolaemia but were less likely to be prescribed lipid-lowering medication and antihypertensive drugs, especially if they had cardiovascular disease in comparison to men \cite{Wright2019}. Moreover it was shown that admissions for acute myocardial infarctions steadily increased in the last two decades especially in younger patients whose history of diabetes and hypertension increased in parallel \cite{Arora2019}. The proportion attributable to younger women was particularly high who also had lower probability of  receiving guideline-directed therapy. 

Limitations of this work include that we do not know the health history of patients prior to the observation period. 
Also, the data only contains health information from hospital stays but not from outpatient contacts or doctor visits. Moreover, the fact that the data was originally recorded for billing purposes means that diagnoses which do not lead to financial reimbursement were often not reported. The actual disease burden of patients might therefore be underrepresented in our data. 
Simulated trajectories might occur that are logically not possible.
The model might produce a patient trajectory starting in cluster $A$, with disease $d$ as an inclusion criterion, stepping to cluster $B$, where $d$ is neither an exclusion nor an inclusion criterion, and from there to cluster $C$, where $d$ is now an exclusion criterion. 
To assess the quality of the trajectory model, we compared our results with different benchmark models that assign patients to clusters based on single diseases (either the most recent or least frequent one) rather than using exclusion criteria.
These benchmarks perform either comparable or only slightly worse than the DIVCLUS-T approach in terms of cluster inertia.
This can be understood from the fact that the benchmark models nevertheless capture disease--disease correlations (note that for a given cluster, the cluster-specific frequencies of other diseases are the marginal frequencies of the diseases computed over all patients fulfilling the criteria for that cluster).
However, if we take the longitudinal component of the data into account and compare the statistics of simulated and real patient trajectories, we find that our approach clearly outperforms the benchmark models.
This confirms that the issue of logically impossible trajectories does not substantially impact our results.
In addition, these results reveal that our multiplex network approach based on multimorbidity clusters provides a more adequate framework to capture the path-dependent nature of long-term health trajectories compared to the single-disease benchmark models.
It will be of interest to extend the current framework with additional clinical variables or to focus on more specific subsets of patients for a more precise identification of disease gatekeepers.

\appendix

\footnotesize

\section*{Data and Methods}

\renewcommand\thesubsection{\Alph{subsection}}

\subsection{Data}

We analyze a data set provided by the Austrian Federal Ministry for Health, covering all approx. 45M hospital stays of about 9M individuals in Austria during the 17 years from 1997 until 2014. Here, the term hospital also includes facilities of long-term care such as rehabilitation centers or psychiatric hospitals \cite{Sozialministerium2019}.
For each stay of a patient, the data includes sex, age group (5y interval), admission and release date, release type (release, transfer, death), as well as in total about 120M main or side diagnoses in the form of level-3 ICD-10 codes \cite{WHO2016}.
We restrict our analysis to 1,074 codes from A00 to N99, as the remaining codes relate to medical conditions that do not necessarily reflect diseases.
Due to the similarity of the conditions described by many codes, we group them into 131 blocks as defined by the WHO. 
For example, the codes from E10 to E14 form the block `Diabetes Mellitus'. A list of all diagnosis blocks can be found in section 9 of the Supporting Information (SI) to this article.
To make sure that the patients are in a well defined health state at the beginning of the observed time period, we restrict the analysis to all patients not assigned a diagnosis from the range A00 to N99 during a hospital stay in Austria in the period from 1997 to 2002 inclusively.

\subsection{Clustering}

The (observed) health state of a patient at a given point in time consists of all diseases he or she has been diagnosed with until that point. 
We represent the health state of a patient $i$ ($1\leq i \leq M$) at time $t$ by a binary vector $\mathbf{x}^{(i)}(t)=(x_1^{(i)}(t),\dots,x_N^{(i)}(t))$, where $N$ denotes the total number of disease codes, and $x_j^{(i)}(t)$ is one if patient $i$ has been diagnosed with disease $j$ until time $t$ and zero else.
We choose a set of equidistant time values $T=\{t_0,\dots,t_n\}$, where $t_{i+1}-t_i=\Delta~(1\leq i < n)$, at which we sample the health state of each patient. The parameter $\Delta$ controls the temporal resolution of the analysis. The values $t_0$ and $t_n$ mark the start and end point of the observation period, respectively.

We use a divisive clustering algorithm called DIVCLUS-T \cite{Chavent07}, to partition the set of observed health states of patients with at least one diagnosis into $K=131$ clusters; see SI, section 2. An additional cluster contains all health states of patients who did not receive any diagnose yet. The DIVCLUS-T algorithm aims to minimize the same objective function as the well-known Ward's method and the $k$-means algorithm, but produces results which are easier to interpret. More precisely, each cluster is defined by a set of diseases which all patients in that cluster have been diagnosed with and a set of diseases which all patients in that cluster have not been diagnosed with. We refer to these sets as the inclusion, respectively the exclusion criteria for that cluster.

\subsection{Analysis of cluster transitions}
\label{subsection_analysis_of_transitions}

Denote for $t\in T$ by $C_i(t)$ the cluster to which observation $\mathbf{x}^{(i)}(t)$ is assigned. The sequence $\mathcal{S}_i=(C_i(t_0),\dots,C_i(t_n))$ describes the health trajectory of patient $i$ throughout the observation period. We also define the reduced health trajectory of a patient as the health trajectory with all repetitions removed. 
If for a given $1\leq i \leq M$ and $1\leq s < n$, $C_i(t_s) = k$ and $C_i(t_{s+1}) = j$, then we say that patient $i$ steps from cluster $k$ to cluster $j$. For each pair of clusters $1\leq k,j \leq K$, there exists at least one disease which is an exclusion criterion in one and an inclusion criterion in the other cluster. Since we represent patients by all diseases which they have been diagnosed with so far, a patient cannot step from a cluster in which a certain disease is an inclusion criterion to one where this disease is an exclusion criterion, and therefore, steps between two clusters are only possible in one direction.  

The cohort is stratified into groups according to sex and age. For a given sex $g$ and age group $a$, we define $s_{g,a,k,j}$ as the number of patients of sex $g$ and age group $g$ who step from cluster $k$ to cluster $j$; the value
\begin{equation}
q_{g,a,k,j} = \f{s_{g,a,k,j}}{\sum_{j=1}^{K} s_{g,a,k,j}}
\end{equation}
is the rate at which patients of this sex and age group make this step.

\subsection{Modeling health trajectories}
\label{modeling health trajectories}

To simulate the future trajectory of a patient, we first determine his or her starting position according to age and history of medical diagnoses. Let at time $t_0$ the patient be in cluster $k$ on time slice $a$ of the network. Then we randomly choose the position of this patient at time $t_1=t_0+\Delta$, such that with probability $q_{g,a',k,j}$, his or her next position is in cluster $j$ on time slice $a'$; the time slice $a'$ is determined from the age of the patient at time $t_1$.
To compare results of our method to simulate random health trajectories with the health trajectories observed, we simulate as many synthetic health trajectories as there are patients in our study cohort, with the same sex and age distribution as our cohort. 
The initial condition of the model is taken from the data, i.e., all patients start their trajectory in the ``healthy'' cluster where all diseases are exclusion criteria (cluster zero). 
To assess the quality of the simulated health trajectories, we compute several metrics for comparing disease frequencies according to the simulated trajectories with the data, and compare the performance of our method in terms of these metrics with two benchmark models, see SI. Fig. S4 (b) in the SI shows the forecasted distribution of patients of the study cohort over the different clusters at the end of their lives.

\normalsize

\section*{Acknowledgements}

We acknowledge support from the WWTF "Mathematics and ..." Project MA16-045,  European Commission, H2020 SmartResilience No. 700621 and FFG Project 857136. NH would like to thank Rudolf Hanel, Markus Strauss and Sarah Etter for helpful discussions.

\section*{Author Contributions}
N.H. and P.K. conceived the study and devised the analytic methods, N.H. wrote the manuscript with contributions from P.K. and A.K.-W., N.H. carried out the analysis and produced plots and graphics, S.T. made critical comments regarding the manuscript, A.K.-W. contributed expertise regarding the medical interpretation of the findings, M.G., P.K. and N.H. researched and prepared the data. All authors read and approved the manuscript.




\section*{Supplementary material}

# Supplementary Information for
# Identification of gatekeeper diseases on the way to cardiovascular mortality

Nils Haug, Stefan Thurner, Alexandra Kautzky-Willer,
Michael Gyimesi and Peter Klimek

August 2, 2019

## 1 Data

Fig. 1 shows a two-dimensional histogram of the number of distinct diagnoses which patients in the study cohort acquire during the observation period from 2003 to 2014, depending on their age. The number of distinct diagnoses is larger for older people, indicating an increase in comorbidity with age.

The distribution of the dates of birth of patients in the study cohort is shown in Fig. 2.

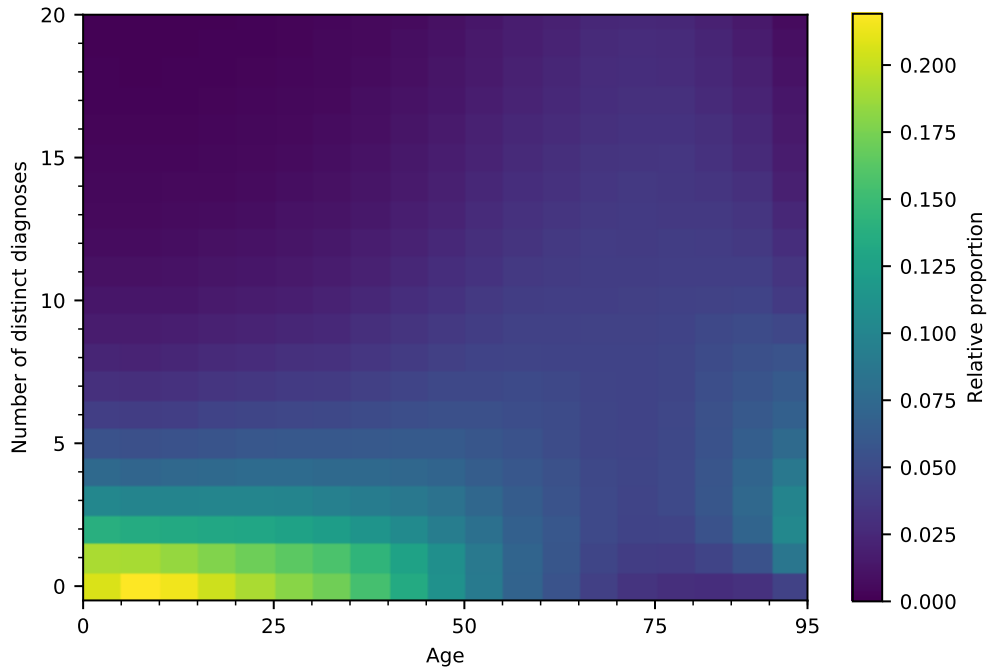

;

Figure 1: Two-dimensional histogram of the number of distinct diagnoses acquired by patients of the study cohort in the observation period from 2003 to 2014, depending on the 5-year age group they belong to in 2014. The data is normalized such that values sum to one separately for each age group.

## 2 Clustering algorithm

The divisive hierarchical clustering algorithm which we use in this work is called DIVCLUS-T, and was presented in [1]. It is similar to the well established Ward's minimum variance method [2] and the $k$-means algorithm [3, chap. 20] in that it aims to optimize the same objective function as these two clustering methods, but increases interpretability due to the fact that its results can be read like a decision tree.

The input data passed to the algorithm is in the form of a binary matrix $\mathbf{X} \in \{0,1\}^{M \times N}$, recording $M$ observations of $N$ different features. The element of $\mathbf{X}$ which stands in the $i$th row and $j$th column is denoted by $X_{i,j}$, where $1 \leq i \leq M$ and $1 \leq j \leq N$. We also denote for $1 \leq i \leq M$ the binary vector representing

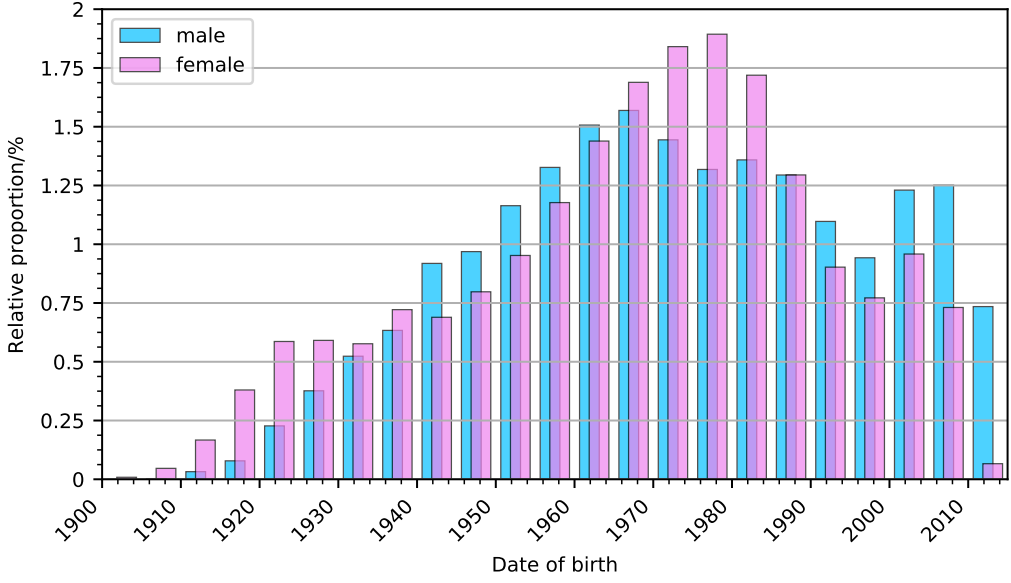

Figure 2: Distribution of the dates of birth of patients in the study cohort. The data is normalized such that the bar heights for males and females sum to one separately.

one particular observation by $\mathbf{x}^{(i)} = (X_{i,1}, \ldots, X_{i,N})$. In our application, one row of $\mathbf{X}$ corresponds to the health state of one patient, observed at a certain time, and each column corresponds to a particular block of diagnoses. For example, if column $j$ represents the diagnosis block 'Diabetes Mellitus', then $X_{i,j} = 1$ if the patient represented by row $i$ has been diagnosed with diabetes until the point in time the observation was made, and 0 otherwise. In our implementation of the algorithm, the user further specifies a binary vector $\mathbf{v} \in \{0,1\}^N$, a real, non-negative vector $\mathbf{w} \in [0,\infty)^N$, as well as two positive integers $\theta$ and $K$.

Starting from the root node, the algorithm iteratively constructs a binary tree where each node $k$ represents one particular subset $S_k$ of the total set of observations $S_0 = \{1, \ldots, M\}$, and $S_k \supset S_j$ if and only if node $k$ is an ancestor of node $j$; the set $S_0$ is represented by the root node. To each node $k$ is further associated a vector $\mathbf{p}^{(k)} = (p_1^{(k)}, \ldots, p_N^{(k)}) \in [0,1]^N$, defined as

$$\mathbf{p}^{(k)} = \frac{1}{|S_k|} \sum_{i \in S_k} \mathbf{x}^{(i)}. \tag{1}$$

Geometrically, $\mathbf{p}^{(k)}$ is the centroid of the $|S_k|$ binary vectors represented by node $k$, embedded into $\mathbb{R}^N$. For $1 \leq j \leq N$, the value $p_j^{(k)}$ is the probability that an observation, randomly selected from $S_k$, has feature $j$. The inertia of a node $k$ is defined as

$$I_k = \sum_{i \in S_k} \|\mathbf{w} \circ (\mathbf{x}_i - \mathbf{p}^{(k)})\|_2^2 = |S_k| \mathbf{w} \cdot \left(\mathbf{p}^{(k)} \circ (\mathbf{1} - \mathbf{p}^{(k)})\right), \tag{2}$$

where '$\circ$' denotes the coefficientwise product, '$\cdot$' is the Euclidean scalar product, $\mathbf{1} = (1, \ldots, 1)$ and $\|\cdot\|_2$ denotes the Euclidean norm. For a given feature $1 \leq j \leq N$, the value $p_j^{(k)}(1 - p_j^{(k)})$ is the variance of the set of points $\{\mathbf{x}_i \mid i \in S_k\}$, projected onto the $j$th axis. It can therefore be regarded as a measure for how homogeneous the observations belonging to $S_k$ are with respect to feature $j$. The inertia $I_k$ quantifies the overall inhomogeneity of the observations represented by note $k$, averaged over all the features, weighted according to the entries of $\mathbf{w}$. Note that our use of the letter $\mathbf{w}$ differs from [1], where weights were assigned to observations.

Let after any number of iterations of the algorithm node $k$ be a leaf node of the tree constructed thus far. To simplify the notation, we drop the index $k$ and let $S$ be the set of observations represented by that node. Define for a given feature $1 \leq j \leq N$ the subsets $S_0^j = \{i \in S \mid X_{i,j} = 0\}$ and $S_1^j = \{i \in S \mid X_{i,j} = 1\}$. Hence, the sets $S_1^j$ and $S_0^j$ partition the set $S$ into those observations which do or do not have feature $j$, respectively. Among the features $1 \leq j \leq N$ for which $v_j = 1$, the algorithm now selects the feature

$$j_0 = \operatorname*{argmin}_{j \in J} \left(I_0^j + I_1^j\right), \tag{3}$$

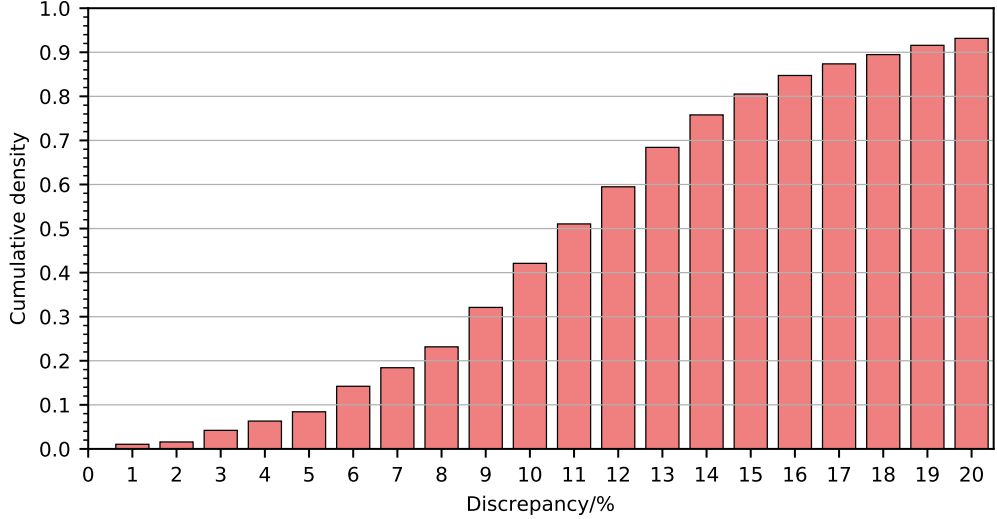

Figure 3: Discrepancy between synthetic and real trajectory counts in % of the real trajectory counts for all trajectories followed by more than 1000 patients.

where $I_0^j$ and $I_1^j$ are the inertias of $S_1^j$ and $S_0^j$, respectively, and

$$J = \{1 \leq j \leq N \,|\, v_j = 1 \wedge \min(|S_0^j|, |S_1^j|) > \theta\}. \tag{4}$$

Node $k$ now becomes the parent node of two newly created leaf nodes representing the subsets $S_0^{j_0}$ and $S_1^{j_0}$, respectively. The feature $j_0$ is hence the feature used to split of node $k$; the vector $\mathbf{v}$ specifies which features are allowed to be used to split up nodes.

After a certain number of iterations, the set $J$ is empty for each leaf node, and the algorithm therefore stops. Assume that the number of leaves of the generated tree is $L$ and denote the sum of the inertias of all leaves of this tree by $F$. By removing two leaves of the tree with common parent node, we obtain a pruned tree of $L-1$ leaves where the parent node is now a leaf. Denote the sum of the inertias of all leaves of this new tree by $F'$. We remove the pair of nodes with common parent node such that the difference $F' - F$ is minimal. By repeating this procedure $L - K$ times, we arrive at a tree with $K$ nodes. We denote the sets represented by the leaves of this tree by $C_1, \ldots, C_K$. They form a partition of the set $S_0 = \{1, \ldots, M\}$, and hence give a clustering of the observations contained in $\mathbf{X}$ into $K$ disjoint subsets.

Suppose we are given a certain observation but do not know which cluster it belongs to. To determine the cluster membership of that observation, we proceed as follows. Starting from the root node, we ask whether the feature which was used to split up the root node is present in the observation or not, and, accordingly, we move on to the child node of the root which has or does not have this feature, respectively, and so forth, until we arrive at a leaf node (i.e., a cluster). The cluster membership of any observation can thus be determined by a finite sequence of binary questions. It is this property which distinguishes it from Ward's method and the $k$-means algorithm, and which increases the interpretability of the results obtained. However, this comes at the cost of restricting the set of possible cluster partitions, and therefore potentially less optimal solutions.

Specifically in this work, we set $\theta = 1000$, $\mathbf{v} = (1, 1, \ldots, 1)$, and the entries of the weight vector $\mathbf{w} = (w_1, \ldots, w_N)$ are set as

$$w_j = -\ln(p_j) \quad \text{for } 1 \leq j \leq N, \tag{5}$$

where $p_j = \frac{1}{M} \sum_{i=1}^{M} X_{i,j}$ is the probability that a randomly selected patient of the cohort, observed at a uniformly selected time in $T$, has been diagnosed with disease $j$ so far. The quantity on the right of Eq. (5) is known as the inverse document frequency (idf), a commonly used measure in text analytics [4]. The idea behind idf is that terms appearing less frequently in a corpus of documents are more distinctive to a text which contains them. In our context, documents correspond to observations and words correspond to individual diseases.

The objective function aimed to be minimized by DIVCLUS-T is

$$F = \sum_{k=1}^{K} I_k,$$

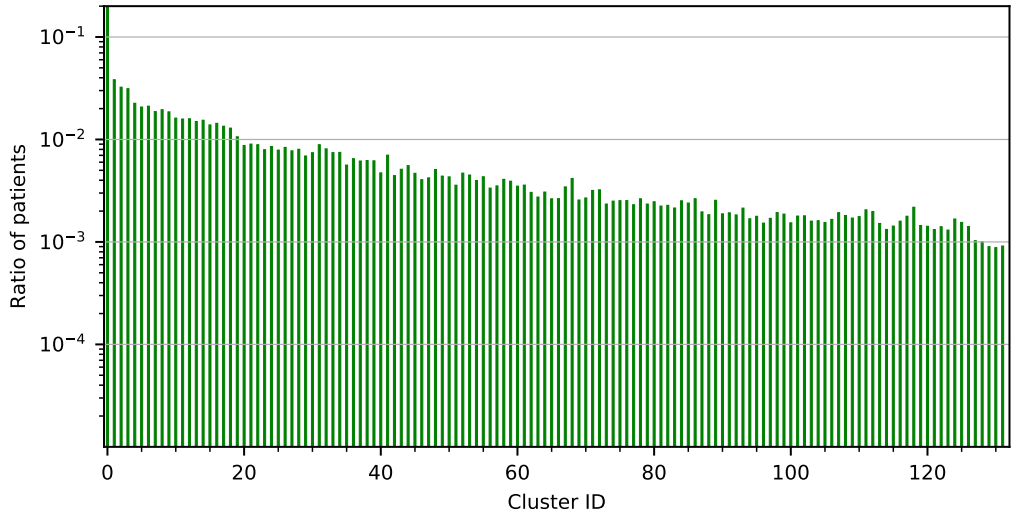

(a)

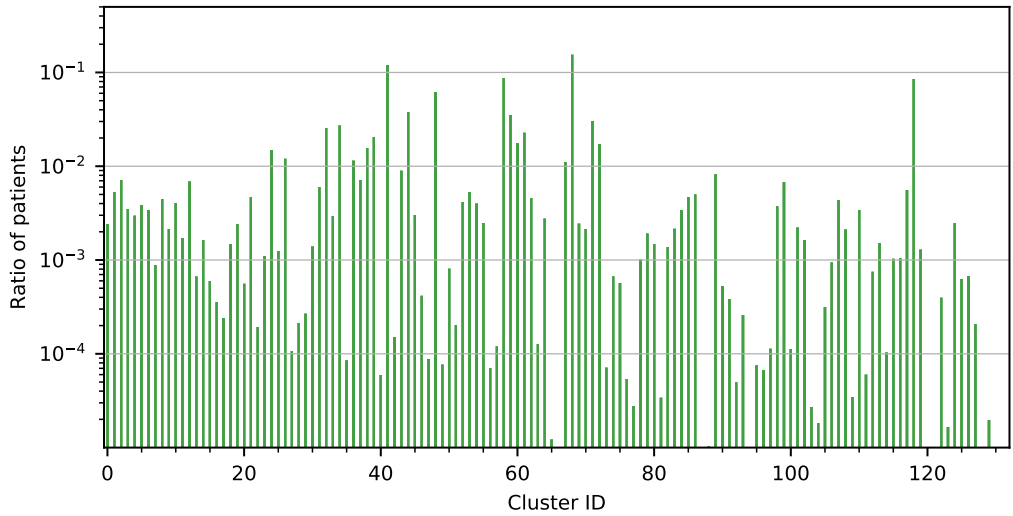

(b)

Figure 4: Distribution of clusters occupied by patients at the end of the observation period (a). Forecasted distribution of clusters occupied by patients at the end of their lives (b).

where for $1 \leq k \leq K$, $I_k$ is the inertia of cluster $k$. In the case where $\mathbf{w} = (1, \ldots, 1)$, $F$ is the same objective function aimed to be minimized by Ward's method and the $k$-means algorithm.

Note that DIVCLUS-T, as described here, can also be used for prediction purposes. For example, if we set $\mathbf{v} = (1, 1, \ldots, 1, 0)$ and $\mathbf{w} = (0, 0, \ldots, 0, 1)$, then the algorithm generates a decision tree predicting the feature represented by the last column of $\mathbf{X}$ based on all other features.

## 3 Selection of the number of clusters

As in the last section, we denote by $F$ the sum of the inertias of the leaf nodes of a clustering tree. The mean cluster inertia of a clustering tree of $K$ leaves is defined as $f = \frac{F}{K}$. In order to find out whether our set of data points admits a natural partitioning into a certain number of clusters, in Fig. 5 we plot the mean cluster inertia $f$ against the number $K$ of clusters. In case such a natural partitioning exists, one expects that the mean cluster inertia decreases strongly as the number of clusters increases from one to the natural number of clusters in the data, and less strongly as the number of clusters increases further.

This plot shows that the mean cluster inertia decreases strongly for small values of the number of clusters and less strongly for larger numbers of clusters. We make the heuristic choice $K = 132$, since this value enables us to directly compare the performance of the algorithm with two simple approaches used as benchmarks, as discussed below.

In order to quantify the effect of correlations in the data, as in the previous section we compute for each disease block $j$ the marginal probability $p_j = \frac{1}{M}\sum_{i=1}^{M} X_{i,j}$. Then we create a ranking $r : \{1, \ldots, N\} \to \{1, \ldots, N\}$ such that $w_i p_i (1 - p_i) > w_j p_j (1 - p_j)$ if $r(i) < r(j)$. For a given number of clusters $K$ we compute

$$\tilde{f} = \sum_{d=K}^{N} w_{r(d)} p_{r(d)} (1 - p_{r(d)}). \tag{6}$$

The value $\tilde{f}$ is the expectation value of the mean cluster inertia obtained by randomly shuffling the ones of the matrix $\mathbf{X}$ independently in each column. The difference $\tilde{f} - f$ indicates how strongly different diseases in the data are correlated. The pink line in Fig. 5 shows the value of $\tilde{f}$ as a function of $K$. We see that there is a clear effect of correlations between different diagnoses in the data, which is made use of by our clustering algorithm.

## 4 Robustness of our results

As explained in section C of the methods section in the main article, the weight $q_{s,g,i,j}$ is the rate at which patients of sex $g$ and age group $a$ step from cluster $i$ to cluster $j$. This rate is computed by dividing the number $s_{g,a,i,j}$ of steps of patients with sex $g$ from cluster $i$ to cluster $j$ in layer $a$ by the number of observations $\sum_{j=1}^{K} s_{g,a,i,j}$ (the size) of cluster $i$ in layer $a$ for sex $g$. If we consider our dataset as a realization of a certain stochastic process, then $q_{g,a,i,j}$ is an empirical estimation of the transition probability $\hat{q}_{g,a,i,j}$ inherent to this process. We call an edge weight robust if the width of the 95% confidence interval for the estimation of $\hat{q}_{g,a,i,j}$ is smaller than $\lambda\, q_{g,a,i,j}$, where $\lambda = 0.2$. To check if a given edge weight $q_{g,i,j}$ is robust, we proceed as follows.

Let $g, a, i$ and $j$ be fixed and simplify the notation to $p = q_{g,a,i,j}$, $\hat{p} = \hat{q}_{g,a,i,j}$ and $s = \sum_{j=1}^{K} s_{g,a,i,j}$. The empirical probability is a random variable

$$p = \frac{1}{s}\sum_{i=1}^{s} X_i, \tag{7}$$

where $X_1, \ldots, X_s$ are iid random variables with $X_i \sim \text{Bernoulli}(\hat{p})$. It follows from the central limit theorem that for $s \to \infty$,

$$\sqrt{\frac{s}{\hat{p}(1-\hat{p})}}(p - \hat{p}) \xrightarrow{d} \mathcal{N}(0,1), \tag{8}$$

where $\xrightarrow{d}$ denotes convergence in distribution. This means that for $\delta > 0$,

$$\lim_{s \to \infty} \mathbb{P}\left(\hat{p} \in \left[p - \delta\sqrt{\frac{\hat{p}(1-\hat{p})}{s}}, p + \delta\sqrt{\frac{\hat{p}(1-\hat{p})}{s}}\right]\right) = 2(C(\delta) - 1), \tag{9}$$

where $C$ is the cumulative distribution function of the standard normal distribution. The condition that

$$\delta\sqrt{\frac{\hat{p}(1-\hat{p})}{s}} \leq \lambda p$$

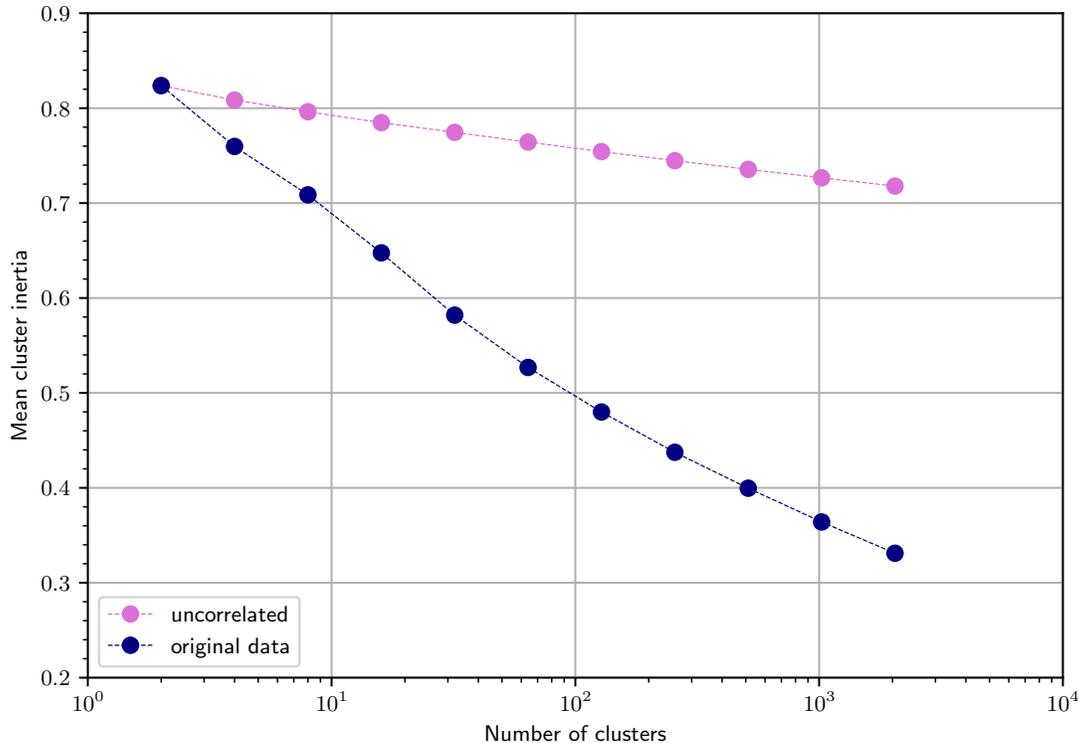

Figure 5: Mean cluster inertia as a function of the number of clusters for original and uncorrelated data.

is for sufficiently large $s$ equivalent to $p \geq \delta^2/(s\lambda^2)$.

For $\delta \simeq 1.96$, $2(C(\delta) - 1) = 0.95$. Our condition for an edge to be robust is therefore that

$$q_{g,a,i,j} \geq \frac{1.96^2}{0.2^2 \sum_{j=1}^{K} s_{g,a,i,j}}. \tag{10}$$

## 5 Simulated patient trajectories

We compare for each observed reduced health trajectory how much the number of patients following that reduced trajectory as predicted from the random walk simulation differs from the actual observed number of patients with that reduced trajectory, relative to that number. The relative discrepancy between the count of reduced trajectories in the observed and synthetic cohort differs by less than 15% for more than 90% of the reduced trajectories followed by more than 1,000 patients; see Fig. 3.

## 6 Gatekeeper clusters to the sink state

As explained in the main text, by gatekeeper clusters to a cluster $j$ we mean clusters which tend to be visited by patients early in their trajectory and where patients entering that cluster have a high probability of visiting cluster $j$ during their lifetime.

In Fig. 6 we show a two-dimensional scatter plot of the reachability of each cluster from the healthy state (cluster 0) against the reachability of the sink state, cluster 68, from that cluster. Here, the reachability of a cluster $j$ from the healthy state is defined as the rate at which patients step from cluster 0 to cluster $j$; the reachability of the sink state from cluster $j$ is the probability that, according to our random walk model, a patient entering cluster $j$ will reach the sink state before he or she reaches the age of 100 years.

As before we denote by $s_{g,a,k,j}$ the number of patients of sex $g$ and age group $a$ who step from cluster $k$ to cluster $j$; the value $s_{k,j} = \sum_{g=0}^{1} \sum_{a=0}^{9} s_{g,a,k,j}$ is the number of all patients stepping from $k$ to $j$, and

$$q_{k,j} = \frac{s_{k,j}}{\sum_{j=1}^{K} s_{k,j}} \tag{11}$$

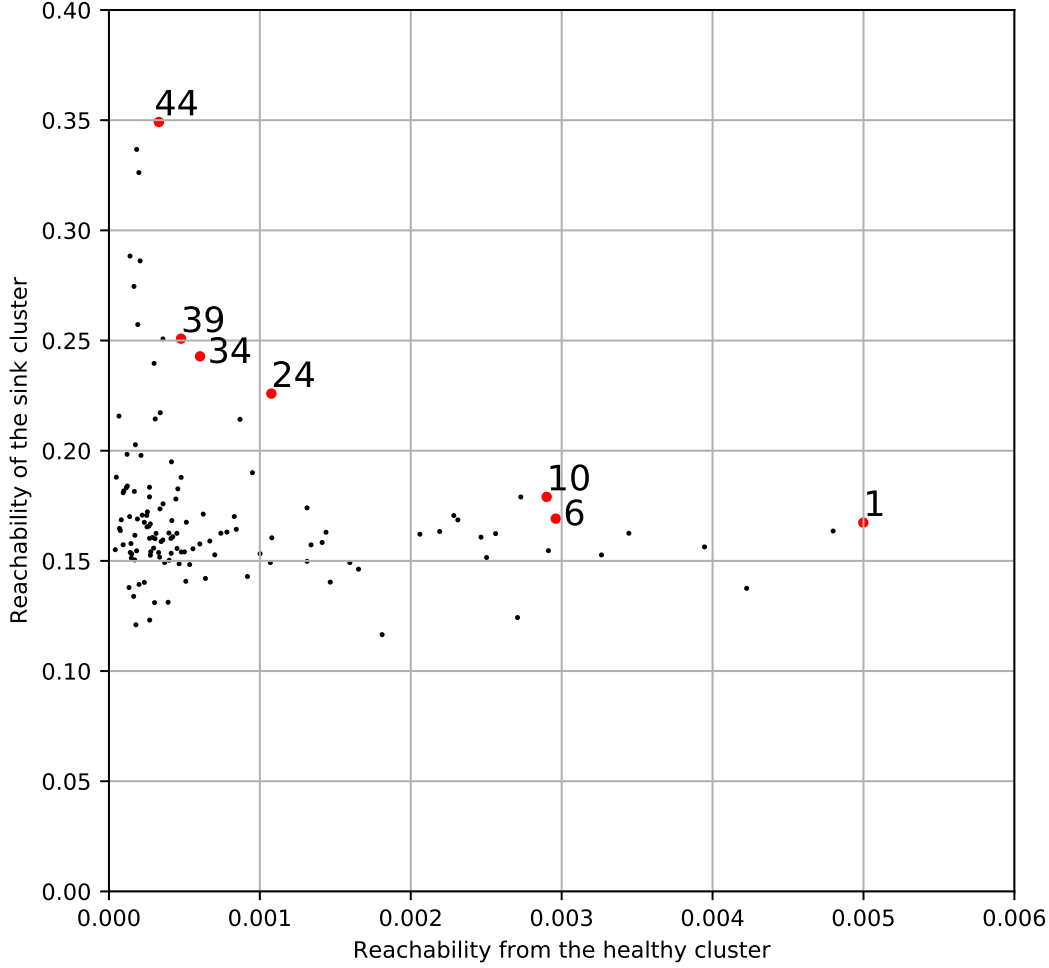

Figure 6: Scatter plot of the reachability of a cluster from the healthy state versus the reachability of the sink state from that cluster. Those clusters for which there is no other cluster with higher values of both quantities are highlighted red.

with $k = 0$ gives the reachability of cluster $j$ from cluster 0.

We compute the reachability of the sink state $k = 68$ from a cluster $j$ for a patient of sex $s$ and age group $g$ as

$$r_{g,a,j} = \left(\mathbf{Q}_{g,a}^{10} \cdot \mathbf{Q}_{g,a+1}^{10} \cdots \mathbf{Q}_{g,9}^{10}\right)_{j,k}, \qquad (12)$$

where for $g \in \{0,1\}$ and $a \in \{0,1,\ldots,9\}$, the elements of the $(K \times K)$-matrix $\mathbf{Q}_{s,g}$ are given by

$$(\mathbf{Q}_{g,a})_{k,j} = q_{g,a,k,j}, \qquad (13)$$

and where as in the main text, $q_{g,a,k,j}$ is defined as the transition rate of patients of sex $g$ and age group $a$ from cluster $k$ to cluster $j$. To obtain an aggregated measure over sex and age groups, we take the weighted mean

$$r_j = \frac{\sum_{g=0}^{1} \sum_{a=0}^{9} r_{g,a,j} n_{g,a,j}}{\sum_{g=0}^{1} \sum_{a=0}^{9} n_{g,a,j}}, \qquad (14)$$

where $n_{g,a,j} = \left(\sum_{k=1}^{K} s_{g,a,k,j}\right) - s_{g,a,j,j}$ is the number of patients of sex $g$ and age group $a$ who enter into cluster $j$.

## 7 Performance of the method and comparison with benchmarks

To quantify how well the cluster transitions shown in Fig. 3 of the main text characterize the progression of actual multimorbid health states, we evaluate the performance of our method for simulating patient

trajectories in terms of several metrics and compare these metrics to two benchmark models that replace DIVCLUS-T by a clustering scheme which assigns patients to clusters based on a single disease.

We denote for $1 \leq d \leq N$ by $p_d$ the probability that a randomly selected patient of the study cohort has been assigned a diagnosis from the ICD-10 code block $d$ until the end of the observation period. The total disease burden of the study cohort at the end of the observation period is given by

$$N_{\text{obs}} = M \cdot \|\mathbf{p}\|_1, \tag{15}$$

where $M = 5,112,811$ is the size of the cohort and $\mathbf{p} = (p_1, \ldots, p_N)$. Here, for $p \geq 1$, $\|\|_p$ denotes the $L^p$ norm, which, for a vector $\mathbf{x} = (x_1, \ldots, x_L)$, is defined as

$$\|\mathbf{x}\|_p = \left( \sum_{\ell=1}^{L} x_\ell^p \right)^{\frac{1}{p}}. \tag{16}$$

Let $\hat{p}_d$ be the marginal probability for a randomly selected patient to have been diagnosed with disease $d$ until the end of the observation period according to our random walk model. The vector $\hat{\mathbf{p}} = (\hat{p}_1, \ldots, \hat{p}_N)$ is computed as

$$\hat{\mathbf{p}} = \sum_{k=1}^{K} \hat{c}_k \mathbf{p}^{(k)}, \tag{17}$$

where $\mathbf{p}^{(k)} = (p_k^{(1)}, \ldots, p_k^{(N)})$, with $p_d^{(k)}$ for $1 \leq d \leq N$ being the probability for a randomly selected patient in cluster $k$ to have disease $d$; $\hat{c}^{(k)}$ is the probability that a randomly selected patient of the cohort is assigned to cluster $k$ at the end of the observation period. The total disease burden of the study population according to our model is

$$N_{\text{exp}} = M \cdot \|\hat{\mathbf{p}}\|_1. \tag{18}$$

The quantity $\|\hat{\mathbf{p}}\|_1 / \|\mathbf{p}\|_1$ serves us as a measure for how well our model reflects the observed total disease burden of our study cohort at the end of the observation period. We also compute the quantities $\|\Delta\mathbf{p}\|_1$ and $\|\Delta\mathbf{p}\|_2$, where $\Delta\mathbf{p} = \mathbf{p} - \hat{\mathbf{p}}$. Together with the mean cluster inertia $f$, these quantities measure the performance of our algorithm in describing patient trajectories.

Benchmark method 1 consists in clustering the health states of patients according to their most recent main diagnosis. For example, if a patient had one hospital stay with main diagnosis 'Malignant neoplasms' in 2008, then her or his health state at the end of that year is assigned to the cluster where all patients had malignant neoplasms as their most recent main diagnosis. If a patient received more than one main diagnosis in a given year, then her or his health state at the end of this year is assigned to the cluster of patients with the less frequent main diagnosis. An additional cluster contains all patients who have not had any diagnoses yet.

Benchmark method 2 consists in clustering patients according to the rarest disease they have been diagnosed with so far. For example, if a patient has been diagnosed with diagnoses from the block 'Hypertensive diseases' and 'Ischaemic heart diseases', then he or she is a assigned to the cluster of patients who have 'Ischaemic heart diseases' as their rarest disease, because diagnoses of this block are rarer than 'Hypertensive diseases' diagnoses. As for Benchmark method 1, an additional cluster contains all patients who have not had any diagnoses yet.

The quantities defined above are compared in table 0. While both with DIVCLUS-T and Benchmark method 2 we attain a mean cluster inertia of 0.48, DIVCLUS-T outperforms the latter method in terms of the other metrics. Benchmark method 1 performs worse than DIVCLUS-T in all metrics.

|  | $\|\hat{\mathbf{p}}\|_1 / \|\mathbf{p}\|_1$ | $\|\Delta\mathbf{p}\|_1$ | $\|\Delta\mathbf{p}\|_2$ | $f$ |
| --- | --- | --- | --- | --- |
| DIVCLUS-T | 98% | 0.14 | 0.021 | 0.48 |
| Benchmark 1 | 88% | 0.30 | 0.039 | 0.55 |
| Benchmark 2 | 94% | 0.17 | 0.029 | 0.48 |

Table 0: Comparison of the performance of our approach to modeling patient trajectories based on clustering patients using DIVCLUS-T and two other benchmark algorithms. The meaning of the different quantities is explained in the text.

# 8 Inclusion and exclusion criteria for all clusters

In tables 1 to 131, the symbols ✓ and ✗ next to a diagnosis block indicate that this diagnosis block is an inclusion or exclusion criterion for the given cluster, respectively. In the captions we also detail the sex

distribution and the mean age of patients assigned to the given cluster. Note that we compute both means on the level of observations, and not on the level of unique patients. This means that if a patient stays in a cluster for $n$ time steps (years), then in the computation of the means, their age and sex weighs $n$ times as much as the age and sex of a patient who only stays in the cluster for one year. For cluster 0 (the "healthy cluster"), all disease blocks are exclusion criteria.

| | |
|---|---|
| Intestinal infectious diseases (A00-A09) | ✗ |
| Benign neoplasms (D10-D36) | ✗ |
| Episodic and paroxysmal disorders (G40-G47) | ✗ |
| Diseases of middle ear and mastoid (H65-H75) | ✗ |
| Hypertensive diseases (I10-I15) | ✗ |
| Acute upper respiratory infections (J00-J06) | ✗ |
| Other diseases of upper respiratory tract (J30-J39) | ✓ |
| Chronic lower respiratory diseases (J40-J47) | ✗ |
| Diseases of oral cavity, salivary glands and jaws (K00-K14) | ✗ |
| Other disorders of the musculoskeletal system and connective tissue (M95-M99) | ✗ |

Table 1: Inclusion and exclusion criteria for cluster 1. Female ratio: 46%, mean age of patients: 27.

| | |
|---|---|
| Metabolic disorders (E70-E90) | ✗ |
| Hypertensive diseases (I10-I15) | ✗ |
| Diseases of veins, lymphatic vessels and lymph nodes, not elsewhere classified (I80-I89) | ✗ |
| Other diseases of upper respiratory tract (J30-J39) | ✗ |
| Arthropathies (M00-M25) | ✓ |
| Dorsopathies (M40-M54) | ✗ |
| Soft tissue disorders (M60-M79) | ✗ |
| Osteopathies and chondropathies (M80-M94) | ✗ |

Table 2: Inclusion and exclusion criteria for cluster 2. Female ratio: 47%, mean age of patients: 50.

| | |
|---|---|
| Benign neoplasms (D10-D36) | ✗ |
| Hypertensive diseases (I10-I15) | ✗ |
| Other diseases of upper respiratory tract (J30-J39) | ✗ |
| Arthropathies (M00-M25) | ✗ |
| Other diseases of urinary system (N30-N39) | ✗ |
| Inflammatory diseases of female pelvic organs (N70-N77) | ✗ |
| Noninflammatory disorders of female genital tract (N80-N98) | ✓ |

Table 3: Inclusion and exclusion criteria for cluster 3. Female ratio: 100%, mean age of patients: 43.

| | |
|---|---|
| Intestinal infectious diseases (A00-A09) | ✗ |
| Metabolic disorders (E70-E90) | ✗ |
| Mental and behavioural disorders due to psychoactive substance use (F10-F19) | ✗ |
| Mood [affective] disorders (F30-F39) | ✗ |
| Hypertensive diseases (I10-I15) | ✗ |
| Other diseases of upper respiratory tract (J30-J39) | ✗ |
| Other diseases of intestines (K55-K63) | ✗ |
| Arthropathies (M00-M25) | ✗ |
| Dorsopathies (M40-M54) | ✓ |
| Noninflammatory disorders of female genital tract (N80-N98) | ✗ |

Table 4: Inclusion and exclusion criteria for cluster 4. Female ratio: 48%, mean age of patients: 48.

| | |
|---|---|
| Intestinal infectious diseases (A00-A09) | ✗ |
| Mental and behavioural disorders due to psychoactive substance use (F10-F19) | ✗ |
| Hypertensive diseases (I10-I15) | ✗ |
| Diseases of veins, lymphatic vessels and lymph nodes, not elsewhere classified (I80-I89) | ✓ |
| Other diseases of upper respiratory tract (J30-J39) | ✗ |
| Diseases of oesophagus, stomach and duodenum (K20-K31) | ✗ |
| Other diseases of intestines (K55-K63) | ✗ |
| Arthropathies (M00-M25) | ✗ |
| Dorsopathies (M40-M54) | ✗ |
| Noninflammatory disorders of female genital tract (N80-N98) | ✗ |

Table 5: Inclusion and exclusion criteria for cluster 5. Female ratio: 55%, mean age of patients: 48.

| | |
|---|---|
| Intestinal infectious diseases (A00-A09) | ✓ |
| Metabolic disorders (E70-E90) | ✗ |
| Hypertensive diseases (I10-I15) | ✗ |
| Acute upper respiratory infections (J00-J06) | ✗ |
| Other acute lower respiratory infections (J20-J22) | ✗ |
| Other diseases of upper respiratory tract (J30-J39) | ✗ |
| Arthropathies (M00-M25) | ✗ |
| Noninflammatory disorders of female genital tract (N80-N98) | ✗ |

Table 6: Inclusion and exclusion criteria for cluster 6. Female ratio: 52%, mean age of patients: 21.

| | |
|---|---|
| Intestinal infectious diseases (A00-A09) | ✗ |
| Other bacterial diseases (A30-A49) | ✗ |
| Infections with a predominantly sexual mode of transmission (A50-A64) | ✗ |
| Viral infections characterized by skin and mucous membrane lesions (B00-B09) | ✗ |
| Other viral diseases (B25-B34) | ✗ |
| Malignant neoplasms (C00-C97) | ✗ |
| In situ neoplasms (D00-D09) | ✗ |
| Benign neoplasms (D10-D36) | ✗ |
| Neoplasms of uncertain or unknown behaviour (D37-D48) | ✗ |
| Disorders of thyroid gland (E00-E07) | ✗ |
| Diabetes mellitus (E10-E14) | ✗ |
| Obesity and other hyperalimentation (E65-E68) | ✗ |
| Metabolic disorders (E70-E90) | ✗ |
| Mental and behavioural disorders due to psychoactive substance use (F10-F19) | ✗ |
| Schizophrenia, schizotypal and delusional disorders (F20-F29) | ✗ |
| Mood [affective] disorders (F30-F39) | ✗ |
| Neurotic, stress-related and somatoform disorders (F40-F48) | ✗ |
| Behavioural and emotional disorders with onset usually occurring in childhood and adolescence (F90-F98) | ✗ |
| Episodic and paroxysmal disorders (G40-G47) | ✗ |
| Nerve, nerve root and plexus disorders (G50-G59) | ✗ |
| Disorders of eyelid, lacrimal system and orbit (H00-H06) | ✗ |
| Disorders of lens (H25-H28) | ✗ |
| Disorders of choroid and retina (H30-H36) | ✗ |
| Disorders of ocular muscles, binocular movement, accommodation and refraction (H49-H52) | ✗ |
| Diseases of middle ear and mastoid (H65-H75) | ✗ |
| Diseases of inner ear (H80-H83) | ✗ |
| Other disorders of ear (H90-H95) | ✗ |
| Hypertensive diseases (I10-I15) | ✗ |
| Ischaemic heart diseases (I20-I25) | ✗ |
| Other forms of heart disease (I30-I52) | ✗ |
| Cerebrovascular diseases (I60-I69) | ✗ |
| Diseases of veins, lymphatic vessels and lymph nodes, not elsewhere classified (I80-I89) | ✗ |
| Other and unspecified disorders of the circulatory system (I95-I99) | ✗ |
| Acute upper respiratory infections (J00-J06) | ✗ |
| Influenza and pneumonia (J09-J18) | ✗ |
| Other acute lower respiratory infections (J20-J22) | ✗ |
| Other diseases of upper respiratory tract (J30-J39) | ✗ |
| Chronic lower respiratory diseases (J40-J47) | ✗ |
| Diseases of oral cavity, salivary glands and jaws (K00-K14) | ✗ |
| Diseases of oesophagus, stomach and duodenum (K20-K31) | ✗ |
| Diseases of appendix (K35-K38) | ✗ |
| Hernia (K40-K46) | ✗ |
| Noninfective enteritis and colitis (K50-K52) | ✗ |
| Other diseases of intestines (K55-K63) | ✗ |
| Disorders of gallbladder, biliary tract and pancreas (K80-K87) | ✗ |
| Infections of the skin and subcutaneous tissue (L00-L08) | ✗ |
| Dermatitis and eczema (L20-L30) | ✗ |
| Urticaria and erythema (L50-L54) | ✗ |
| Disorders of skin appendages (L60-L75) | ✗ |
| Other disorders of the skin and subcutaneous tissue (L80-L99) | ✗ |
| Arthropathies (M00-M25) | ✗ |
| Dorsopathies (M40-M54) | ✗ |
| Soft tissue disorders (M60-M79) | ✗ |
| Osteopathies and chondropathies (M80-M94) | ✗ |
| Renal tubulo-interstitial diseases (N10-N16) | ✗ |
| Urolithiasis (N20-N23) | ✗ |
| Other diseases of urinary system (N30-N39) | ✗ |
| Diseases of male genital organs (N40-N51) | ✗ |
| Disorders of breast (N60-N64) | ✗ |
| Inflammatory diseases of female pelvic organs (N70-N77) | ✗ |
| Noninflammatory disorders of female genital tract (N80-N98) | ✗ |

Table 7: Inclusion and exclusion criteria for cluster 7. Female ratio: 50%, mean age of patients: 35.

| | |
|---|---|
| Intestinal infectious diseases (A00-A09) | ✗ |
| Malignant neoplasms (C00-C97) | ✗ |
| Hypertensive diseases (I10-I15) | ✗ |
| Diseases of veins, lymphatic vessels and lymph nodes, not elsewhere classified (I80-I89) | ✗ |
| Other diseases of upper respiratory tract (J30-J39) | ✗ |
| Diseases of oesophagus, stomach and duodenum (K20-K31) | ✗ |
| Hernia (K40-K46) | ✗ |
| Other diseases of intestines (K55-K63) | ✓ |
| Arthropathies (M00-M25) | ✗ |
| Noninflammatory disorders of female genital tract (N80-N98) | ✗ |

Table 8: Inclusion and exclusion criteria for cluster 8. Female ratio: 47%, mean age of patients: 45.

| | |
|---|---|
| Intestinal infectious diseases (A00-A09) | ✗ |
| Mental and behavioural disorders due to psychoactive substance use (F10-F19) | ✗ |
| Hypertensive diseases (I10-I15) | ✗ |
| Diseases of veins, lymphatic vessels and lymph nodes, not elsewhere classified (I80-I89) | ✗ |
| Other diseases of upper respiratory tract (J30-J39) | ✗ |
| Other diseases of intestines (K55-K63) | ✗ |
| Arthropathies (M00-M25) | ✗ |
| Dorsopathies (M40-M54) | ✗ |
| Other diseases of urinary system (N30-N39) | ✗ |
| Diseases of male genital organs (N40-N51) | ✓ |
| Noninflammatory disorders of female genital tract (N80-N98) | ✗ |

Table 9: Inclusion and exclusion criteria for cluster 9. Female ratio: 0%, mean age of patients: 30.

| | |
|---|---|
| Malignant neoplasms (C00-C97) | ✗ |
| Diabetes mellitus (E10-E14) | ✗ |
| Obesity and other hyperalimentation (E65-E68) | ✗ |
| Metabolic disorders (E70-E90) | ✗ |
| Disorders of lens (H25-H28) | ✗ |
| Hypertensive diseases (I10-I15) | ✓ |
| Ischaemic heart diseases (I20-I25) | ✗ |
| Other forms of heart disease (I30-I52) | ✗ |
| Cerebrovascular diseases (I60-I69) | ✗ |
| Diseases of oesophagus, stomach and duodenum (K20-K31) | ✗ |
| Arthropathies (M00-M25) | ✗ |
| Dorsopathies (M40-M54) | ✗ |

Table 10: Inclusion and exclusion criteria for cluster 10. Female ratio: 53%, mean age of patients: 61.

| | |
|---|---|
| Intestinal infectious diseases (A00-A09) | ✗ |
| Mental and behavioural disorders due to psychoactive substance use (F10-F19) | ✗ |
| Hypertensive diseases (I10-I15) | ✗ |
| Diseases of veins, lymphatic vessels and lymph nodes, not elsewhere classified (I80-I89) | ✗ |
| Other diseases of upper respiratory tract (J30-J39) | ✗ |
| Diseases of oesophagus, stomach and duodenum (K20-K31) | ✗ |
| Hernia (K40-K46) | ✓ |
| Other diseases of intestines (K55-K63) | ✗ |
| Arthropathies (M00-M25) | ✗ |
| Dorsopathies (M40-M54) | ✗ |
| Diseases of male genital organs (N40-N51) | ✗ |
| Noninflammatory disorders of female genital tract (N80-N98) | ✗ |

Table 11: Inclusion and exclusion criteria for cluster 11. Female ratio: 24%, mean age of patients: 43.

| | |
|---|---|
| Intestinal infectious diseases (A00-A09) | ✗ |
| Malignant neoplasms (C00-C97) | ✓ |
| Neoplasms of uncertain or unknown behaviour (D37-D48) | ✗ |
| Mental and behavioural disorders due to psychoactive substance use (F10-F19) | ✗ |
| Hypertensive diseases (I10-I15) | ✗ |
| Diseases of veins, lymphatic vessels and lymph nodes, not elsewhere classified (I80-I89) | ✗ |
| Other diseases of upper respiratory tract (J30-J39) | ✗ |
| Hernia (K40-K46) | ✗ |
| Other diseases of intestines (K55-K63) | ✗ |
| Arthropathies (M00-M25) | ✗ |
| Dorsopathies (M40-M54) | ✗ |
| Diseases of male genital organs (N40-N51) | ✗ |
| Noninflammatory disorders of female genital tract (N80-N98) | ✗ |

Table 12: Inclusion and exclusion criteria for cluster 12. Female ratio: 51%, mean age of patients: 61.

| | |
|---|---|
| Intestinal infectious diseases (A00-A09) | ✗ |
| Malignant neoplasms (C00-C97) | ✗ |
| Benign neoplasms (D10-D36) | ✓ |
| Mental and behavioural disorders due to psychoactive substance use (F10-F19) | ✗ |
| Hypertensive diseases (I10-I15) | ✗ |
| Diseases of veins, lymphatic vessels and lymph nodes, not elsewhere classified (I80-I89) | ✗ |
| Acute upper respiratory infections (J00-J06) | ✗ |
| Other diseases of upper respiratory tract (J30-J39) | ✗ |
| Diseases of appendix (K35-K38) | ✗ |
| Hernia (K40-K46) | ✗ |
| Other diseases of intestines (K55-K63) | ✗ |
| Arthropathies (M00-M25) | ✗ |
| Dorsopathies (M40-M54) | ✗ |
| Diseases of male genital organs (N40-N51) | ✗ |
| Noninflammatory disorders of female genital tract (N80-N98) | ✗ |

Table 13: Inclusion and exclusion criteria for cluster 13. Female ratio: 67%, mean age of patients: 42.

| | |
|---|---|
| Intestinal infectious diseases (A00-A09) | ✗ |
| Metabolic disorders (E70-E90) | ✗ |
| Mental and behavioural disorders due to psychoactive substance use (F10-F19) | ✓ |
| Mood [affective] disorders (F30-F39) | ✗ |
| Neurotic, stress-related and somatoform disorders (F40-F48) | ✗ |
| Hypertensive diseases (I10-I15) | ✗ |
| Other diseases of upper respiratory tract (J30-J39) | ✗ |
| Chronic lower respiratory diseases (J40-J47) | ✗ |
| Other diseases of intestines (K55-K63) | ✗ |
| Diseases of liver (K70-K77) | ✗ |
| Arthropathies (M00-M25) | ✗ |
| Noninflammatory disorders of female genital tract (N80-N98) | ✗ |

Table 14: Inclusion and exclusion criteria for cluster 14. Female ratio: 30%, mean age of patients: 37.

| | |
|---|---|
| Intestinal infectious diseases (A00-A09) | ✗ |
| Malignant neoplasms (C00-C97) | ✗ |
| Benign neoplasms (D10-D36) | ✗ |
| Mental and behavioural disorders due to psychoactive substance use (F10-F19) | ✗ |
| Hypertensive diseases (I10-I15) | ✗ |
| Diseases of veins, lymphatic vessels and lymph nodes, not elsewhere classified (I80-I89) | ✗ |
| Acute upper respiratory infections (J00-J06) | ✗ |
| Other diseases of upper respiratory tract (J30-J39) | ✗ |
| Diseases of appendix (K35-K38) | ✗ |
| Hernia (K40-K46) | ✗ |
| Other diseases of intestines (K55-K63) | ✗ |
| Arthropathies (M00-M25) | ✗ |
| Dorsopathies (M40-M54) | ✗ |
| Soft tissue disorders (M60-M79) | ✓ |
| Diseases of male genital organs (N40-N51) | ✗ |
| Noninflammatory disorders of female genital tract (N80-N98) | ✗ |

Table 15: Inclusion and exclusion criteria for cluster 15. Female ratio: 45%, mean age of patients: 45.

| | |
|---|---|
| Intestinal infectious diseases (A00-A09) | ✗ |
| Malignant neoplasms (C00-C97) | ✗ |
| Mental and behavioural disorders due to psychoactive substance use (F10-F19) | ✗ |
| Hypertensive diseases (I10-I15) | ✗ |
| Diseases of veins, lymphatic vessels and lymph nodes, not elsewhere classified (I80-I89) | ✗ |
| Acute upper respiratory infections (J00-J06) | ✓ |
| Other acute lower respiratory infections (J20-J22) | ✗ |
| Other diseases of upper respiratory tract (J30-J39) | ✗ |
| Hernia (K40-K46) | ✗ |
| Other diseases of intestines (K55-K63) | ✗ |
| Arthropathies (M00-M25) | ✗ |
| Dorsopathies (M40-M54) | ✗ |
| Diseases of male genital organs (N40-N51) | ✗ |
| Noninflammatory disorders of female genital tract (N80-N98) | ✗ |

Table 16: Inclusion and exclusion criteria for cluster 16. Female ratio: 50%, mean age of patients: 18.

| | |
|---|---|
| Intestinal infectious diseases (A00-A09) | ✗ |
| Malignant neoplasms (C00-C97) | ✗ |
| Mental and behavioural disorders due to psychoactive substance use (F10-F19) | ✗ |
| Hypertensive diseases (I10-I15) | ✗ |
| Diseases of veins, lymphatic vessels and lymph nodes, not elsewhere classified (I80-I89) | ✗ |
| Acute upper respiratory infections (J00-J06) | ✗ |
| Other diseases of upper respiratory tract (J30-J39) | ✗ |
| Diseases of appendix (K35-K38) | ✓ |
| Hernia (K40-K46) | ✗ |
| Other diseases of intestines (K55-K63) | ✗ |
| Arthropathies (M00-M25) | ✗ |
| Dorsopathies (M40-M54) | ✗ |
| Diseases of male genital organs (N40-N51) | ✗ |
| Noninflammatory disorders of female genital tract (N80-N98) | ✗ |

Table 17: Inclusion and exclusion criteria for cluster 17. Female ratio: 52%, mean age of patients: 26.

| | |
|---|---|
| Intestinal infectious diseases (A00-A09) | ✗ |
| Malignant neoplasms (C00-C97) | ✗ |
| Benign neoplasms (D10-D36) | ✗ |
| Mental and behavioural disorders due to psychoactive substance use (F10-F19) | ✗ |
| Episodic and paroxysmal disorders (G40-G47) | ✓ |
| Hypertensive diseases (I10-I15) | ✗ |
| Diseases of veins, lymphatic vessels and lymph nodes, not elsewhere classified (I80-I89) | ✗ |
| Acute upper respiratory infections (J00-J06) | ✗ |
| Other diseases of upper respiratory tract (J30-J39) | ✗ |
| Diseases of appendix (K35-K38) | ✗ |
| Hernia (K40-K46) | ✗ |
| Other diseases of intestines (K55-K63) | ✗ |
| Arthropathies (M00-M25) | ✗ |
| Dorsopathies (M40-M54) | ✗ |
| Soft tissue disorders (M60-M79) | ✗ |
| Diseases of male genital organs (N40-N51) | ✗ |
| Noninflammatory disorders of female genital tract (N80-N98) | ✗ |

Table 18: Inclusion and exclusion criteria for cluster 18. Female ratio: 46%, mean age of patients: 40.

| | |
|---|---|
| Diseases of middle ear and mastoid (H65-H75) | ✓ |
| Hypertensive diseases (I10-I15) | ✗ |
| Acute upper respiratory infections (J00-J06) | ✗ |
| Other diseases of upper respiratory tract (J30-J39) | ✓ |

Table 19: Inclusion and exclusion criteria for cluster 19. Female ratio: 42%, mean age of patients: 10.

| | |
|---|---|
| Intestinal infectious diseases (A00-A09) | ✗ |
| Malignant neoplasms (C00-C97) | ✗ |
| Benign neoplasms (D10-D36) | ✗ |
| Mental and behavioural disorders due to psychoactive substance use (F10-F19) | ✗ |
| Episodic and paroxysmal disorders (G40-G47) | ✗ |
| Disorders of lens (H25-H28) | ✗ |
| Hypertensive diseases (I10-I15) | ✗ |
| Other forms of heart disease (I30-I52) | ✗ |
| Diseases of veins, lymphatic vessels and lymph nodes, not elsewhere classified (I80-I89) | ✗ |
| Acute upper respiratory infections (J00-J06) | ✗ |
| Other diseases of upper respiratory tract (J30-J39) | ✗ |
| Diseases of oral cavity, salivary glands and jaws (K00-K14) | ✗ |
| Diseases of appendix (K35-K38) | ✗ |
| Hernia (K40-K46) | ✗ |
| Other diseases of intestines (K55-K63) | ✗ |
| Disorders of gallbladder, biliary tract and pancreas (K80-K87) | ✓ |
| Infections of the skin and subcutaneous tissue (L00-L08) | ✗ |
| Arthropathies (M00-M25) | ✗ |
| Dorsopathies (M40-M54) | ✗ |
| Soft tissue disorders (M60-M79) | ✗ |
| Diseases of male genital organs (N40-N51) | ✗ |
| Noninflammatory disorders of female genital tract (N80-N98) | ✗ |

Table 20: Inclusion and exclusion criteria for cluster 20. Female ratio: 65%, mean age of patients: 49.

| | |
|---|---|
| Intestinal infectious diseases (A00-A09) | ✗ |
| Malignant neoplasms (C00-C97) | ✗ |
| Benign neoplasms (D10-D36) | ✗ |
| Mental and behavioural disorders due to psychoactive substance use (F10-F19) | ✗ |
| Episodic and paroxysmal disorders (G40-G47) | ✗ |
| Hypertensive diseases (I10-I15) | ✗ |
| Ischaemic heart diseases (I20-I25) | ✗ |
| Other forms of heart disease (I30-I52) | ✓ |
| Diseases of veins, lymphatic vessels and lymph nodes, not elsewhere classified (I80-I89) | ✗ |
| Acute upper respiratory infections (J00-J06) | ✗ |
| Other diseases of upper respiratory tract (J30-J39) | ✗ |
| Diseases of appendix (K35-K38) | ✗ |
| Hernia (K40-K46) | ✗ |
| Other diseases of intestines (K55-K63) | ✗ |
| Arthropathies (M00-M25) | ✗ |
| Dorsopathies (M40-M54) | ✗ |
| Soft tissue disorders (M60-M79) | ✗ |
| Diseases of male genital organs (N40-N51) | ✗ |
| Noninflammatory disorders of female genital tract (N80-N98) | ✗ |

Table 21: Inclusion and exclusion criteria for cluster 21. Female ratio: 51%, mean age of patients: 61.

| | |
|---|---|
| Intestinal infectious diseases (A00-A09) | ✗ |
| Malignant neoplasms (C00-C97) | ✗ |
| Benign neoplasms (D10-D36) | ✗ |
| Mental and behavioural disorders due to psychoactive substance use (F10-F19) | ✗ |
| Episodic and paroxysmal disorders (G40-G47) | ✗ |
| Hypertensive diseases (I10-I15) | ✗ |
| Other forms of heart disease (I30-I52) | ✗ |
| Diseases of veins, lymphatic vessels and lymph nodes, not elsewhere classified (I80-I89) | ✗ |
| Acute upper respiratory infections (J00-J06) | ✗ |
| Other diseases of upper respiratory tract (J30-J39) | ✗ |
| Diseases of oral cavity, salivary glands and jaws (K00-K14) | ✓ |
| Diseases of appendix (K35-K38) | ✗ |
| Hernia (K40-K46) | ✗ |
| Other diseases of intestines (K55-K63) | ✗ |
| Arthropathies (M00-M25) | ✗ |
| Dorsopathies (M40-M54) | ✗ |
| Soft tissue disorders (M60-M79) | ✗ |
| Diseases of male genital organs (N40-N51) | ✗ |
| Noninflammatory disorders of female genital tract (N80-N98) | ✗ |

Table 22: Inclusion and exclusion criteria for cluster 22. Female ratio: 52%, mean age of patients: 25.

| | |
|---|---|
| Intestinal infectious diseases (A00-A09) | ✗ |
| Malignant neoplasms (C00-C97) | ✗ |
| Benign neoplasms (D10-D36) | ✗ |
| Mental and behavioural disorders due to psychoactive substance use (F10-F19) | ✗ |
| Neurotic, stress-related and somatoform disorders (F40-F48) | ✗ |
| Episodic and paroxysmal disorders (G40-G47) | ✗ |
| Disorders of lens (H25-H28) | ✗ |
| Hypertensive diseases (I10-I15) | ✗ |
| Other forms of heart disease (I30-I52) | ✗ |
| Diseases of veins, lymphatic vessels and lymph nodes, not elsewhere classified (I80-I89) | ✗ |
| Acute upper respiratory infections (J00-J06) | ✗ |
| Other diseases of upper respiratory tract (J30-J39) | ✗ |
| Diseases of oral cavity, salivary glands and jaws (K00-K14) | ✗ |
| Diseases of appendix (K35-K38) | ✗ |
| Hernia (K40-K46) | ✗ |
| Other diseases of intestines (K55-K63) | ✗ |
| Disorders of gallbladder, biliary tract and pancreas (K80-K87) | ✗ |
| Infections of the skin and subcutaneous tissue (L00-L08) | ✗ |
| Arthropathies (M00-M25) | ✗ |
| Dorsopathies (M40-M54) | ✗ |
| Soft tissue disorders (M60-M79) | ✗ |
| Urolithiasis (N20-N23) | ✗ |
| Other diseases of urinary system (N30-N39) | ✓ |
| Diseases of male genital organs (N40-N51) | ✗ |
| Noninflammatory disorders of female genital tract (N80-N98) | ✗ |

Table 23: Inclusion and exclusion criteria for cluster 23. Female ratio: 79%, mean age of patients: 37.

| | |
|---|---|
| Diabetes mellitus (E10-E14) | ✓ |
| Obesity and other hyperalimentation (E65-E68) | ✗ |
| Metabolic disorders (E70-E90) | ✗ |
| Hypertensive diseases (I10-I15) | ✓ |
| Other forms of heart disease (I30-I52) | ✗ |

Table 24: Inclusion and exclusion criteria for cluster 24. Female ratio: 49%, mean age of patients: 71.

| | |
|---|---|
| Intestinal infectious diseases (A00-A09) | ✗ |
| Malignant neoplasms (C00-C97) | ✗ |
| Benign neoplasms (D10-D36) | ✗ |
| Mental and behavioural disorders due to psychoactive substance use (F10-F19) | ✗ |
| Episodic and paroxysmal disorders (G40-G47) | ✗ |
| Disorders of lens (H25-H28) | ✓ |
| Disorders of choroid and retina (H30-H36) | ✗ |
| Hypertensive diseases (I10-I15) | ✗ |
| Other forms of heart disease (I30-I52) | ✗ |
| Diseases of veins, lymphatic vessels and lymph nodes, not elsewhere classified (I80-I89) | ✗ |
| Acute upper respiratory infections (J00-J06) | ✗ |
| Other diseases of upper respiratory tract (J30-J39) | ✗ |
| Diseases of oral cavity, salivary glands and jaws (K00-K14) | ✗ |
| Diseases of appendix (K35-K38) | ✗ |
| Hernia (K40-K46) | ✗ |
| Other diseases of intestines (K55-K63) | ✗ |
| Arthropathies (M00-M25) | ✗ |
| Dorsopathies (M40-M54) | ✗ |
| Soft tissue disorders (M60-M79) | ✗ |
| Diseases of male genital organs (N40-N51) | ✗ |
| Noninflammatory disorders of female genital tract (N80-N98) | ✗ |

Table 25: Inclusion and exclusion criteria for cluster 25. Female ratio: 58%, mean age of patients: 71.

| | |
|---|---|
| Diabetes mellitus (E10-E14) | ✗ |
| Metabolic disorders (E70-E90) | ✓ |
| Hypertensive diseases (I10-I15) | ✓ |
| Ischaemic heart diseases (I20-I25) | ✗ |
| Other forms of heart disease (I30-I52) | ✗ |
| Cerebrovascular diseases (I60-I69) | ✗ |
| Diseases of oesophagus, stomach and duodenum (K20-K31) | ✗ |
| Dorsopathies (M40-M54) | ✗ |

Table 26: Inclusion and exclusion criteria for cluster 26. Female ratio: 51%, mean age of patients: 65.

| | |
|---|---|
| Intestinal infectious diseases (A00-A09) | ✗ |
| Malignant neoplasms (C00-C97) | ✗ |
| Benign neoplasms (D10-D36) | ✗ |
| Mental and behavioural disorders due to psychoactive substance use (F10-F19) | ✗ |
| Episodic and paroxysmal disorders (G40-G47) | ✗ |
| Disorders of lens (H25-H28) | ✗ |
| Hypertensive diseases (I10-I15) | ✗ |
| Other forms of heart disease (I30-I52) | ✗ |
| Diseases of veins, lymphatic vessels and lymph nodes, not elsewhere classified (I80-I89) | ✗ |
| Acute upper respiratory infections (J00-J06) | ✗ |
| Other diseases of upper respiratory tract (J30-J39) | ✗ |
| Diseases of oral cavity, salivary glands and jaws (K00-K14) | ✗ |
| Diseases of appendix (K35-K38) | ✗ |
| Hernia (K40-K46) | ✗ |
| Other diseases of intestines (K55-K63) | ✗ |
| Infections of the skin and subcutaneous tissue (L00-L08) | ✓ |
| Arthropathies (M00-M25) | ✗ |
| Dorsopathies (M40-M54) | ✗ |
| Soft tissue disorders (M60-M79) | ✗ |
| Diseases of male genital organs (N40-N51) | ✗ |
| Noninflammatory disorders of female genital tract (N80-N98) | ✗ |

Table 27: Inclusion and exclusion criteria for cluster 27. Female ratio: 34%, mean age of patients: 31.

| | |
|---|---|
| Intestinal infectious diseases (A00-A09) | ✗ |
| Malignant neoplasms (C00-C97) | ✗ |
| Benign neoplasms (D10-D36) | ✗ |
| Mental and behavioural disorders due to psychoactive substance use (F10-F19) | ✗ |
| Neurotic, stress-related and somatoform disorders (F40-F48) | ✓ |
| Episodic and paroxysmal disorders (G40-G47) | ✗ |
| Disorders of lens (H25-H28) | ✗ |
| Hypertensive diseases (I10-I15) | ✗ |
| Other forms of heart disease (I30-I52) | ✗ |
| Diseases of veins, lymphatic vessels and lymph nodes, not elsewhere classified (I80-I89) | ✗ |
| Acute upper respiratory infections (J00-J06) | ✗ |
| Other diseases of upper respiratory tract (J30-J39) | ✗ |
| Diseases of oral cavity, salivary glands and jaws (K00-K14) | ✗ |
| Diseases of appendix (K35-K38) | ✗ |
| Hernia (K40-K46) | ✗ |
| Other diseases of intestines (K55-K63) | ✗ |
| Disorders of gallbladder, biliary tract and pancreas (K80-K87) | ✗ |
| Infections of the skin and subcutaneous tissue (L00-L08) | ✗ |
| Arthropathies (M00-M25) | ✗ |
| Dorsopathies (M40-M54) | ✗ |
| Soft tissue disorders (M60-M79) | ✗ |
| Diseases of male genital organs (N40-N51) | ✗ |
| Noninflammatory disorders of female genital tract (N80-N98) | ✗ |

Table 28: Inclusion and exclusion criteria for cluster 28. Female ratio: 60%, mean age of patients: 34.

| | |
|---|---|
| Intestinal infectious diseases (A00-A09) | ✗ |
| Malignant neoplasms (C00-C97) | ✗ |
| Benign neoplasms (D10-D36) | ✗ |
| Mental and behavioural disorders due to psychoactive substance use (F10-F19) | ✗ |
| Neurotic, stress-related and somatoform disorders (F40-F48) | ✗ |
| Episodic and paroxysmal disorders (G40-G47) | ✗ |
| Disorders of lens (H25-H28) | ✗ |
| Hypertensive diseases (I10-I15) | ✗ |
| Other forms of heart disease (I30-I52) | ✗ |
| Diseases of veins, lymphatic vessels and lymph nodes, not elsewhere classified (I80-I89) | ✗ |
| Acute upper respiratory infections (J00-J06) | ✗ |
| Influenza and pneumonia (J09-J18) | ✗ |
| Other diseases of upper respiratory tract (J30-J39) | ✗ |
| Diseases of oral cavity, salivary glands and jaws (K00-K14) | ✗ |
| Diseases of oesophagus, stomach and duodenum (K20-K31) | ✓ |
| Diseases of appendix (K35-K38) | ✗ |
| Hernia (K40-K46) | ✗ |
| Noninfective enteritis and colitis (K50-K52) | ✗ |
| Other diseases of intestines (K55-K63) | ✗ |
| Disorders of gallbladder, biliary tract and pancreas (K80-K87) | ✗ |
| Infections of the skin and subcutaneous tissue (L00-L08) | ✗ |
| Arthropathies (M00-M25) | ✗ |
| Dorsopathies (M40-M54) | ✗ |
| Soft tissue disorders (M60-M79) | ✗ |
| Urolithiasis (N20-N23) | ✗ |
| Other diseases of urinary system (N30-N39) | ✗ |
| Diseases of male genital organs (N40-N51) | ✗ |
| Noninflammatory disorders of female genital tract (N80-N98) | ✗ |

Table 29: Inclusion and exclusion criteria for cluster 29. Female ratio: 48%, mean age of patients: 42.

| | |
|---|---|
| Intestinal infectious diseases (A00-A09) | ✗ |
| Malignant neoplasms (C00-C97) | ✗ |
| Benign neoplasms (D10-D36) | ✗ |
| Mental and behavioural disorders due to psychoactive substance use (F10-F19) | ✗ |
| Neurotic, stress-related and somatoform disorders (F40-F48) | ✗ |
| Episodic and paroxysmal disorders (G40-G47) | ✗ |
| Disorders of lens (H25-H28) | ✗ |
| Hypertensive diseases (I10-I15) | ✗ |
| Other forms of heart disease (I30-I52) | ✗ |
| Diseases of veins, lymphatic vessels and lymph nodes, not elsewhere classified (I80-I89) | ✗ |
| Acute upper respiratory infections (J00-J06) | ✗ |
| Influenza and pneumonia (J09-J18) | ✓ |
| Other diseases of upper respiratory tract (J30-J39) | ✗ |
| Diseases of oral cavity, salivary glands and jaws (K00-K14) | ✗ |
| Diseases of appendix (K35-K38) | ✗ |
| Hernia (K40-K46) | ✗ |
| Other diseases of intestines (K55-K63) | ✗ |
| Disorders of gallbladder, biliary tract and pancreas (K80-K87) | ✗ |
| Infections of the skin and subcutaneous tissue (L00-L08) | ✗ |
| Arthropathies (M00-M25) | ✗ |
| Dorsopathies (M40-M54) | ✗ |
| Soft tissue disorders (M60-M79) | ✗ |
| Urolithiasis (N20-N23) | ✗ |
| Other diseases of urinary system (N30-N39) | ✗ |
| Diseases of male genital organs (N40-N51) | ✗ |
| Noninflammatory disorders of female genital tract (N80-N98) | ✗ |

Table 30: Inclusion and exclusion criteria for cluster 30. Female ratio: 47%, mean age of patients: 33.

| | |
|---|---|
| Hypertensive diseases (I10-I15) | ✗ |
| Other diseases of upper respiratory tract (J30-J39) | ✗ |
| Arthropathies (M00-M25) | ✓ |
| Soft tissue disorders (M60-M79) | ✓ |
| Osteopathies and chondropathies (M80-M94) | ✗ |

Table 31: Inclusion and exclusion criteria for cluster 31. Female ratio: 47%, mean age of patients: 52.

| | |
|---|---|
| Diabetes mellitus (E10-E14) | ✗ |
| Metabolic disorders (E70-E90) | ✗ |
| Hypertensive diseases (I10-I15) | ✓ |
| Ischaemic heart diseases (I20-I25) | ✗ |
| Other forms of heart disease (I30-I52) | ✓ |
| Cerebrovascular diseases (I60-I69) | ✗ |
| Renal failure (N17-N19) | ✗ |

Table 32: Inclusion and exclusion criteria for cluster 32. Female ratio: 55%, mean age of patients: 74.

| | |
|---|---|
| Benign neoplasms (D10-D36) | ✓ |
| Hypertensive diseases (I10-I15) | ✗ |
| Other diseases of upper respiratory tract (J30-J39) | ✗ |
| Arthropathies (M00-M25) | ✗ |
| Noninflammatory disorders of female genital tract (N80-N98) | ✓ |

Table 33: Inclusion and exclusion criteria for cluster 33. Female ratio: 100%, mean age of patients: 48.

| | |
|---|---|
| Diabetes mellitus (E10-E14) | ✗ |
| Metabolic disorders (E70-E90) | ✓ |
| Hypertensive diseases (I10-I15) | ✓ |
| Ischaemic heart diseases (I20-I25) | ✓ |
| Other forms of heart disease (I30-I52) | ✗ |

Table 34: Inclusion and exclusion criteria for cluster 34. Female ratio: 36%, mean age of patients: 67.

| | |
|---|---|
| Intestinal infectious diseases (A00-A09) | ✗ |
| Malignant neoplasms (C00-C97) | ✗ |
| Benign neoplasms (D10-D36) | ✗ |
| Mental and behavioural disorders due to psychoactive substance use (F10-F19) | ✗ |
| Neurotic, stress-related and somatoform disorders (F40-F48) | ✗ |
| Episodic and paroxysmal disorders (G40-G47) | ✗ |
| Disorders of lens (H25-H28) | ✗ |
| Hypertensive diseases (I10-I15) | ✗ |
| Other forms of heart disease (I30-I52) | ✗ |
| Diseases of veins, lymphatic vessels and lymph nodes, not elsewhere classified (I80-I89) | ✗ |
| Acute upper respiratory infections (J00-J06) | ✗ |
| Influenza and pneumonia (J09-J18) | ✗ |
| Other diseases of upper respiratory tract (J30-J39) | ✗ |
| Diseases of oral cavity, salivary glands and jaws (K00-K14) | ✗ |
| Diseases of appendix (K35-K38) | ✗ |
| Hernia (K40-K46) | ✗ |
| Noninfective enteritis and colitis (K50-K52) | ✓ |
| Other diseases of intestines (K55-K63) | ✗ |
| Disorders of gallbladder, biliary tract and pancreas (K80-K87) | ✗ |
| Infections of the skin and subcutaneous tissue (L00-L08) | ✗ |
| Arthropathies (M00-M25) | ✗ |
| Dorsopathies (M40-M54) | ✗ |
| Soft tissue disorders (M60-M79) | ✗ |
| Urolithiasis (N20-N23) | ✗ |
| Other diseases of urinary system (N30-N39) | ✗ |
| Diseases of male genital organs (N40-N51) | ✗ |
| Noninflammatory disorders of female genital tract (N80-N98) | ✗ |

Table 35: Inclusion and exclusion criteria for cluster 35. Female ratio: 55%, mean age of patients: 30.

| | |
|---|---|
| Diabetes mellitus (E10-E14) | ✗ |
| Metabolic disorders (E70-E90) | ✗ |
| Hypertensive diseases (I10-I15) | ✓ |
| Other forms of heart disease (I30-I52) | ✗ |
| Arthropathies (M00-M25) | ✓ |
| Dorsopathies (M40-M54) | ✗ |

Table 36: Inclusion and exclusion criteria for cluster 36. Female ratio: 59%, mean age of patients: 68.

| | |
|---|---|
| Intestinal infectious diseases (A00-A09) | ✗ |
| Hypertensive diseases (I10-I15) | ✗ |
| Other diseases of upper respiratory tract (J30-J39) | ✗ |
| Diseases of oesophagus, stomach and duodenum (K20-K31) | ✓ |
| Other diseases of intestines (K55-K63) | ✓ |
| Arthropathies (M00-M25) | ✗ |
| Noninflammatory disorders of female genital tract (N80-N98) | ✗ |

Table 37: Inclusion and exclusion criteria for cluster 37. Female ratio: 42%, mean age of patients: 59.

| | |
|---|---|
| Malignant neoplasms (C00-C97) | ✓ |
| Diabetes mellitus (E10-E14) | ✗ |
| Metabolic disorders (E70-E90) | ✗ |
| Hypertensive diseases (I10-I15) | ✓ |
| Other forms of heart disease (I30-I52) | ✗ |
| Arthropathies (M00-M25) | ✗ |

Table 38: Inclusion and exclusion criteria for cluster 38. Female ratio: 48%, mean age of patients: 71.

| | |
|---|---|
| Diabetes mellitus (E10-E14) | ✓ |
| Metabolic disorders (E70-E90) | ✓ |
| Hypertensive diseases (I10-I15) | ✓ |
| Ischaemic heart diseases (I20-I25) | ✗ |
| Other forms of heart disease (I30-I52) | ✗ |

Table 39: Inclusion and exclusion criteria for cluster 39. Female ratio: 50%, mean age of patients: 68.

| | |
|---|---|
| Intestinal infectious diseases (A00-A09) | ✗ |
| Malignant neoplasms (C00-C97) | ✗ |
| Benign neoplasms (D10-D36) | ✗ |
| Mental and behavioural disorders due to psychoactive substance use (F10-F19) | ✗ |
| Neurotic, stress-related and somatoform disorders (F40-F48) | ✗ |
| Episodic and paroxysmal disorders (G40-G47) | ✗ |
| Nerve, nerve root and plexus disorders (G50-G59) | ✓ |
| Disorders of lens (H25-H28) | ✗ |
| Hypertensive diseases (I10-I15) | ✗ |
| Other forms of heart disease (I30-I52) | ✗ |
| Diseases of veins, lymphatic vessels and lymph nodes, not elsewhere classified (I80-I89) | ✗ |
| Acute upper respiratory infections (J00-J06) | ✗ |
| Influenza and pneumonia (J09-J18) | ✗ |
| Other diseases of upper respiratory tract (J30-J39) | ✗ |
| Diseases of oral cavity, salivary glands and jaws (K00-K14) | ✗ |
| Diseases of oesophagus, stomach and duodenum (K20-K31) | ✗ |
| Diseases of appendix (K35-K38) | ✗ |
| Hernia (K40-K46) | ✗ |
| Noninfective enteritis and colitis (K50-K52) | ✗ |
| Other diseases of intestines (K55-K63) | ✗ |
| Disorders of gallbladder, biliary tract and pancreas (K80-K87) | ✗ |
| Infections of the skin and subcutaneous tissue (L00-L08) | ✗ |
| Arthropathies (M00-M25) | ✗ |
| Dorsopathies (M40-M54) | ✗ |
| Soft tissue disorders (M60-M79) | ✗ |
| Urolithiasis (N20-N23) | ✗ |
| Other diseases of urinary system (N30-N39) | ✗ |
| Diseases of male genital organs (N40-N51) | ✗ |
| Noninflammatory disorders of female genital tract (N80-N98) | ✗ |

Table 40: Inclusion and exclusion criteria for cluster 40. Female ratio: 58%, mean age of patients: 45.

| | |
|---|---|
| Organic, including symptomatic, mental disorders (F00-F09) | ✗ |
| Hypertensive diseases (I10-I15) | ✓ |
| Other forms of heart disease (I30-I52) | ✓ |
| Diseases of arteries, arterioles and capillaries (I70-I79) | ✗ |
| Renal failure (N17-N19) | ✓ |

Table 41: Inclusion and exclusion criteria for cluster 41. Female ratio: 52%, mean age of patients: 80.

| | |
|---|---|
| Intestinal infectious diseases (A00-A09) | ✗ |
| Malignant neoplasms (C00-C97) | ✗ |
| Benign neoplasms (D10-D36) | ✗ |
| Disorders of thyroid gland (E00-E07) | ✓ |
| Mental and behavioural disorders due to psychoactive substance use (F10-F19) | ✗ |
| Neurotic, stress-related and somatoform disorders (F40-F48) | ✗ |
| Episodic and paroxysmal disorders (G40-G47) | ✗ |
| Nerve, nerve root and plexus disorders (G50-G59) | ✗ |
| Disorders of lens (H25-H28) | ✗ |
| Hypertensive diseases (I10-I15) | ✗ |
| Other forms of heart disease (I30-I52) | ✗ |
| Diseases of veins, lymphatic vessels and lymph nodes, not elsewhere classified (I80-I89) | ✗ |
| Acute upper respiratory infections (J00-J06) | ✗ |
| Influenza and pneumonia (J09-J18) | ✗ |
| Other acute lower respiratory infections (J20-J22) | ✗ |
| Other diseases of upper respiratory tract (J30-J39) | ✗ |
| Diseases of oral cavity, salivary glands and jaws (K00-K14) | ✗ |
| Diseases of oesophagus, stomach and duodenum (K20-K31) | ✗ |
| Diseases of appendix (K35-K38) | ✗ |
| Hernia (K40-K46) | ✗ |
| Noninfective enteritis and colitis (K50-K52) | ✗ |
| Other diseases of intestines (K55-K63) | ✗ |
| Disorders of gallbladder, biliary tract and pancreas (K80-K87) | ✗ |
| Infections of the skin and subcutaneous tissue (L00-L08) | ✗ |
| Arthropathies (M00-M25) | ✗ |
| Dorsopathies (M40-M54) | ✗ |
| Soft tissue disorders (M60-M79) | ✗ |
| Urolithiasis (N20-N23) | ✗ |
| Other diseases of urinary system (N30-N39) | ✗ |
| Diseases of male genital organs (N40-N51) | ✗ |
| Noninflammatory disorders of female genital tract (N80-N98) | ✗ |

Table 42: Inclusion and exclusion criteria for cluster 42. Female ratio: 78%, mean age of patients: 48.

| | |
|---|---|
| Hypertensive diseases (I10-I15) | ✗ |
| Other diseases of upper respiratory tract (J30-J39) | ✗ |
| Arthropathies (M00-M25) | ✓ |
| Soft tissue disorders (M60-M79) | ✗ |
| Osteopathies and chondropathies (M80-M94) | ✓ |

Table 43: Inclusion and exclusion criteria for cluster 43. Female ratio: 55%, mean age of patients: 57.

| | |
|---|---|
| Metabolic disorders (E70-E90) | ✓ |
| Hypertensive diseases (I10-I15) | ✓ |
| Ischaemic heart diseases (I20-I25) | ✗ |
| Other forms of heart disease (I30-I52) | ✓ |
| Cerebrovascular diseases (I60-I69) | ✗ |
| Renal failure (N17-N19) | ✗ |

Table 44: Inclusion and exclusion criteria for cluster 44. Female ratio: 55%, mean age of patients: 73.

| | |
|---|---|
| Malignant neoplasms (C00-C97) | ✗ |
| Diabetes mellitus (E10-E14) | ✗ |
| Metabolic disorders (E70-E90) | ✗ |
| Hypertensive diseases (I10-I15) | ✓ |
| Other forms of heart disease (I30-I52) | ✗ |
| Arthropathies (M00-M25) | ✗ |
| Dorsopathies (M40-M54) | ✓ |

Table 45: Inclusion and exclusion criteria for cluster 45. Female ratio: 59%, mean age of patients: 64.

| | |
|---|---|
| Intestinal infectious diseases (A00-A09) | ✗ |
| Malignant neoplasms (C00-C97) | ✗ |
| Benign neoplasms (D10-D36) | ✗ |
| Disorders of thyroid gland (E00-E07) | ✗ |
| Metabolic disorders (E70-E90) | ✓ |
| Mental and behavioural disorders due to psychoactive substance use (F10-F19) | ✗ |
| Mood [affective] disorders (F30-F39) | ✗ |
| Neurotic, stress-related and somatoform disorders (F40-F48) | ✗ |
| Episodic and paroxysmal disorders (G40-G47) | ✗ |
| Nerve, nerve root and plexus disorders (G50-G59) | ✗ |
| Disorders of lens (H25-H28) | ✗ |
| Other disorders of ear (H90-H95) | ✗ |
| Hypertensive diseases (I10-I15) | ✗ |
| Other forms of heart disease (I30-I52) | ✗ |
| Diseases of veins, lymphatic vessels and lymph nodes, not elsewhere classified (I80-I89) | ✗ |
| Acute upper respiratory infections (J00-J06) | ✗ |
| Influenza and pneumonia (J09-J18) | ✗ |
| Other acute lower respiratory infections (J20-J22) | ✗ |
| Other diseases of upper respiratory tract (J30-J39) | ✗ |
| Chronic lower respiratory diseases (J40-J47) | ✗ |
| Diseases of oral cavity, salivary glands and jaws (K00-K14) | ✗ |
| Diseases of oesophagus, stomach and duodenum (K20-K31) | ✗ |
| Diseases of appendix (K35-K38) | ✗ |
| Hernia (K40-K46) | ✗ |
| Noninfective enteritis and colitis (K50-K52) | ✗ |
| Other diseases of intestines (K55-K63) | ✗ |
| Disorders of gallbladder, biliary tract and pancreas (K80-K87) | ✗ |
| Infections of the skin and subcutaneous tissue (L00-L08) | ✗ |
| Disorders of skin appendages (L60-L75) | ✗ |
| Arthropathies (M00-M25) | ✗ |
| Dorsopathies (M40-M54) | ✗ |
| Soft tissue disorders (M60-M79) | ✗ |
| Urolithiasis (N20-N23) | ✗ |
| Other diseases of urinary system (N30-N39) | ✗ |
| Diseases of male genital organs (N40-N51) | ✗ |
| Inflammatory diseases of female pelvic organs (N70-N77) | ✗ |
| Noninflammatory disorders of female genital tract (N80-N98) | ✗ |

Table 46: Inclusion and exclusion criteria for cluster 46. Female ratio: 44%, mean age of patients: 44.

| | |
|---|---|
| Intestinal infectious diseases (A00-A09) | ✗ |
| Malignant neoplasms (C00-C97) | ✗ |
| Benign neoplasms (D10-D36) | ✗ |
| Mental and behavioural disorders due to psychoactive substance use (F10-F19) | ✗ |
| Neurotic, stress-related and somatoform disorders (F40-F48) | ✗ |
| Episodic and paroxysmal disorders (G40-G47) | ✗ |
| Disorders of lens (H25-H28) | ✗ |
| Hypertensive diseases (I10-I15) | ✗ |
| Other forms of heart disease (I30-I52) | ✗ |
| Diseases of veins, lymphatic vessels and lymph nodes, not elsewhere classified (I80-I89) | ✗ |
| Acute upper respiratory infections (J00-J06) | ✗ |
| Other diseases of upper respiratory tract (J30-J39) | ✗ |
| Diseases of oral cavity, salivary glands and jaws (K00-K14) | ✗ |
| Diseases of appendix (K35-K38) | ✗ |
| Hernia (K40-K46) | ✗ |
| Other diseases of intestines (K55-K63) | ✗ |
| Disorders of gallbladder, biliary tract and pancreas (K80-K87) | ✗ |
| Infections of the skin and subcutaneous tissue (L00-L08) | ✗ |
| Arthropathies (M00-M25) | ✗ |
| Dorsopathies (M40-M54) | ✗ |
| Soft tissue disorders (M60-M79) | ✗ |
| Renal tubulo-interstitial diseases (N10-N16) | ✗ |
| Urolithiasis (N20-N23) | ✓ |
| Diseases of male genital organs (N40-N51) | ✗ |
| Noninflammatory disorders of female genital tract (N80-N98) | ✗ |

Table 47: Inclusion and exclusion criteria for cluster 47. Female ratio: 30%, mean age of patients: 44.

| | |
|---|---|
| Hypertensive diseases (I10-I15) | ✓ |
| Ischaemic heart diseases (I20-I25) | ✗ |
| Other forms of heart disease (I30-I52) | ✓ |
| Cerebrovascular diseases (I60-I69) | ✓ |
| Renal failure (N17-N19) | ✗ |

Table 48: Inclusion and exclusion criteria for cluster 48. Female ratio: 59%, mean age of patients: 80.

| | |
|---|---|
| Intestinal infectious diseases (A00-A09) | ✗ |
| Malignant neoplasms (C00-C97) | ✗ |
| Benign neoplasms (D10-D36) | ✗ |
| Mental and behavioural disorders due to psychoactive substance use (F10-F19) | ✗ |
| Neurotic, stress-related and somatoform disorders (F40-F48) | ✗ |
| Episodic and paroxysmal disorders (G40-G47) | ✗ |
| Nerve, nerve root and plexus disorders (G50-G59) | ✗ |
| Disorders of lens (H25-H28) | ✗ |
| Hypertensive diseases (I10-I15) | ✗ |
| Other forms of heart disease (I30-I52) | ✗ |
| Diseases of veins, lymphatic vessels and lymph nodes, not elsewhere classified (I80-I89) | ✗ |
| Acute upper respiratory infections (J00-J06) | ✗ |
| Influenza and pneumonia (J09-J18) | ✗ |
| Other acute lower respiratory infections (J20-J22) | ✓ |
| Other diseases of upper respiratory tract (J30-J39) | ✗ |
| Diseases of oral cavity, salivary glands and jaws (K00-K14) | ✗ |
| Diseases of oesophagus, stomach and duodenum (K20-K31) | ✗ |
| Diseases of appendix (K35-K38) | ✗ |
| Hernia (K40-K46) | ✗ |
| Noninfective enteritis and colitis (K50-K52) | ✗ |
| Other diseases of intestines (K55-K63) | ✗ |
| Disorders of gallbladder, biliary tract and pancreas (K80-K87) | ✗ |
| Infections of the skin and subcutaneous tissue (L00-L08) | ✗ |
| Arthropathies (M00-M25) | ✗ |
| Dorsopathies (M40-M54) | ✗ |
| Soft tissue disorders (M60-M79) | ✗ |
| Urolithiasis (N20-N23) | ✗ |
| Other diseases of urinary system (N30-N39) | ✗ |
| Diseases of male genital organs (N40-N51) | ✗ |
| Noninflammatory disorders of female genital tract (N80-N98) | ✗ |

Table 49: Inclusion and exclusion criteria for cluster 49. Female ratio: 45%, mean age of patients: 14.

| | |
|---|---|
| Intestinal infectious diseases (A00-A09) | ✓ |
| Hypertensive diseases (I10-I15) | ✗ |
| Acute upper respiratory infections (J00-J06) | ✓ |
| Other diseases of upper respiratory tract (J30-J39) | ✗ |
| Arthropathies (M00-M25) | ✗ |
| Noninflammatory disorders of female genital tract (N80-N98) | ✗ |

Table 50: Inclusion and exclusion criteria for cluster 50. Female ratio: 46%, mean age of patients: 9.

| | |
|---|---|
| Intestinal infectious diseases (A00-A09) | ✗ |
| Malignant neoplasms (C00-C97) | ✗ |
| Benign neoplasms (D10-D36) | ✗ |
| Disorders of thyroid gland (E00-E07) | ✗ |
| Mental and behavioural disorders due to psychoactive substance use (F10-F19) | ✗ |
| Neurotic, stress-related and somatoform disorders (F40-F48) | ✗ |
| Episodic and paroxysmal disorders (G40-G47) | ✗ |
| Nerve, nerve root and plexus disorders (G50-G59) | ✗ |
| Disorders of lens (H25-H28) | ✗ |
| Other disorders of ear (H90-H95) | ✗ |
| Hypertensive diseases (I10-I15) | ✗ |
| Other forms of heart disease (I30-I52) | ✗ |
| Diseases of veins, lymphatic vessels and lymph nodes, not elsewhere classified (I80-I89) | ✗ |
| Acute upper respiratory infections (J00-J06) | ✗ |
| Influenza and pneumonia (J09-J18) | ✗ |
| Other acute lower respiratory infections (J20-J22) | ✗ |
| Other diseases of upper respiratory tract (J30-J39) | ✗ |
| Chronic lower respiratory diseases (J40-J47) | ✓ |
| Diseases of oral cavity, salivary glands and jaws (K00-K14) | ✗ |
| Diseases of oesophagus, stomach and duodenum (K20-K31) | ✗ |
| Diseases of appendix (K35-K38) | ✗ |
| Hernia (K40-K46) | ✗ |
| Noninfective enteritis and colitis (K50-K52) | ✗ |
| Other diseases of intestines (K55-K63) | ✗ |
| Disorders of gallbladder, biliary tract and pancreas (K80-K87) | ✗ |
| Infections of the skin and subcutaneous tissue (L00-L08) | ✗ |
| Disorders of skin appendages (L60-L75) | ✗ |
| Arthropathies (M00-M25) | ✗ |
| Dorsopathies (M40-M54) | ✗ |
| Soft tissue disorders (M60-M79) | ✗ |
| Urolithiasis (N20-N23) | ✗ |
| Other diseases of urinary system (N30-N39) | ✗ |
| Diseases of male genital organs (N40-N51) | ✗ |
| Inflammatory diseases of female pelvic organs (N70-N77) | ✗ |
| Noninflammatory disorders of female genital tract (N80-N98) | ✗ |

Table 51: Inclusion and exclusion criteria for cluster 51. Female ratio: 44%, mean age of patients: 33.

| | |
|---|---|
| Diseases of middle ear and mastoid (H65-H75) | ✗ |
| Hypertensive diseases (I10-I15) | ✗ |
| Acute upper respiratory infections (J00-J06) | ✓ |
| Other diseases of upper respiratory tract (J30-J39) | ✓ |

Table 52: Inclusion and exclusion criteria for cluster 52. Female ratio: 45%, mean age of patients: 21.

| | |
|---|---|
| Hypertensive diseases (I10-I15) | ✗ |
| Other diseases of upper respiratory tract (J30-J39) | ✗ |
| Arthropathies (M00-M25) | ✓ |
| Dorsopathies (M40-M54) | ✓ |
| Soft tissue disorders (M60-M79) | ✗ |
| Osteopathies and chondropathies (M80-M94) | ✗ |

Table 53: Inclusion and exclusion criteria for cluster 53. Female ratio: 54%, mean age of patients: 59.

| | |
|---|---|
| Intestinal infectious diseases (A00-A09) | ✓ |
| Metabolic disorders (E70-E90) | ✓ |
| Hypertensive diseases (I10-I15) | ✗ |
| Acute upper respiratory infections (J00-J06) | ✗ |
| Other diseases of upper respiratory tract (J30-J39) | ✗ |
| Arthropathies (M00-M25) | ✗ |
| Noninflammatory disorders of female genital tract (N80-N98) | ✗ |

Table 54: Inclusion and exclusion criteria for cluster 54. Female ratio: 51%, mean age of patients: 22.

| | |
|---|---|
| Intestinal infectious diseases (A00-A09) | ✗ |
| Mental and behavioural disorders due to psychoactive substance use (F10-F19) | ✓ |
| Mood [affective] disorders (F30-F39) | ✓ |
| Hypertensive diseases (I10-I15) | ✗ |
| Other diseases of upper respiratory tract (J30-J39) | ✗ |
| Other diseases of intestines (K55-K63) | ✗ |
| Arthropathies (M00-M25) | ✗ |
| Noninflammatory disorders of female genital tract (N80-N98) | ✗ |

Table 55: Inclusion and exclusion criteria for cluster 55. Female ratio: 41%, mean age of patients: 43.

| | |
|---|---|
| Intestinal infectious diseases (A00-A09) | ✗ |
| Malignant neoplasms (C00-C97) | ✗ |
| Benign neoplasms (D10-D36) | ✗ |
| Disorders of thyroid gland (E00-E07) | ✗ |
| Mental and behavioural disorders due to psychoactive substance use (F10-F19) | ✗ |
| Neurotic, stress-related and somatoform disorders (F40-F48) | ✗ |
| Episodic and paroxysmal disorders (G40-G47) | ✗ |
| Nerve, nerve root and plexus disorders (G50-G59) | ✗ |
| Disorders of lens (H25-H28) | ✗ |
| Other disorders of ear (H90-H95) | ✓ |
| Hypertensive diseases (I10-I15) | ✗ |
| Other forms of heart disease (I30-I52) | ✗ |
| Diseases of veins, lymphatic vessels and lymph nodes, not elsewhere classified (I80-I89) | ✗ |
| Acute upper respiratory infections (J00-J06) | ✗ |
| Influenza and pneumonia (J09-J18) | ✗ |
| Other acute lower respiratory infections (J20-J22) | ✗ |
| Other diseases of upper respiratory tract (J30-J39) | ✗ |
| Diseases of oral cavity, salivary glands and jaws (K00-K14) | ✗ |
| Diseases of oesophagus, stomach and duodenum (K20-K31) | ✗ |
| Diseases of appendix (K35-K38) | ✗ |
| Hernia (K40-K46) | ✗ |
| Noninfective enteritis and colitis (K50-K52) | ✗ |
| Other diseases of intestines (K55-K63) | ✗ |
| Disorders of gallbladder, biliary tract and pancreas (K80-K87) | ✗ |
| Infections of the skin and subcutaneous tissue (L00-L08) | ✗ |
| Arthropathies (M00-M25) | ✗ |
| Dorsopathies (M40-M54) | ✗ |
| Soft tissue disorders (M60-M79) | ✗ |
| Urolithiasis (N20-N23) | ✗ |
| Other diseases of urinary system (N30-N39) | ✗ |
| Diseases of male genital organs (N40-N51) | ✗ |
| Noninflammatory disorders of female genital tract (N80-N98) | ✗ |

Table 56: Inclusion and exclusion criteria for cluster 56. Female ratio: 43%, mean age of patients: 40.

| | |
|---|---|
| Intestinal infectious diseases (A00-A09) | ✗ |
| Malignant neoplasms (C00-C97) | ✗ |
| Benign neoplasms (D10-D36) | ✗ |
| Disorders of thyroid gland (E00-E07) | ✗ |
| Mental and behavioural disorders due to psychoactive substance use (F10-F19) | ✗ |
| Mood [affective] disorders (F30-F39) | ✓ |
| Neurotic, stress-related and somatoform disorders (F40-F48) | ✗ |
| Episodic and paroxysmal disorders (G40-G47) | ✗ |
| Nerve, nerve root and plexus disorders (G50-G59) | ✗ |
| Disorders of lens (H25-H28) | ✗ |
| Other disorders of ear (H90-H95) | ✗ |
| Hypertensive diseases (I10-I15) | ✗ |
| Other forms of heart disease (I30-I52) | ✗ |
| Diseases of veins, lymphatic vessels and lymph nodes, not elsewhere classified (I80-I89) | ✗ |
| Acute upper respiratory infections (J00-J06) | ✗ |
| Influenza and pneumonia (J09-J18) | ✗ |
| Other acute lower respiratory infections (J20-J22) | ✗ |
| Other diseases of upper respiratory tract (J30-J39) | ✗ |
| Chronic lower respiratory diseases (J40-J47) | ✗ |
| Diseases of oral cavity, salivary glands and jaws (K00-K14) | ✗ |
| Diseases of oesophagus, stomach and duodenum (K20-K31) | ✗ |
| Diseases of appendix (K35-K38) | ✗ |
| Hernia (K40-K46) | ✗ |
| Noninfective enteritis and colitis (K50-K52) | ✗ |
| Other diseases of intestines (K55-K63) | ✗ |
| Disorders of gallbladder, biliary tract and pancreas (K80-K87) | ✗ |
| Infections of the skin and subcutaneous tissue (L00-L08) | ✗ |
| Disorders of skin appendages (L60-L75) | ✗ |
| Arthropathies (M00-M25) | ✗ |
| Dorsopathies (M40-M54) | ✗ |
| Soft tissue disorders (M60-M79) | ✗ |
| Urolithiasis (N20-N23) | ✗ |
| Other diseases of urinary system (N30-N39) | ✗ |
| Diseases of male genital organs (N40-N51) | ✗ |
| Inflammatory diseases of female pelvic organs (N70-N77) | ✗ |
| Noninflammatory disorders of female genital tract (N80-N98) | ✗ |

Table 57: Inclusion and exclusion criteria for cluster 57. Female ratio: 61%, mean age of patients: 42.

| | |
|---|---|
| Hypertensive diseases (I10-I15) | ✓ |
| Ischaemic heart diseases (I20-I25) | ✓ |
| Other forms of heart disease (I30-I52) | ✓ |
| Cerebrovascular diseases (I60-I69) | ✓ |
| Renal failure (N17-N19) | ✗ |

Table 58: Inclusion and exclusion criteria for cluster 58. Female ratio: 52%, mean age of patients: 80.

| | |
|---|---|
| Diabetes mellitus (E10-E14) | ✗ |
| Metabolic disorders (E70-E90) | ✓ |
| Hypertensive diseases (I10-I15) | ✓ |
| Ischaemic heart diseases (I20-I25) | ✓ |
| Other forms of heart disease (I30-I52) | ✓ |
| Cerebrovascular diseases (I60-I69) | ✗ |
| Renal failure (N17-N19) | ✗ |

Table 59: Inclusion and exclusion criteria for cluster 59. Female ratio: 43%, mean age of patients: 73.

| | |
|---|---|
| Diabetes mellitus (E10-E14) | ✗ |
| Metabolic disorders (E70-E90) | ✗ |
| Hypertensive diseases (I10-I15) | ✓ |
| Ischaemic heart diseases (I20-I25) | ✓ |
| Other forms of heart disease (I30-I52) | ✓ |
| Cerebrovascular diseases (I60-I69) | ✗ |
| Renal failure (N17-N19) | ✗ |

Table 60: Inclusion and exclusion criteria for cluster 60. Female ratio: 50%, mean age of patients: 77.

| | |
|---|---|
| Diabetes mellitus (E10-E14) | ✗ |
| Metabolic disorders (E70-E90) | ✓ |
| Hypertensive diseases (I10-I15) | ✓ |
| Ischaemic heart diseases (I20-I25) | ✗ |
| Other forms of heart disease (I30-I52) | ✗ |
| Cerebrovascular diseases (I60-I69) | ✓ |

Table 61: Inclusion and exclusion criteria for cluster 61. Female ratio: 50%, mean age of patients: 72.

| | |
|---|---|
| Malignant neoplasms (C00-C97) | ✗ |
| Diabetes mellitus (E10-E14) | ✗ |
| Metabolic disorders (E70-E90) | ✗ |
| Hypertensive diseases (I10-I15) | ✓ |
| Other forms of heart disease (I30-I52) | ✗ |
| Cerebrovascular diseases (I60-I69) | ✓ |
| Arthropathies (M00-M25) | ✗ |
| Dorsopathies (M40-M54) | ✗ |

Table 62: Inclusion and exclusion criteria for cluster 62. Female ratio: 50%, mean age of patients: 73.

| | |
|---|---|
| Intestinal infectious diseases (A00-A09) | ✗ |
| Viral infections characterized by skin and mucous membrane lesions (B00-B09) | ✗ |
| Malignant neoplasms (C00-C97) | ✗ |
| Benign neoplasms (D10-D36) | ✗ |
| Disorders of thyroid gland (E00-E07) | ✗ |
| Metabolic disorders (E70-E90) | ✗ |
| Mental and behavioural disorders due to psychoactive substance use (F10-F19) | ✗ |
| Mood [affective] disorders (F30-F39) | ✗ |
| Neurotic, stress-related and somatoform disorders (F40-F48) | ✗ |
| Episodic and paroxysmal disorders (G40-G47) | ✗ |
| Nerve, nerve root and plexus disorders (G50-G59) | ✗ |
| Disorders of lens (H25-H28) | ✗ |
| Other disorders of ear (H90-H95) | ✗ |
| Hypertensive diseases (I10-I15) | ✗ |
| Other forms of heart disease (I30-I52) | ✗ |
| Diseases of veins, lymphatic vessels and lymph nodes, not elsewhere classified (I80-I89) | ✗ |
| Acute upper respiratory infections (J00-J06) | ✗ |
| Influenza and pneumonia (J09-J18) | ✗ |
| Other acute lower respiratory infections (J20-J22) | ✗ |
| Other diseases of upper respiratory tract (J30-J39) | ✗ |
| Chronic lower respiratory diseases (J40-J47) | ✗ |
| Diseases of oral cavity, salivary glands and jaws (K00-K14) | ✗ |
| Diseases of oesophagus, stomach and duodenum (K20-K31) | ✗ |
| Diseases of appendix (K35-K38) | ✗ |
| Hernia (K40-K46) | ✗ |
| Noninfective enteritis and colitis (K50-K52) | ✗ |
| Other diseases of intestines (K55-K63) | ✗ |
| Disorders of gallbladder, biliary tract and pancreas (K80-K87) | ✗ |
| Infections of the skin and subcutaneous tissue (L00-L08) | ✗ |
| Disorders of skin appendages (L60-L75) | ✗ |
| Arthropathies (M00-M25) | ✗ |
| Dorsopathies (M40-M54) | ✗ |
| Soft tissue disorders (M60-M79) | ✗ |
| Osteopathies and chondropathies (M80-M94) | ✓ |
| Urolithiasis (N20-N23) | ✗ |
| Other diseases of urinary system (N30-N39) | ✗ |
| Diseases of male genital organs (N40-N51) | ✗ |
| Inflammatory diseases of female pelvic organs (N70-N77) | ✗ |
| Noninflammatory disorders of female genital tract (N80-N98) | ✗ |

Table 63: Inclusion and exclusion criteria for cluster 63. Female ratio: 43%, mean age of patients: 43.

| | |
|---|---|
| Intestinal infectious diseases (A00-A09) | ✗ |
| Hypertensive diseases (I10-I15) | ✗ |
| Diseases of veins, lymphatic vessels and lymph nodes, not elsewhere classified (I80-I89) | ✓ |
| Other diseases of upper respiratory tract (J30-J39) | ✗ |
| Diseases of oesophagus, stomach and duodenum (K20-K31) | ✗ |
| Other diseases of intestines (K55-K63) | ✓ |
| Arthropathies (M00-M25) | ✗ |
| Noninflammatory disorders of female genital tract (N80-N98) | ✗ |

Table 64: Inclusion and exclusion criteria for cluster 64. Female ratio: 44%, mean age of patients: 55.

| | |
|---|---|
| Intestinal infectious diseases (A00-A09) | ✗ |
| Malignant neoplasms (C00-C97) | ✗ |
| Benign neoplasms (D10-D36) | ✗ |
| Disorders of thyroid gland (E00-E07) | ✗ |
| Mental and behavioural disorders due to psychoactive substance use (F10-F19) | ✗ |
| Neurotic, stress-related and somatoform disorders (F40-F48) | ✗ |
| Episodic and paroxysmal disorders (G40-G47) | ✗ |
| Nerve, nerve root and plexus disorders (G50-G59) | ✗ |
| Disorders of lens (H25-H28) | ✗ |
| Other disorders of ear (H90-H95) | ✗ |
| Hypertensive diseases (I10-I15) | ✗ |
| Other forms of heart disease (I30-I52) | ✗ |
| Diseases of veins, lymphatic vessels and lymph nodes, not elsewhere classified (I80-I89) | ✗ |
| Acute upper respiratory infections (J00-J06) | ✗ |
| Influenza and pneumonia (J09-J18) | ✗ |
| Other acute lower respiratory infections (J20-J22) | ✗ |
| Other diseases of upper respiratory tract (J30-J39) | ✗ |
| Diseases of oral cavity, salivary glands and jaws (K00-K14) | ✗ |
| Diseases of oesophagus, stomach and duodenum (K20-K31) | ✗ |
| Diseases of appendix (K35-K38) | ✗ |
| Hernia (K40-K46) | ✗ |
| Noninfective enteritis and colitis (K50-K52) | ✗ |
| Other diseases of intestines (K55-K63) | ✗ |
| Disorders of gallbladder, biliary tract and pancreas (K80-K87) | ✗ |
| Infections of the skin and subcutaneous tissue (L00-L08) | ✗ |
| Disorders of skin appendages (L60-L75) | ✗ |
| Arthropathies (M00-M25) | ✗ |
| Dorsopathies (M40-M54) | ✗ |
| Soft tissue disorders (M60-M79) | ✗ |
| Urolithiasis (N20-N23) | ✗ |
| Other diseases of urinary system (N30-N39) | ✗ |
| Diseases of male genital organs (N40-N51) | ✗ |
| Inflammatory diseases of female pelvic organs (N70-N77) | ✓ |
| Noninflammatory disorders of female genital tract (N80-N98) | ✗ |

Table 65: Inclusion and exclusion criteria for cluster 65. Female ratio: 100%, mean age of patients: 35.

| | |
|---|---|
| Intestinal infectious diseases (A00-A09) | ✗ |
| Malignant neoplasms (C00-C97) | ✗ |
| Benign neoplasms (D10-D36) | ✗ |
| Disorders of thyroid gland (E00-E07) | ✗ |
| Mental and behavioural disorders due to psychoactive substance use (F10-F19) | ✗ |
| Neurotic, stress-related and somatoform disorders (F40-F48) | ✗ |
| Episodic and paroxysmal disorders (G40-G47) | ✗ |
| Nerve, nerve root and plexus disorders (G50-G59) | ✗ |
| Disorders of lens (H25-H28) | ✗ |
| Other disorders of ear (H90-H95) | ✗ |
| Hypertensive diseases (I10-I15) | ✗ |
| Other forms of heart disease (I30-I52) | ✗ |
| Diseases of veins, lymphatic vessels and lymph nodes, not elsewhere classified (I80-I89) | ✗ |
| Acute upper respiratory infections (J00-J06) | ✗ |
| Influenza and pneumonia (J09-J18) | ✗ |
| Other acute lower respiratory infections (J20-J22) | ✗ |
| Other diseases of upper respiratory tract (J30-J39) | ✗ |
| Diseases of oral cavity, salivary glands and jaws (K00-K14) | ✗ |
| Diseases of oesophagus, stomach and duodenum (K20-K31) | ✗ |
| Diseases of appendix (K35-K38) | ✗ |
| Hernia (K40-K46) | ✗ |
| Noninfective enteritis and colitis (K50-K52) | ✗ |
| Other diseases of intestines (K55-K63) | ✗ |
| Disorders of gallbladder, biliary tract and pancreas (K80-K87) | ✗ |
| Infections of the skin and subcutaneous tissue (L00-L08) | ✗ |
| Disorders of skin appendages (L60-L75) | ✓ |
| Arthropathies (M00-M25) | ✗ |
| Dorsopathies (M40-M54) | ✗ |
| Soft tissue disorders (M60-M79) | ✗ |
| Urolithiasis (N20-N23) | ✗ |
| Other diseases of urinary system (N30-N39) | ✗ |
| Diseases of male genital organs (N40-N51) | ✗ |
| Noninflammatory disorders of female genital tract (N80-N98) | ✗ |

Table 66: Inclusion and exclusion criteria for cluster 66. Female ratio: 38%, mean age of patients: 34.

| | |
|---|---|
| Diabetes mellitus (E10-E14) | ✗ |
| Metabolic disorders (E70-E90) | ✓ |
| Hypertensive diseases (I10-I15) | ✓ |
| Ischaemic heart diseases (I20-I25) | ✗ |
| Other forms of heart disease (I30-I52) | ✗ |
| Cerebrovascular diseases (I60-I69) | ✗ |
| Dorsopathies (M40-M54) | ✓ |

Table 67: Inclusion and exclusion criteria for cluster 67. Female ratio: 59%, mean age of patients: 66.

| | |
|---|---|
| Hypertensive diseases (I10-I15) | ✓ |
| Other forms of heart disease (I30-I52) | ✓ |
| Diseases of arteries, arterioles and capillaries (I70-I79) | ✓ |
| Renal failure (N17-N19) | ✓ |

Table 68: Inclusion and exclusion criteria for cluster 68. Female ratio: 49%, mean age of patients: 81.

| | |
|---|---|
| Intestinal infectious diseases (A00-A09) | ✗ |
| Malignant neoplasms (C00-C97) | ✗ |
| Benign neoplasms (D10-D36) | ✗ |
| Mental and behavioural disorders due to psychoactive substance use (F10-F19) | ✗ |
| Episodic and paroxysmal disorders (G40-G47) | ✗ |
| Hypertensive diseases (I10-I15) | ✗ |
| Ischaemic heart diseases (I20-I25) | ✓ |
| Other forms of heart disease (I30-I52) | ✓ |
| Diseases of veins, lymphatic vessels and lymph nodes, not elsewhere classified (I80-I89) | ✗ |
| Acute upper respiratory infections (J00-J06) | ✗ |
| Other diseases of upper respiratory tract (J30-J39) | ✗ |
| Diseases of appendix (K35-K38) | ✗ |
| Hernia (K40-K46) | ✗ |
| Other diseases of intestines (K55-K63) | ✗ |
| Arthropathies (M00-M25) | ✗ |
| Dorsopathies (M40-M54) | ✗ |
| Soft tissue disorders (M60-M79) | ✗ |
| Diseases of male genital organs (N40-N51) | ✗ |
| Noninflammatory disorders of female genital tract (N80-N98) | ✗ |

Table 69: Inclusion and exclusion criteria for cluster 69. Female ratio: 46%, mean age of patients: 76.

| | |
|---|---|
| Malignant neoplasms (C00-C97) | ✗ |
| Diabetes mellitus (E10-E14) | ✗ |
| Obesity and other hyperalimentation (E65-E68) | ✗ |
| Metabolic disorders (E70-E90) | ✗ |
| Hypertensive diseases (I10-I15) | ✓ |
| Ischaemic heart diseases (I20-I25) | ✓ |
| Other forms of heart disease (I30-I52) | ✗ |
| Cerebrovascular diseases (I60-I69) | ✗ |
| Arthropathies (M00-M25) | ✗ |
| Dorsopathies (M40-M54) | ✗ |

Table 70: Inclusion and exclusion criteria for cluster 70. Female ratio: 41%, mean age of patients: 69.

| | |
|---|---|
| Diabetes mellitus (E10-E14) | ✓ |
| Hypertensive diseases (I10-I15) | ✓ |
| Ischaemic heart diseases (I20-I25) | ✓ |
| Other forms of heart disease (I30-I52) | ✓ |
| Cerebrovascular diseases (I60-I69) | ✗ |
| Renal failure (N17-N19) | ✗ |

Table 71: Inclusion and exclusion criteria for cluster 71. Female ratio: 45%, mean age of patients: 75.

| | |
|---|---|
| Diabetes mellitus (E10-E14) | ✓ |
| Metabolic disorders (E70-E90) | ✓ |
| Hypertensive diseases (I10-I15) | ✓ |
| Ischaemic heart diseases (I20-I25) | ✓ |
| Other forms of heart disease (I30-I52) | ✗ |

Table 72: Inclusion and exclusion criteria for cluster 72. Female ratio: 36%, mean age of patients: 69.

| | |
|---|---|
| Intestinal infectious diseases (A00-A09) | ✗ |
| Viral infections characterized by skin and mucous membrane lesions (B00-B09) | ✗ |
| Malignant neoplasms (C00-C97) | ✗ |
| Benign neoplasms (D10-D36) | ✗ |
| Neoplasms of uncertain or unknown behaviour (D37-D48) | ✓ |
| Disorders of thyroid gland (E00-E07) | ✗ |
| Metabolic disorders (E70-E90) | ✗ |
| Mental and behavioural disorders due to psychoactive substance use (F10-F19) | ✗ |
| Mood [affective] disorders (F30-F39) | ✗ |
| Neurotic, stress-related and somatoform disorders (F40-F48) | ✗ |
| Episodic and paroxysmal disorders (G40-G47) | ✗ |
| Nerve, nerve root and plexus disorders (G50-G59) | ✗ |
| Disorders of eyelid, lacrimal system and orbit (H00-H06) | ✗ |
| Disorders of lens (H25-H28) | ✗ |
| Other disorders of ear (H90-H95) | ✗ |
| Hypertensive diseases (I10-I15) | ✗ |
| Other forms of heart disease (I30-I52) | ✗ |
| Diseases of veins, lymphatic vessels and lymph nodes, not elsewhere classified (I80-I89) | ✗ |
| Acute upper respiratory infections (J00-J06) | ✗ |
| Influenza and pneumonia (J09-J18) | ✗ |
| Other acute lower respiratory infections (J20-J22) | ✗ |
| Other diseases of upper respiratory tract (J30-J39) | ✗ |
| Chronic lower respiratory diseases (J40-J47) | ✗ |
| Diseases of oral cavity, salivary glands and jaws (K00-K14) | ✗ |
| Diseases of oesophagus, stomach and duodenum (K20-K31) | ✗ |
| Diseases of appendix (K35-K38) | ✗ |
| Hernia (K40-K46) | ✗ |
| Noninfective enteritis and colitis (K50-K52) | ✗ |
| Other diseases of intestines (K55-K63) | ✗ |
| Disorders of gallbladder, biliary tract and pancreas (K80-K87) | ✗ |
| Infections of the skin and subcutaneous tissue (L00-L08) | ✗ |
| Disorders of skin appendages (L60-L75) | ✗ |
| Arthropathies (M00-M25) | ✗ |
| Dorsopathies (M40-M54) | ✗ |
| Soft tissue disorders (M60-M79) | ✗ |
| Osteopathies and chondropathies (M80-M94) | ✗ |
| Renal tubulo-interstitial diseases (N10-N16) | ✗ |
| Urolithiasis (N20-N23) | ✗ |
| Other diseases of urinary system (N30-N39) | ✗ |
| Diseases of male genital organs (N40-N51) | ✗ |
| Disorders of breast (N60-N64) | ✗ |
| Inflammatory diseases of female pelvic organs (N70-N77) | ✗ |
| Noninflammatory disorders of female genital tract (N80-N98) | ✗ |

Table 73: Inclusion and exclusion criteria for cluster 73. Female ratio: 58%, mean age of patients: 46.

| | |
|---|---|
| Intestinal infectious diseases (A00-A09) | ✗ |
| Mental and behavioural disorders due to psychoactive substance use (F10-F19) | ✗ |
| Hypertensive diseases (I10-I15) | ✗ |
| Diseases of veins, lymphatic vessels and lymph nodes, not elsewhere classified (I80-I89) | ✗ |
| Other diseases of upper respiratory tract (J30-J39) | ✗ |
| Diseases of oesophagus, stomach and duodenum (K20-K31) | ✓ |
| Hernia (K40-K46) | ✓ |
| Other diseases of intestines (K55-K63) | ✗ |
| Arthropathies (M00-M25) | ✗ |
| Dorsopathies (M40-M54) | ✗ |
| Diseases of male genital organs (N40-N51) | ✗ |
| Noninflammatory disorders of female genital tract (N80-N98) | ✗ |

Table 74: Inclusion and exclusion criteria for cluster 74. Female ratio: 40%, mean age of patients: 51.

| | |
|---|---|
| Malignant neoplasms (C00-C97) | ✗ |
| Diabetes mellitus (E10-E14) | ✗ |
| Obesity and other hyperalimentation (E65-E68) | ✓ |
| Metabolic disorders (E70-E90) | ✗ |
| Hypertensive diseases (I10-I15) | ✓ |
| Other forms of heart disease (I30-I52) | ✗ |
| Cerebrovascular diseases (I60-I69) | ✗ |
| Arthropathies (M00-M25) | ✗ |
| Dorsopathies (M40-M54) | ✗ |

Table 75: Inclusion and exclusion criteria for cluster 75. Female ratio: 52%, mean age of patients: 54.

| | |
|---|---|
| Intestinal infectious diseases (A00-A09) | ✗ |
| Viral infections characterized by skin and mucous membrane lesions (B00-B09) | ✗ |
| Malignant neoplasms (C00-C97) | ✗ |
| Benign neoplasms (D10-D36) | ✗ |
| Disorders of thyroid gland (E00-E07) | ✗ |
| Metabolic disorders (E70-E90) | ✗ |
| Mental and behavioural disorders due to psychoactive substance use (F10-F19) | ✗ |
| Mood [affective] disorders (F30-F39) | ✗ |
| Neurotic, stress-related and somatoform disorders (F40-F48) | ✗ |
| Episodic and paroxysmal disorders (G40-G47) | ✗ |
| Nerve, nerve root and plexus disorders (G50-G59) | ✗ |
| Disorders of lens (H25-H28) | ✗ |
| Other disorders of ear (H90-H95) | ✗ |
| Hypertensive diseases (I10-I15) | ✗ |
| Other forms of heart disease (I30-I52) | ✗ |
| Diseases of veins, lymphatic vessels and lymph nodes, not elsewhere classified (I80-I89) | ✗ |
| Acute upper respiratory infections (J00-J06) | ✗ |
| Influenza and pneumonia (J09-J18) | ✗ |
| Other acute lower respiratory infections (J20-J22) | ✗ |
| Other diseases of upper respiratory tract (J30-J39) | ✗ |
| Chronic lower respiratory diseases (J40-J47) | ✗ |
| Diseases of oral cavity, salivary glands and jaws (K00-K14) | ✗ |
| Diseases of oesophagus, stomach and duodenum (K20-K31) | ✗ |
| Diseases of appendix (K35-K38) | ✗ |
| Hernia (K40-K46) | ✗ |
| Noninfective enteritis and colitis (K50-K52) | ✗ |
| Other diseases of intestines (K55-K63) | ✗ |
| Disorders of gallbladder, biliary tract and pancreas (K80-K87) | ✗ |
| Infections of the skin and subcutaneous tissue (L00-L08) | ✗ |
| Disorders of skin appendages (L60-L75) | ✗ |
| Arthropathies (M00-M25) | ✗ |
| Dorsopathies (M40-M54) | ✗ |
| Soft tissue disorders (M60-M79) | ✗ |
| Osteopathies and chondropathies (M80-M94) | ✗ |
| Renal tubulo-interstitial diseases (N10-N16) | ✓ |
| Urolithiasis (N20-N23) | ✗ |
| Other diseases of urinary system (N30-N39) | ✗ |
| Diseases of male genital organs (N40-N51) | ✗ |
| Inflammatory diseases of female pelvic organs (N70-N77) | ✗ |
| Noninflammatory disorders of female genital tract (N80-N98) | ✗ |

Table 76: Inclusion and exclusion criteria for cluster 76. Female ratio: 75%, mean age of patients: 27.

| | |
|---|---|
| Intestinal infectious diseases (A00-A09) | ✗ |
| Viral infections characterized by skin and mucous membrane lesions (B00-B09) | ✓ |
| Malignant neoplasms (C00-C97) | ✗ |
| Benign neoplasms (D10-D36) | ✗ |
| Disorders of thyroid gland (E00-E07) | ✗ |
| Metabolic disorders (E70-E90) | ✗ |
| Mental and behavioural disorders due to psychoactive substance use (F10-F19) | ✗ |
| Mood [affective] disorders (F30-F39) | ✗ |
| Neurotic, stress-related and somatoform disorders (F40-F48) | ✗ |
| Episodic and paroxysmal disorders (G40-G47) | ✗ |
| Nerve, nerve root and plexus disorders (G50-G59) | ✗ |
| Disorders of lens (H25-H28) | ✗ |
| Other disorders of ear (H90-H95) | ✗ |
| Hypertensive diseases (I10-I15) | ✗ |
| Other forms of heart disease (I30-I52) | ✗ |
| Diseases of veins, lymphatic vessels and lymph nodes, not elsewhere classified (I80-I89) | ✗ |
| Acute upper respiratory infections (J00-J06) | ✗ |
| Influenza and pneumonia (J09-J18) | ✗ |
| Other acute lower respiratory infections (J20-J22) | ✗ |
| Other diseases of upper respiratory tract (J30-J39) | ✗ |
| Chronic lower respiratory diseases (J40-J47) | ✗ |
| Diseases of oral cavity, salivary glands and jaws (K00-K14) | ✗ |
| Diseases of oesophagus, stomach and duodenum (K20-K31) | ✗ |
| Diseases of appendix (K35-K38) | ✗ |
| Hernia (K40-K46) | ✗ |
| Noninfective enteritis and colitis (K50-K52) | ✗ |
| Other diseases of intestines (K55-K63) | ✗ |
| Disorders of gallbladder, biliary tract and pancreas (K80-K87) | ✗ |
| Infections of the skin and subcutaneous tissue (L00-L08) | ✗ |
| Disorders of skin appendages (L60-L75) | ✗ |
| Arthropathies (M00-M25) | ✗ |
| Dorsopathies (M40-M54) | ✗ |
| Soft tissue disorders (M60-M79) | ✗ |
| Urolithiasis (N20-N23) | ✗ |
| Other diseases of urinary system (N30-N39) | ✗ |
| Diseases of male genital organs (N40-N51) | ✗ |
| Inflammatory diseases of female pelvic organs (N70-N77) | ✗ |
| Noninflammatory disorders of female genital tract (N80-N98) | ✗ |

Table 77: Inclusion and exclusion criteria for cluster 77. Female ratio: 54%, mean age of patients: 22.

| | |
|---|---|
| Benign neoplasms (D10-D36) | ✗ |
| Hypertensive diseases (I10-I15) | ✗ |
| Other diseases of upper respiratory tract (J30-J39) | ✗ |
| Arthropathies (M00-M25) | ✗ |
| Inflammatory diseases of female pelvic organs (N70-N77) | ✓ |
| Noninflammatory disorders of female genital tract (N80-N98) | ✓ |

Table 78: Inclusion and exclusion criteria for cluster 78. Female ratio: 100%, mean age of patients: 39.

| | |
|---|---|
| Intestinal infectious diseases (A00-A09) | ✗ |
| Metabolic disorders (E70-E90) | ✓ |
| Mental and behavioural disorders due to psychoactive substance use (F10-F19) | ✗ |
| Hypertensive diseases (I10-I15) | ✗ |
| Other diseases of upper respiratory tract (J30-J39) | ✗ |
| Other diseases of intestines (K55-K63) | ✗ |
| Arthropathies (M00-M25) | ✗ |
| Dorsopathies (M40-M54) | ✓ |
| Noninflammatory disorders of female genital tract (N80-N98) | ✗ |

Table 79: Inclusion and exclusion criteria for cluster 79. Female ratio: 45%, mean age of patients: 57.

| | |
|---|---|
| Intestinal infectious diseases (A00-A09) | ✗ |
| Malignant neoplasms (C00-C97) | ✓ |
| Neoplasms of uncertain or unknown behaviour (D37-D48) | ✓ |
| Mental and behavioural disorders due to psychoactive substance use (F10-F19) | ✗ |
| Hypertensive diseases (I10-I15) | ✗ |
| Diseases of veins, lymphatic vessels and lymph nodes, not elsewhere classified (I80-I89) | ✗ |
| Other diseases of upper respiratory tract (J30-J39) | ✗ |
| Hernia (K40-K46) | ✗ |
| Other diseases of intestines (K55-K63) | ✗ |
| Arthropathies (M00-M25) | ✗ |
| Dorsopathies (M40-M54) | ✗ |
| Diseases of male genital organs (N40-N51) | ✗ |
| Noninflammatory disorders of female genital tract (N80-N98) | ✗ |

Table 80: Inclusion and exclusion criteria for cluster 80. Female ratio: 47%, mean age of patients: 58.

| | |
|---|---|
| Intestinal infectious diseases (A00-A09) | ✗ |
| Viral infections characterized by skin and mucous membrane lesions (B00-B09) | ✗ |
| Malignant neoplasms (C00-C97) | ✗ |
| Benign neoplasms (D10-D36) | ✗ |
| Disorders of thyroid gland (E00-E07) | ✗ |
| Metabolic disorders (E70-E90) | ✗ |
| Mental and behavioural disorders due to psychoactive substance use (F10-F19) | ✗ |
| Mood [affective] disorders (F30-F39) | ✗ |
| Neurotic, stress-related and somatoform disorders (F40-F48) | ✗ |
| Episodic and paroxysmal disorders (G40-G47) | ✗ |
| Nerve, nerve root and plexus disorders (G50-G59) | ✗ |
| Disorders of eyelid, lacrimal system and orbit (H00-H06) | ✓ |
| Disorders of lens (H25-H28) | ✗ |
| Other disorders of ear (H90-H95) | ✗ |
| Hypertensive diseases (I10-I15) | ✗ |
| Other forms of heart disease (I30-I52) | ✗ |
| Diseases of veins, lymphatic vessels and lymph nodes, not elsewhere classified (I80-I89) | ✗ |
| Acute upper respiratory infections (J00-J06) | ✗ |
| Influenza and pneumonia (J09-J18) | ✗ |
| Other acute lower respiratory infections (J20-J22) | ✗ |
| Other diseases of upper respiratory tract (J30-J39) | ✗ |
| Chronic lower respiratory diseases (J40-J47) | ✗ |
| Diseases of oral cavity, salivary glands and jaws (K00-K14) | ✗ |
| Diseases of oesophagus, stomach and duodenum (K20-K31) | ✗ |
| Diseases of appendix (K35-K38) | ✗ |
| Hernia (K40-K46) | ✗ |
| Noninfective enteritis and colitis (K50-K52) | ✗ |
| Other diseases of intestines (K55-K63) | ✗ |
| Disorders of gallbladder, biliary tract and pancreas (K80-K87) | ✗ |
| Infections of the skin and subcutaneous tissue (L00-L08) | ✗ |
| Disorders of skin appendages (L60-L75) | ✗ |
| Arthropathies (M00-M25) | ✗ |
| Dorsopathies (M40-M54) | ✗ |
| Soft tissue disorders (M60-M79) | ✗ |
| Osteopathies and chondropathies (M80-M94) | ✗ |
| Renal tubulo-interstitial diseases (N10-N16) | ✗ |
| Urolithiasis (N20-N23) | ✗ |
| Other diseases of urinary system (N30-N39) | ✗ |
| Diseases of male genital organs (N40-N51) | ✗ |
| Inflammatory diseases of female pelvic organs (N70-N77) | ✗ |
| Noninflammatory disorders of female genital tract (N80-N98) | ✗ |

Table 81: Inclusion and exclusion criteria for cluster 81. Female ratio: 60%, mean age of patients: 41.

| | |
|---|---|
| Malignant neoplasms (C00-C97) | ✗ |
| Diabetes mellitus (E10-E14) | ✗ |
| Obesity and other hyperalimentation (E65-E68) | ✗ |
| Metabolic disorders (E70-E90) | ✗ |
| Hypertensive diseases (I10-I15) | ✓ |
| Ischaemic heart diseases (I20-I25) | ✗ |
| Other forms of heart disease (I30-I52) | ✗ |
| Cerebrovascular diseases (I60-I69) | ✗ |
| Diseases of oesophagus, stomach and duodenum (K20-K31) | ✓ |
| Arthropathies (M00-M25) | ✗ |
| Dorsopathies (M40-M54) | ✗ |

Table 82: Inclusion and exclusion criteria for cluster 82. Female ratio: 52%, mean age of patients: 64.

| | |
|---|---|
| Intestinal infectious diseases (A00-A09) | ✗ |
| Mental and behavioural disorders due to psychoactive substance use (F10-F19) | ✓ |
| Mood [affective] disorders (F30-F39) | ✗ |
| Hypertensive diseases (I10-I15) | ✗ |
| Other diseases of upper respiratory tract (J30-J39) | ✗ |
| Other diseases of intestines (K55-K63) | ✗ |
| Diseases of liver (K70-K77) | ✓ |
| Arthropathies (M00-M25) | ✗ |
| Noninflammatory disorders of female genital tract (N80-N98) | ✗ |

Table 83: Inclusion and exclusion criteria for cluster 83. Female ratio: 18%, mean age of patients: 52.

| | |
|---|---|
| Hypertensive diseases (I10-I15) | ✗ |
| Diseases of veins, lymphatic vessels and lymph nodes, not elsewhere classified (I80-I89) | ✓ |
| Other diseases of upper respiratory tract (J30-J39) | ✗ |
| Arthropathies (M00-M25) | ✓ |
| Dorsopathies (M40-M54) | ✗ |
| Soft tissue disorders (M60-M79) | ✗ |
| Osteopathies and chondropathies (M80-M94) | ✗ |

Table 84: Inclusion and exclusion criteria for cluster 84. Female ratio: 57%, mean age of patients: 59.

| | |
|---|---|
| Intestinal infectious diseases (A00-A09) | ✗ |
| Malignant neoplasms (C00-C97) | ✓ |
| Hypertensive diseases (I10-I15) | ✗ |
| Diseases of veins, lymphatic vessels and lymph nodes, not elsewhere classified (I80-I89) | ✗ |
| Other diseases of upper respiratory tract (J30-J39) | ✗ |
| Diseases of oesophagus, stomach and duodenum (K20-K31) | ✗ |
| Other diseases of intestines (K55-K63) | ✓ |
| Arthropathies (M00-M25) | ✗ |
| Noninflammatory disorders of female genital tract (N80-N98) | ✗ |

Table 85: Inclusion and exclusion criteria for cluster 85. Female ratio: 42%, mean age of patients: 68.

| | |
|---|---|
| Episodic and paroxysmal disorders (G40-G47) | ✓ |
| Diseases of middle ear and mastoid (H65-H75) | ✗ |
| Hypertensive diseases (I10-I15) | ✗ |
| Acute upper respiratory infections (J00-J06) | ✗ |
| Other diseases of upper respiratory tract (J30-J39) | ✓ |

Table 86: Inclusion and exclusion criteria for cluster 86. Female ratio: 34%, mean age of patients: 38.

| | |
|---|---|
| Intestinal infectious diseases (A00-A09) | ✗ |
| Infections with a predominantly sexual mode of transmission (A50-A64) | ✗ |
| Viral infections characterized by skin and mucous membrane lesions (B00-B09) | ✗ |
| Malignant neoplasms (C00-C97) | ✗ |
| Benign neoplasms (D10-D36) | ✗ |
| Neoplasms of uncertain or unknown behaviour (D37-D48) | ✗ |
| Disorders of thyroid gland (E00-E07) | ✗ |
| Metabolic disorders (E70-E90) | ✗ |
| Mental and behavioural disorders due to psychoactive substance use (F10-F19) | ✗ |
| Schizophrenia, schizotypal and delusional disorders (F20-F29) | ✗ |
| Mood [affective] disorders (F30-F39) | ✗ |
| Neurotic, stress-related and somatoform disorders (F40-F48) | ✗ |
| Episodic and paroxysmal disorders (G40-G47) | ✗ |
| Nerve, nerve root and plexus disorders (G50-G59) | ✗ |
| Disorders of eyelid, lacrimal system and orbit (H00-H06) | ✗ |
| Disorders of lens (H25-H28) | ✗ |
| Disorders of ocular muscles, binocular movement, accommodation and refraction (H49-H52) | ✗ |
| Diseases of middle ear and mastoid (H65-H75) | ✓ |
| Other disorders of ear (H90-H95) | ✗ |
| Hypertensive diseases (I10-I15) | ✗ |
| Other forms of heart disease (I30-I52) | ✗ |
| Diseases of veins, lymphatic vessels and lymph nodes, not elsewhere classified (I80-I89) | ✗ |
| Acute upper respiratory infections (J00-J06) | ✗ |
| Influenza and pneumonia (J09-J18) | ✗ |
| Other acute lower respiratory infections (J20-J22) | ✗ |
| Other diseases of upper respiratory tract (J30-J39) | ✗ |
| Chronic lower respiratory diseases (J40-J47) | ✗ |
| Diseases of oral cavity, salivary glands and jaws (K00-K14) | ✗ |
| Diseases of oesophagus, stomach and duodenum (K20-K31) | ✗ |
| Diseases of appendix (K35-K38) | ✗ |
| Hernia (K40-K46) | ✗ |
| Noninfective enteritis and colitis (K50-K52) | ✗ |
| Other diseases of intestines (K55-K63) | ✗ |
| Disorders of gallbladder, biliary tract and pancreas (K80-K87) | ✗ |
| Infections of the skin and subcutaneous tissue (L00-L08) | ✗ |
| Disorders of skin appendages (L60-L75) | ✗ |
| Arthropathies (M00-M25) | ✗ |
| Dorsopathies (M40-M54) | ✗ |
| Soft tissue disorders (M60-M79) | ✗ |
| Osteopathies and chondropathies (M80-M94) | ✗ |
| Renal tubulo-interstitial diseases (N10-N16) | ✗ |
| Urolithiasis (N20-N23) | ✗ |
| Other diseases of urinary system (N30-N39) | ✗ |
| Diseases of male genital organs (N40-N51) | ✗ |
| Disorders of breast (N60-N64) | ✗ |
| Inflammatory diseases of female pelvic organs (N70-N77) | ✗ |
| Noninflammatory disorders of female genital tract (N80-N98) | ✗ |

Table 87: Inclusion and exclusion criteria for cluster 87. Female ratio: 45%, mean age of patients: 24.

| | |
|---|---|
| Intestinal infectious diseases (A00-A09) | ✗ |
| Viral infections characterized by skin and mucous membrane lesions (B00-B09) | ✗ |
| Malignant neoplasms (C00-C97) | ✗ |
| Benign neoplasms (D10-D36) | ✗ |
| Disorders of thyroid gland (E00-E07) | ✗ |
| Metabolic disorders (E70-E90) | ✗ |
| Mental and behavioural disorders due to psychoactive substance use (F10-F19) | ✗ |
| Mood [affective] disorders (F30-F39) | ✗ |
| Neurotic, stress-related and somatoform disorders (F40-F48) | ✗ |
| Episodic and paroxysmal disorders (G40-G47) | ✗ |
| Nerve, nerve root and plexus disorders (G50-G59) | ✗ |
| Disorders of eyelid, lacrimal system and orbit (H00-H06) | ✗ |
| Disorders of lens (H25-H28) | ✗ |
| Other disorders of ear (H90-H95) | ✗ |
| Hypertensive diseases (I10-I15) | ✗ |
| Other forms of heart disease (I30-I52) | ✗ |
| Diseases of veins, lymphatic vessels and lymph nodes, not elsewhere classified (I80-I89) | ✗ |
| Acute upper respiratory infections (J00-J06) | ✗ |
| Influenza and pneumonia (J09-J18) | ✗ |
| Other acute lower respiratory infections (J20-J22) | ✗ |
| Other diseases of upper respiratory tract (J30-J39) | ✗ |
| Chronic lower respiratory diseases (J40-J47) | ✗ |
| Diseases of oral cavity, salivary glands and jaws (K00-K14) | ✗ |
| Diseases of oesophagus, stomach and duodenum (K20-K31) | ✗ |
| Diseases of appendix (K35-K38) | ✗ |
| Hernia (K40-K46) | ✗ |
| Noninfective enteritis and colitis (K50-K52) | ✗ |
| Other diseases of intestines (K55-K63) | ✗ |
| Disorders of gallbladder, biliary tract and pancreas (K80-K87) | ✗ |
| Infections of the skin and subcutaneous tissue (L00-L08) | ✗ |
| Disorders of skin appendages (L60-L75) | ✗ |
| Arthropathies (M00-M25) | ✗ |
| Dorsopathies (M40-M54) | ✗ |
| Soft tissue disorders (M60-M79) | ✗ |
| Osteopathies and chondropathies (M80-M94) | ✗ |
| Renal tubulo-interstitial diseases (N10-N16) | ✗ |
| Urolithiasis (N20-N23) | ✗ |
| Other diseases of urinary system (N30-N39) | ✗ |
| Diseases of male genital organs (N40-N51) | ✗ |
| Disorders of breast (N60-N64) | ✓ |
| Inflammatory diseases of female pelvic organs (N70-N77) | ✗ |
| Noninflammatory disorders of female genital tract (N80-N98) | ✗ |

Table 88: Inclusion and exclusion criteria for cluster 88. Female ratio: 79%, mean age of patients: 37.

| | |
|---|---|
| Diabetes mellitus (E10-E14) | ✗ |
| Metabolic disorders (E70-E90) | ✗ |
| Hypertensive diseases (I10-I15) | ✓ |
| Other forms of heart disease (I30-I52) | ✗ |
| Arthropathies (M00-M25) | ✓ |
| Dorsopathies (M40-M54) | ✓ |

Table 89: Inclusion and exclusion criteria for cluster 89. Female ratio: 67%, mean age of patients: 70.

| | |
|---|---|
| Intestinal infectious diseases (A00-A09) | ✗ |
| Other bacterial diseases (A30-A49) | ✗ |
| Infections with a predominantly sexual mode of transmission (A50-A64) | ✗ |
| Viral infections characterized by skin and mucous membrane lesions (B00-B09) | ✗ |
| Other viral diseases (B25-B34) | ✗ |
| Malignant neoplasms (C00-C97) | ✗ |
| Benign neoplasms (D10-D36) | ✗ |
| Neoplasms of uncertain or unknown behaviour (D37-D48) | ✗ |
| Disorders of thyroid gland (E00-E07) | ✗ |
| Metabolic disorders (E70-E90) | ✗ |
| Mental and behavioural disorders due to psychoactive substance use (F10-F19) | ✗ |
| Schizophrenia, schizotypal and delusional disorders (F20-F29) | ✗ |
| Mood [affective] disorders (F30-F39) | ✗ |
| Neurotic, stress-related and somatoform disorders (F40-F48) | ✗ |
| Episodic and paroxysmal disorders (G40-G47) | ✗ |
| Nerve, nerve root and plexus disorders (G50-G59) | ✗ |
| Disorders of eyelid, lacrimal system and orbit (H00-H06) | ✗ |
| Disorders of lens (H25-H28) | ✗ |
| Disorders of choroid and retina (H30-H36) | ✗ |
| Disorders of ocular muscles, binocular movement, accommodation and refraction (H49-H52) | ✗ |
| Diseases of middle ear and mastoid (H65-H75) | ✗ |
| Diseases of inner ear (H80-H83) | ✗ |
| Other disorders of ear (H90-H95) | ✗ |
| Hypertensive diseases (I10-I15) | ✗ |
| Other forms of heart disease (I30-I52) | ✗ |
| Cerebrovascular diseases (I60-I69) | ✓ |
| Diseases of veins, lymphatic vessels and lymph nodes, not elsewhere classified (I80-I89) | ✗ |
| Acute upper respiratory infections (J00-J06) | ✗ |
| Influenza and pneumonia (J09-J18) | ✗ |
| Other acute lower respiratory infections (J20-J22) | ✗ |
| Other diseases of upper respiratory tract (J30-J39) | ✗ |
| Chronic lower respiratory diseases (J40-J47) | ✗ |
| Diseases of oral cavity, salivary glands and jaws (K00-K14) | ✗ |
| Diseases of oesophagus, stomach and duodenum (K20-K31) | ✗ |
| Diseases of appendix (K35-K38) | ✗ |
| Hernia (K40-K46) | ✗ |
| Noninfective enteritis and colitis (K50-K52) | ✗ |
| Other diseases of intestines (K55-K63) | ✗ |
| Disorders of gallbladder, biliary tract and pancreas (K80-K87) | ✗ |
| Infections of the skin and subcutaneous tissue (L00-L08) | ✗ |
| Disorders of skin appendages (L60-L75) | ✗ |
| Arthropathies (M00-M25) | ✗ |
| Dorsopathies (M40-M54) | ✗ |
| Soft tissue disorders (M60-M79) | ✗ |
| Osteopathies and chondropathies (M80-M94) | ✗ |
| Renal tubulo-interstitial diseases (N10-N16) | ✗ |
| Urolithiasis (N20-N23) | ✗ |
| Other diseases of urinary system (N30-N39) | ✗ |
| Diseases of male genital organs (N40-N51) | ✗ |
| Disorders of breast (N60-N64) | ✗ |
| Inflammatory diseases of female pelvic organs (N70-N77) | ✗ |
| Noninflammatory disorders of female genital tract (N80-N98) | ✗ |

Table 90: Inclusion and exclusion criteria for cluster 90. Female ratio: 47%, mean age of patients: 61.

| | |
|---|---|
| Intestinal infectious diseases (A00-A09) | ✗ |
| Other bacterial diseases (A30-A49) | ✗ |
| Infections with a predominantly sexual mode of transmission (A50-A64) | ✗ |
| Viral infections characterized by skin and mucous membrane lesions (B00-B09) | ✗ |
| Other viral diseases (B25-B34) | ✗ |
| Malignant neoplasms (C00-C97) | ✗ |
| Benign neoplasms (D10-D36) | ✗ |
| Neoplasms of uncertain or unknown behaviour (D37-D48) | ✗ |
| Disorders of thyroid gland (E00-E07) | ✗ |
| Metabolic disorders (E70-E90) | ✗ |
| Mental and behavioural disorders due to psychoactive substance use (F10-F19) | ✗ |
| Schizophrenia, schizotypal and delusional disorders (F20-F29) | ✗ |
| Mood [affective] disorders (F30-F39) | ✗ |
| Neurotic, stress-related and somatoform disorders (F40-F48) | ✗ |
| Episodic and paroxysmal disorders (G40-G47) | ✗ |
| Nerve, nerve root and plexus disorders (G50-G59) | ✗ |
| Disorders of eyelid, lacrimal system and orbit (H00-H06) | ✗ |
| Disorders of lens (H25-H28) | ✗ |
| Disorders of choroid and retina (H30-H36) | ✗ |
| Disorders of ocular muscles, binocular movement, accommodation and refraction (H49-H52) | ✗ |
| Diseases of middle ear and mastoid (H65-H75) | ✗ |
| Diseases of inner ear (H80-H83) | ✗ |
| Other disorders of ear (H90-H95) | ✗ |
| Hypertensive diseases (I10-I15) | ✗ |
| Ischaemic heart diseases (I20-I25) | ✓ |
| Other forms of heart disease (I30-I52) | ✗ |
| Cerebrovascular diseases (I60-I69) | ✗ |
| Diseases of veins, lymphatic vessels and lymph nodes, not elsewhere classified (I80-I89) | ✗ |
| Acute upper respiratory infections (J00-J06) | ✗ |
| Influenza and pneumonia (J09-J18) | ✗ |
| Other acute lower respiratory infections (J20-J22) | ✗ |
| Other diseases of upper respiratory tract (J30-J39) | ✗ |
| Chronic lower respiratory diseases (J40-J47) | ✗ |
| Diseases of oral cavity, salivary glands and jaws (K00-K14) | ✗ |
| Diseases of oesophagus, stomach and duodenum (K20-K31) | ✗ |
| Diseases of appendix (K35-K38) | ✗ |
| Hernia (K40-K46) | ✗ |
| Noninfective enteritis and colitis (K50-K52) | ✗ |
| Other diseases of intestines (K55-K63) | ✗ |
| Disorders of gallbladder, biliary tract and pancreas (K80-K87) | ✗ |
| Infections of the skin and subcutaneous tissue (L00-L08) | ✗ |
| Disorders of skin appendages (L60-L75) | ✗ |
| Other disorders of the skin and subcutaneous tissue (L80-L99) | ✗ |
| Arthropathies (M00-M25) | ✗ |
| Dorsopathies (M40-M54) | ✗ |
| Soft tissue disorders (M60-M79) | ✗ |
| Osteopathies and chondropathies (M80-M94) | ✗ |
| Renal tubulo-interstitial diseases (N10-N16) | ✗ |
| Urolithiasis (N20-N23) | ✗ |
| Other diseases of urinary system (N30-N39) | ✗ |
| Diseases of male genital organs (N40-N51) | ✗ |
| Disorders of breast (N60-N64) | ✗ |
| Inflammatory diseases of female pelvic organs (N70-N77) | ✗ |
| Noninflammatory disorders of female genital tract (N80-N98) | ✗ |

Table 91: Inclusion and exclusion criteria for cluster 91. Female ratio: 32%, mean age of patients: 63.

| | |
|---|---|
| Intestinal infectious diseases (A00-A09) | ✗ |
| Malignant neoplasms (C00-C97) | ✗ |
| Benign neoplasms (D10-D36) | ✗ |
| Mental and behavioural disorders due to psychoactive substance use (F10-F19) | ✗ |
| Neurotic, stress-related and somatoform disorders (F40-F48) | ✗ |
| Episodic and paroxysmal disorders (G40-G47) | ✗ |
| Disorders of lens (H25-H28) | ✗ |
| Hypertensive diseases (I10-I15) | ✗ |
| Other forms of heart disease (I30-I52) | ✗ |
| Diseases of veins, lymphatic vessels and lymph nodes, not elsewhere classified (I80-I89) | ✗ |
| Acute upper respiratory infections (J00-J06) | ✗ |
| Other diseases of upper respiratory tract (J30-J39) | ✗ |
| Diseases of oral cavity, salivary glands and jaws (K00-K14) | ✗ |
| Diseases of appendix (K35-K38) | ✗ |
| Hernia (K40-K46) | ✗ |
| Other diseases of intestines (K55-K63) | ✗ |
| Disorders of gallbladder, biliary tract and pancreas (K80-K87) | ✗ |
| Infections of the skin and subcutaneous tissue (L00-L08) | ✗ |
| Arthropathies (M00-M25) | ✗ |
| Dorsopathies (M40-M54) | ✗ |
| Soft tissue disorders (M60-M79) | ✗ |
| Renal tubulo-interstitial diseases (N10-N16) | ✓ |
| Urolithiasis (N20-N23) | ✓ |
| Diseases of male genital organs (N40-N51) | ✗ |
| Noninflammatory disorders of female genital tract (N80-N98) | ✗ |

Table 92: Inclusion and exclusion criteria for cluster 92. Female ratio: 34%, mean age of patients: 44.

| | |
|---|---|
| Intestinal infectious diseases (A00-A09) | ✗ |
| Mental and behavioural disorders due to psychoactive substance use (F10-F19) | ✓ |
| Mood [affective] disorders (F30-F39) | ✗ |
| Neurotic, stress-related and somatoform disorders (F40-F48) | ✓ |
| Hypertensive diseases (I10-I15) | ✗ |
| Other diseases of upper respiratory tract (J30-J39) | ✗ |
| Other diseases of intestines (K55-K63) | ✗ |
| Diseases of liver (K70-K77) | ✗ |
| Arthropathies (M00-M25) | ✗ |
| Noninflammatory disorders of female genital tract (N80-N98) | ✗ |

Table 93: Inclusion and exclusion criteria for cluster 93. Female ratio: 40%, mean age of patients: 36.

| | |
|---|---|
| Intestinal infectious diseases (A00-A09) | ✗ |
| Viral infections characterized by skin and mucous membrane lesions (B00-B09) | ✗ |
| Malignant neoplasms (C00-C97) | ✗ |
| Benign neoplasms (D10-D36) | ✗ |
| Neoplasms of uncertain or unknown behaviour (D37-D48) | ✗ |
| Disorders of thyroid gland (E00-E07) | ✗ |
| Metabolic disorders (E70-E90) | ✗ |
| Mental and behavioural disorders due to psychoactive substance use (F10-F19) | ✗ |
| Mood [affective] disorders (F30-F39) | ✗ |
| Neurotic, stress-related and somatoform disorders (F40-F48) | ✗ |
| Episodic and paroxysmal disorders (G40-G47) | ✗ |
| Nerve, nerve root and plexus disorders (G50-G59) | ✗ |
| Disorders of eyelid, lacrimal system and orbit (H00-H06) | ✗ |
| Disorders of lens (H25-H28) | ✗ |
| Disorders of ocular muscles, binocular movement, accommodation and refraction (H49-H52) | ✓ |
| Other disorders of ear (H90-H95) | ✗ |
| Hypertensive diseases (I10-I15) | ✗ |
| Other forms of heart disease (I30-I52) | ✗ |
| Diseases of veins, lymphatic vessels and lymph nodes, not elsewhere classified (I80-I89) | ✗ |
| Acute upper respiratory infections (J00-J06) | ✗ |
| Influenza and pneumonia (J09-J18) | ✗ |
| Other acute lower respiratory infections (J20-J22) | ✗ |
| Other diseases of upper respiratory tract (J30-J39) | ✗ |
| Chronic lower respiratory diseases (J40-J47) | ✗ |
| Diseases of oral cavity, salivary glands and jaws (K00-K14) | ✗ |
| Diseases of oesophagus, stomach and duodenum (K20-K31) | ✗ |
| Diseases of appendix (K35-K38) | ✗ |
| Hernia (K40-K46) | ✗ |
| Noninfective enteritis and colitis (K50-K52) | ✗ |
| Other diseases of intestines (K55-K63) | ✗ |
| Disorders of gallbladder, biliary tract and pancreas (K80-K87) | ✗ |
| Infections of the skin and subcutaneous tissue (L00-L08) | ✗ |
| Disorders of skin appendages (L60-L75) | ✗ |
| Arthropathies (M00-M25) | ✗ |
| Dorsopathies (M40-M54) | ✗ |
| Soft tissue disorders (M60-M79) | ✗ |
| Osteopathies and chondropathies (M80-M94) | ✗ |
| Renal tubulo-interstitial diseases (N10-N16) | ✗ |
| Urolithiasis (N20-N23) | ✗ |
| Other diseases of urinary system (N30-N39) | ✗ |
| Diseases of male genital organs (N40-N51) | ✗ |
| Disorders of breast (N60-N64) | ✗ |
| Inflammatory diseases of female pelvic organs (N70-N77) | ✗ |
| Noninflammatory disorders of female genital tract (N80-N98) | ✗ |

Table 94: Inclusion and exclusion criteria for cluster 94. Female ratio: 52%, mean age of patients: 28.

| | |
|---|---|
| Intestinal infectious diseases (A00-A09) | ✗ |
| Other bacterial diseases (A30-A49) | ✓ |
| Infections with a predominantly sexual mode of transmission (A50-A64) | ✗ |
| Viral infections characterized by skin and mucous membrane lesions (B00-B09) | ✗ |
| Malignant neoplasms (C00-C97) | ✗ |
| Benign neoplasms (D10-D36) | ✗ |
| Neoplasms of uncertain or unknown behaviour (D37-D48) | ✗ |
| Disorders of thyroid gland (E00-E07) | ✗ |
| Metabolic disorders (E70-E90) | ✗ |
| Mental and behavioural disorders due to psychoactive substance use (F10-F19) | ✗ |
| Schizophrenia, schizotypal and delusional disorders (F20-F29) | ✗ |
| Mood [affective] disorders (F30-F39) | ✗ |
| Neurotic, stress-related and somatoform disorders (F40-F48) | ✗ |
| Episodic and paroxysmal disorders (G40-G47) | ✗ |
| Nerve, nerve root and plexus disorders (G50-G59) | ✗ |
| Disorders of eyelid, lacrimal system and orbit (H00-H06) | ✗ |
| Disorders of lens (H25-H28) | ✗ |
| Disorders of choroid and retina (H30-H36) | ✗ |
| Disorders of ocular muscles, binocular movement, accommodation and refraction (H49-H52) | ✗ |
| Diseases of middle ear and mastoid (H65-H75) | ✗ |
| Other disorders of ear (H90-H95) | ✗ |
| Hypertensive diseases (I10-I15) | ✗ |
| Other forms of heart disease (I30-I52) | ✗ |
| Diseases of veins, lymphatic vessels and lymph nodes, not elsewhere classified (I80-I89) | ✗ |
| Acute upper respiratory infections (J00-J06) | ✗ |
| Influenza and pneumonia (J09-J18) | ✗ |
| Other acute lower respiratory infections (J20-J22) | ✗ |
| Other diseases of upper respiratory tract (J30-J39) | ✗ |
| Chronic lower respiratory diseases (J40-J47) | ✗ |
| Diseases of oral cavity, salivary glands and jaws (K00-K14) | ✗ |
| Diseases of oesophagus, stomach and duodenum (K20-K31) | ✗ |
| Diseases of appendix (K35-K38) | ✗ |
| Hernia (K40-K46) | ✗ |
| Noninfective enteritis and colitis (K50-K52) | ✗ |
| Other diseases of intestines (K55-K63) | ✗ |
| Disorders of gallbladder, biliary tract and pancreas (K80-K87) | ✗ |
| Infections of the skin and subcutaneous tissue (L00-L08) | ✗ |
| Disorders of skin appendages (L60-L75) | ✗ |
| Arthropathies (M00-M25) | ✗ |
| Dorsopathies (M40-M54) | ✗ |
| Soft tissue disorders (M60-M79) | ✗ |
| Osteopathies and chondropathies (M80-M94) | ✗ |
| Renal tubulo-interstitial diseases (N10-N16) | ✗ |
| Urolithiasis (N20-N23) | ✗ |
| Other diseases of urinary system (N30-N39) | ✗ |
| Diseases of male genital organs (N40-N51) | ✗ |
| Disorders of breast (N60-N64) | ✗ |
| Inflammatory diseases of female pelvic organs (N70-N77) | ✗ |
| Noninflammatory disorders of female genital tract (N80-N98) | ✗ |

Table 95: Inclusion and exclusion criteria for cluster 95. Female ratio: 45%, mean age of patients: 36.

| | |
|---|---|
| Intestinal infectious diseases (A00-A09) | ✗ |
| Infections with a predominantly sexual mode of transmission (A50-A64) | ✗ |
| Viral infections characterized by skin and mucous membrane lesions (B00-B09) | ✗ |
| Malignant neoplasms (C00-C97) | ✗ |
| Benign neoplasms (D10-D36) | ✗ |
| Neoplasms of uncertain or unknown behaviour (D37-D48) | ✗ |
| Disorders of thyroid gland (E00-E07) | ✗ |
| Metabolic disorders (E70-E90) | ✗ |
| Mental and behavioural disorders due to psychoactive substance use (F10-F19) | ✗ |
| Schizophrenia, schizotypal and delusional disorders (F20-F29) | ✗ |
| Mood [affective] disorders (F30-F39) | ✗ |
| Neurotic, stress-related and somatoform disorders (F40-F48) | ✗ |
| Episodic and paroxysmal disorders (G40-G47) | ✗ |
| Nerve, nerve root and plexus disorders (G50-G59) | ✗ |
| Disorders of eyelid, lacrimal system and orbit (H00-H06) | ✗ |
| Disorders of lens (H25-H28) | ✗ |
| Disorders of choroid and retina (H30-H36) | ✓ |
| Disorders of ocular muscles, binocular movement, accommodation and refraction (H49-H52) | ✗ |
| Diseases of middle ear and mastoid (H65-H75) | ✗ |
| Other disorders of ear (H90-H95) | ✗ |
| Hypertensive diseases (I10-I15) | ✗ |
| Other forms of heart disease (I30-I52) | ✗ |
| Diseases of veins, lymphatic vessels and lymph nodes, not elsewhere classified (I80-I89) | ✗ |
| Acute upper respiratory infections (J00-J06) | ✗ |
| Influenza and pneumonia (J09-J18) | ✗ |
| Other acute lower respiratory infections (J20-J22) | ✗ |
| Other diseases of upper respiratory tract (J30-J39) | ✗ |
| Chronic lower respiratory diseases (J40-J47) | ✗ |
| Diseases of oral cavity, salivary glands and jaws (K00-K14) | ✗ |
| Diseases of oesophagus, stomach and duodenum (K20-K31) | ✗ |
| Diseases of appendix (K35-K38) | ✗ |
| Hernia (K40-K46) | ✗ |
| Noninfective enteritis and colitis (K50-K52) | ✗ |
| Other diseases of intestines (K55-K63) | ✗ |
| Disorders of gallbladder, biliary tract and pancreas (K80-K87) | ✗ |
| Infections of the skin and subcutaneous tissue (L00-L08) | ✗ |
| Disorders of skin appendages (L60-L75) | ✗ |
| Arthropathies (M00-M25) | ✗ |
| Dorsopathies (M40-M54) | ✗ |
| Soft tissue disorders (M60-M79) | ✗ |
| Osteopathies and chondropathies (M80-M94) | ✗ |
| Renal tubulo-interstitial diseases (N10-N16) | ✗ |
| Urolithiasis (N20-N23) | ✗ |
| Other diseases of urinary system (N30-N39) | ✗ |
| Diseases of male genital organs (N40-N51) | ✗ |
| Disorders of breast (N60-N64) | ✗ |
| Inflammatory diseases of female pelvic organs (N70-N77) | ✗ |
| Noninflammatory disorders of female genital tract (N80-N98) | ✗ |

Table 96: Inclusion and exclusion criteria for cluster 96. Female ratio: 45%, mean age of patients: 53.

| | |
|---|---|
| Intestinal infectious diseases (A00-A09) | ✗ |
| Infections with a predominantly sexual mode of transmission (A50-A64) | ✗ |
| Viral infections characterized by skin and mucous membrane lesions (B00-B09) | ✗ |
| Malignant neoplasms (C00-C97) | ✗ |
| Benign neoplasms (D10-D36) | ✗ |
| Neoplasms of uncertain or unknown behaviour (D37-D48) | ✗ |
| Disorders of thyroid gland (E00-E07) | ✗ |
| Metabolic disorders (E70-E90) | ✗ |
| Mental and behavioural disorders due to psychoactive substance use (F10-F19) | ✗ |
| Schizophrenia, schizotypal and delusional disorders (F20-F29) | ✓ |
| Mood [affective] disorders (F30-F39) | ✗ |
| Neurotic, stress-related and somatoform disorders (F40-F48) | ✗ |
| Episodic and paroxysmal disorders (G40-G47) | ✗ |
| Nerve, nerve root and plexus disorders (G50-G59) | ✗ |
| Disorders of eyelid, lacrimal system and orbit (H00-H06) | ✗ |
| Disorders of lens (H25-H28) | ✗ |
| Disorders of ocular muscles, binocular movement, accommodation and refraction (H49-H52) | ✗ |
| Other disorders of ear (H90-H95) | ✗ |
| Hypertensive diseases (I10-I15) | ✗ |
| Other forms of heart disease (I30-I52) | ✗ |
| Diseases of veins, lymphatic vessels and lymph nodes, not elsewhere classified (I80-I89) | ✗ |
| Acute upper respiratory infections (J00-J06) | ✗ |
| Influenza and pneumonia (J09-J18) | ✗ |
| Other acute lower respiratory infections (J20-J22) | ✗ |
| Other diseases of upper respiratory tract (J30-J39) | ✗ |
| Chronic lower respiratory diseases (J40-J47) | ✗ |
| Diseases of oral cavity, salivary glands and jaws (K00-K14) | ✗ |
| Diseases of oesophagus, stomach and duodenum (K20-K31) | ✗ |
| Diseases of appendix (K35-K38) | ✗ |
| Hernia (K40-K46) | ✗ |
| Noninfective enteritis and colitis (K50-K52) | ✗ |
| Other diseases of intestines (K55-K63) | ✗ |
| Disorders of gallbladder, biliary tract and pancreas (K80-K87) | ✗ |
| Infections of the skin and subcutaneous tissue (L00-L08) | ✗ |
| Disorders of skin appendages (L60-L75) | ✗ |
| Arthropathies (M00-M25) | ✗ |
| Dorsopathies (M40-M54) | ✗ |
| Soft tissue disorders (M60-M79) | ✗ |
| Osteopathies and chondropathies (M80-M94) | ✗ |
| Renal tubulo-interstitial diseases (N10-N16) | ✗ |
| Urolithiasis (N20-N23) | ✗ |
| Other diseases of urinary system (N30-N39) | ✗ |
| Diseases of male genital organs (N40-N51) | ✗ |
| Disorders of breast (N60-N64) | ✗ |
| Inflammatory diseases of female pelvic organs (N70-N77) | ✗ |
| Noninflammatory disorders of female genital tract (N80-N98) | ✗ |

Table 97: Inclusion and exclusion criteria for cluster 97. Female ratio: 44%, mean age of patients: 41.

| | |
|---|---|
| Metabolic disorders (E70-E90) | ✓ |
| Hypertensive diseases (I10-I15) | ✗ |
| Diseases of veins, lymphatic vessels and lymph nodes, not elsewhere classified (I80-I89) | ✗ |
| Other diseases of upper respiratory tract (J30-J39) | ✗ |
| Arthropathies (M00-M25) | ✓ |
| Dorsopathies (M40-M54) | ✗ |
| Soft tissue disorders (M60-M79) | ✗ |
| Osteopathies and chondropathies (M80-M94) | ✗ |

Table 98: Inclusion and exclusion criteria for cluster 98. Female ratio: 43%, mean age of patients: 61.

| | |
|---|---|
| Diabetes mellitus (E10-E14) | ✓ |
| Metabolic disorders (E70-E90) | ✗ |
| Hypertensive diseases (I10-I15) | ✓ |
| Ischaemic heart diseases (I20-I25) | ✗ |
| Other forms of heart disease (I30-I52) | ✓ |
| Cerebrovascular diseases (I60-I69) | ✗ |
| Renal failure (N17-N19) | ✗ |

Table 99: Inclusion and exclusion criteria for cluster 99. Female ratio: 55%, mean age of patients: 77.

| | |
|---|---|
| Intestinal infectious diseases (A00-A09) | ✗ |
| Other bacterial diseases (A30-A49) | ✗ |
| Infections with a predominantly sexual mode of transmission (A50-A64) | ✗ |
| Viral infections characterized by skin and mucous membrane lesions (B00-B09) | ✗ |
| Other viral diseases (B25-B34) | ✗ |
| Malignant neoplasms (C00-C97) | ✗ |
| In situ neoplasms (D00-D09) | ✗ |
| Benign neoplasms (D10-D36) | ✗ |
| Neoplasms of uncertain or unknown behaviour (D37-D48) | ✗ |
| Disorders of thyroid gland (E00-E07) | ✗ |
| Diabetes mellitus (E10-E14) | ✓ |
| Metabolic disorders (E70-E90) | ✗ |
| Mental and behavioural disorders due to psychoactive substance use (F10-F19) | ✗ |
| Schizophrenia, schizotypal and delusional disorders (F20-F29) | ✗ |
| Mood [affective] disorders (F30-F39) | ✗ |
| Neurotic, stress-related and somatoform disorders (F40-F48) | ✗ |
| Episodic and paroxysmal disorders (G40-G47) | ✗ |
| Nerve, nerve root and plexus disorders (G50-G59) | ✗ |
| Disorders of eyelid, lacrimal system and orbit (H00-H06) | ✗ |
| Disorders of lens (H25-H28) | ✗ |
| Disorders of choroid and retina (H30-H36) | ✗ |
| Disorders of ocular muscles, binocular movement, accommodation and refraction (H49-H52) | ✗ |
| Diseases of middle ear and mastoid (H65-H75) | ✗ |
| Diseases of inner ear (H80-H83) | ✗ |
| Other disorders of ear (H90-H95) | ✗ |
| Hypertensive diseases (I10-I15) | ✗ |
| Ischaemic heart diseases (I20-I25) | ✗ |
| Other forms of heart disease (I30-I52) | ✗ |
| Cerebrovascular diseases (I60-I69) | ✗ |
| Diseases of veins, lymphatic vessels and lymph nodes, not elsewhere classified (I80-I89) | ✗ |
| Acute upper respiratory infections (J00-J06) | ✗ |
| Influenza and pneumonia (J09-J18) | ✗ |
| Other acute lower respiratory infections (J20-J22) | ✗ |
| Other diseases of upper respiratory tract (J30-J39) | ✗ |
| Chronic lower respiratory diseases (J40-J47) | ✗ |
| Diseases of oral cavity, salivary glands and jaws (K00-K14) | ✗ |
| Diseases of oesophagus, stomach and duodenum (K20-K31) | ✗ |
| Diseases of appendix (K35-K38) | ✗ |
| Hernia (K40-K46) | ✗ |
| Noninfective enteritis and colitis (K50-K52) | ✗ |
| Other diseases of intestines (K55-K63) | ✗ |
| Disorders of gallbladder, biliary tract and pancreas (K80-K87) | ✗ |
| Infections of the skin and subcutaneous tissue (L00-L08) | ✗ |
| Dermatitis and eczema (L20-L30) | ✗ |
| Disorders of skin appendages (L60-L75) | ✗ |
| Other disorders of the skin and subcutaneous tissue (L80-L99) | ✗ |
| Arthropathies (M00-M25) | ✗ |
| Dorsopathies (M40-M54) | ✗ |
| Soft tissue disorders (M60-M79) | ✗ |
| Osteopathies and chondropathies (M80-M94) | ✗ |
| Renal tubulo-interstitial diseases (N10-N16) | ✗ |
| Urolithiasis (N20-N23) | ✗ |
| Other diseases of urinary system (N30-N39) | ✗ |
| Diseases of male genital organs (N40-N51) | ✗ |
| Disorders of breast (N60-N64) | ✗ |
| Inflammatory diseases of female pelvic organs (N70-N77) | ✗ |
| Noninflammatory disorders of female genital tract (N80-N98) | ✗ |

Table 100: Inclusion and exclusion criteria for cluster 100. Female ratio: 42%, mean age of patients: 47.

| | |
|---|---|
| Benign neoplasms (D10-D36) | ✗ |
| Hypertensive diseases (I10-I15) | ✗ |
| Other diseases of upper respiratory tract (J30-J39) | ✗ |
| Arthropathies (M00-M25) | ✗ |
| Other diseases of urinary system (N30-N39) | ✓ |
| Inflammatory diseases of female pelvic organs (N70-N77) | ✗ |
| Noninflammatory disorders of female genital tract (N80-N98) | ✓ |

Table 101: Inclusion and exclusion criteria for cluster 101. Female ratio: 100%, mean age of patients: 51.

| | |
|---|---|
| Malignant neoplasms (C00-C97) | ✗ |
| Diabetes mellitus (E10-E14) | ✗ |
| Obesity and other hyperalimentation (E65-E68) | ✗ |
| Metabolic disorders (E70-E90) | ✗ |
| Disorders of lens (H25-H28) | ✓ |
| Hypertensive diseases (I10-I15) | ✓ |
| Ischaemic heart diseases (I20-I25) | ✗ |
| Other forms of heart disease (I30-I52) | ✗ |
| Cerebrovascular diseases (I60-I69) | ✗ |
| Diseases of oesophagus, stomach and duodenum (K20-K31) | ✗ |
| Arthropathies (M00-M25) | ✗ |
| Dorsopathies (M40-M54) | ✗ |

Table 102: Inclusion and exclusion criteria for cluster 102. Female ratio: 65%, mean age of patients: 77.

| | |
|---|---|
| Intestinal infectious diseases (A00-A09) | ✗ |
| Infections with a predominantly sexual mode of transmission (A50-A64) | ✓ |
| Viral infections characterized by skin and mucous membrane lesions (B00-B09) | ✗ |
| Malignant neoplasms (C00-C97) | ✗ |
| Benign neoplasms (D10-D36) | ✗ |
| Neoplasms of uncertain or unknown behaviour (D37-D48) | ✗ |
| Disorders of thyroid gland (E00-E07) | ✗ |
| Metabolic disorders (E70-E90) | ✗ |
| Mental and behavioural disorders due to psychoactive substance use (F10-F19) | ✗ |
| Mood [affective] disorders (F30-F39) | ✗ |
| Neurotic, stress-related and somatoform disorders (F40-F48) | ✗ |
| Episodic and paroxysmal disorders (G40-G47) | ✗ |
| Nerve, nerve root and plexus disorders (G50-G59) | ✗ |
| Disorders of eyelid, lacrimal system and orbit (H00-H06) | ✗ |
| Disorders of lens (H25-H28) | ✗ |
| Disorders of ocular muscles, binocular movement, accommodation and refraction (H49-H52) | ✗ |
| Other disorders of ear (H90-H95) | ✗ |
| Hypertensive diseases (I10-I15) | ✗ |
| Other forms of heart disease (I30-I52) | ✗ |
| Diseases of veins, lymphatic vessels and lymph nodes, not elsewhere classified (I80-I89) | ✗ |
| Acute upper respiratory infections (J00-J06) | ✗ |
| Influenza and pneumonia (J09-J18) | ✗ |
| Other acute lower respiratory infections (J20-J22) | ✗ |
| Other diseases of upper respiratory tract (J30-J39) | ✗ |
| Chronic lower respiratory diseases (J40-J47) | ✗ |
| Diseases of oral cavity, salivary glands and jaws (K00-K14) | ✗ |
| Diseases of oesophagus, stomach and duodenum (K20-K31) | ✗ |
| Diseases of appendix (K35-K38) | ✗ |
| Hernia (K40-K46) | ✗ |
| Noninfective enteritis and colitis (K50-K52) | ✗ |
| Other diseases of intestines (K55-K63) | ✗ |
| Disorders of gallbladder, biliary tract and pancreas (K80-K87) | ✗ |
| Infections of the skin and subcutaneous tissue (L00-L08) | ✗ |
| Disorders of skin appendages (L60-L75) | ✗ |
| Arthropathies (M00-M25) | ✗ |
| Dorsopathies (M40-M54) | ✗ |
| Soft tissue disorders (M60-M79) | ✗ |
| Osteopathies and chondropathies (M80-M94) | ✗ |
| Renal tubulo-interstitial diseases (N10-N16) | ✗ |
| Urolithiasis (N20-N23) | ✗ |
| Other diseases of urinary system (N30-N39) | ✗ |
| Diseases of male genital organs (N40-N51) | ✗ |
| Disorders of breast (N60-N64) | ✗ |
| Inflammatory diseases of female pelvic organs (N70-N77) | ✗ |
| Noninflammatory disorders of female genital tract (N80-N98) | ✗ |

Table 103: Inclusion and exclusion criteria for cluster 103. Female ratio: 73%, mean age of patients: 30.

| | |
|---|---|
| Intestinal infectious diseases (A00-A09) | ✗ |
| Other bacterial diseases (A30-A49) | ✗ |
| Infections with a predominantly sexual mode of transmission (A50-A64) | ✗ |
| Viral infections characterized by skin and mucous membrane lesions (B00-B09) | ✗ |
| Malignant neoplasms (C00-C97) | ✗ |
| Benign neoplasms (D10-D36) | ✗ |
| Neoplasms of uncertain or unknown behaviour (D37-D48) | ✗ |
| Disorders of thyroid gland (E00-E07) | ✗ |
| Metabolic disorders (E70-E90) | ✗ |
| Mental and behavioural disorders due to psychoactive substance use (F10-F19) | ✗ |
| Schizophrenia, schizotypal and delusional disorders (F20-F29) | ✗ |
| Mood [affective] disorders (F30-F39) | ✗ |
| Neurotic, stress-related and somatoform disorders (F40-F48) | ✗ |
| Episodic and paroxysmal disorders (G40-G47) | ✗ |
| Nerve, nerve root and plexus disorders (G50-G59) | ✗ |
| Disorders of eyelid, lacrimal system and orbit (H00-H06) | ✗ |
| Disorders of lens (H25-H28) | ✗ |
| Disorders of choroid and retina (H30-H36) | ✗ |
| Disorders of ocular muscles, binocular movement, accommodation and refraction (H49-H52) | ✗ |
| Diseases of middle ear and mastoid (H65-H75) | ✗ |
| Diseases of inner ear (H80-H83) | ✓ |
| Other disorders of ear (H90-H95) | ✗ |
| Hypertensive diseases (I10-I15) | ✗ |
| Other forms of heart disease (I30-I52) | ✗ |
| Diseases of veins, lymphatic vessels and lymph nodes, not elsewhere classified (I80-I89) | ✗ |
| Acute upper respiratory infections (J00-J06) | ✗ |
| Influenza and pneumonia (J09-J18) | ✗ |
| Other acute lower respiratory infections (J20-J22) | ✗ |
| Other diseases of upper respiratory tract (J30-J39) | ✗ |
| Chronic lower respiratory diseases (J40-J47) | ✗ |
| Diseases of oral cavity, salivary glands and jaws (K00-K14) | ✗ |
| Diseases of oesophagus, stomach and duodenum (K20-K31) | ✗ |
| Diseases of appendix (K35-K38) | ✗ |
| Hernia (K40-K46) | ✗ |
| Noninfective enteritis and colitis (K50-K52) | ✗ |
| Other diseases of intestines (K55-K63) | ✗ |
| Disorders of gallbladder, biliary tract and pancreas (K80-K87) | ✗ |
| Infections of the skin and subcutaneous tissue (L00-L08) | ✗ |
| Disorders of skin appendages (L60-L75) | ✗ |
| Arthropathies (M00-M25) | ✗ |
| Dorsopathies (M40-M54) | ✗ |
| Soft tissue disorders (M60-M79) | ✗ |
| Osteopathies and chondropathies (M80-M94) | ✗ |
| Renal tubulo-interstitial diseases (N10-N16) | ✗ |
| Urolithiasis (N20-N23) | ✗ |
| Other diseases of urinary system (N30-N39) | ✗ |
| Diseases of male genital organs (N40-N51) | ✗ |
| Disorders of breast (N60-N64) | ✗ |
| Inflammatory diseases of female pelvic organs (N70-N77) | ✗ |
| Noninflammatory disorders of female genital tract (N80-N98) | ✗ |

Table 104: Inclusion and exclusion criteria for cluster 104. Female ratio: 51%, mean age of patients: 45.

| | |
|---|---|
| Intestinal infectious diseases (A00-A09) | ✗ |
| Malignant neoplasms (C00-C97) | ✗ |
| Benign neoplasms (D10-D36) | ✗ |
| Mental and behavioural disorders due to psychoactive substance use (F10-F19) | ✗ |
| Episodic and paroxysmal disorders (G40-G47) | ✗ |
| Disorders of lens (H25-H28) | ✓ |
| Disorders of choroid and retina (H30-H36) | ✓ |
| Hypertensive diseases (I10-I15) | ✗ |
| Other forms of heart disease (I30-I52) | ✗ |
| Diseases of veins, lymphatic vessels and lymph nodes, not elsewhere classified (I80-I89) | ✗ |
| Acute upper respiratory infections (J00-J06) | ✗ |
| Other diseases of upper respiratory tract (J30-J39) | ✗ |
| Diseases of oral cavity, salivary glands and jaws (K00-K14) | ✗ |
| Diseases of appendix (K35-K38) | ✗ |
| Hernia (K40-K46) | ✗ |
| Other diseases of intestines (K55-K63) | ✗ |
| Arthropathies (M00-M25) | ✗ |
| Dorsopathies (M40-M54) | ✗ |
| Soft tissue disorders (M60-M79) | ✗ |
| Diseases of male genital organs (N40-N51) | ✗ |
| Noninflammatory disorders of female genital tract (N80-N98) | ✗ |

Table 105: Inclusion and exclusion criteria for cluster 105. Female ratio: 57%, mean age of patients: 71.

| | |
|---|---|
| Intestinal infectious diseases (A00-A09) | ✗ |
| Metabolic disorders (E70-E90) | ✓ |
| Mental and behavioural disorders due to psychoactive substance use (F10-F19) | ✓ |
| Mood [affective] disorders (F30-F39) | ✗ |
| Neurotic, stress-related and somatoform disorders (F40-F48) | ✗ |
| Hypertensive diseases (I10-I15) | ✗ |
| Other diseases of upper respiratory tract (J30-J39) | ✗ |
| Chronic lower respiratory diseases (J40-J47) | ✗ |
| Other diseases of intestines (K55-K63) | ✗ |
| Diseases of liver (K70-K77) | ✗ |
| Arthropathies (M00-M25) | ✗ |
| Noninflammatory disorders of female genital tract (N80-N98) | ✗ |

Table 106: Inclusion and exclusion criteria for cluster 106. Female ratio: 22%, mean age of patients: 50.

| | |
|---|---|
| Diabetes mellitus (E10-E14) | ✗ |
| Metabolic disorders (E70-E90) | ✓ |
| Hypertensive diseases (I10-I15) | ✓ |
| Ischaemic heart diseases (I20-I25) | ✗ |
| Other forms of heart disease (I30-I52) | ✗ |
| Cerebrovascular diseases (I60-I69) | ✗ |
| Diseases of oesophagus, stomach and duodenum (K20-K31) | ✓ |
| Dorsopathies (M40-M54) | ✗ |

Table 107: Inclusion and exclusion criteria for cluster 107. Female ratio: 51%, mean age of patients: 65.

| | |
|---|---|
| Diabetes mellitus (E10-E14) | ✓ |
| Obesity and other hyperalimentation (E65-E68) | ✓ |
| Metabolic disorders (E70-E90) | ✗ |
| Hypertensive diseases (I10-I15) | ✓ |
| Other forms of heart disease (I30-I52) | ✗ |

Table 108: Inclusion and exclusion criteria for cluster 108. Female ratio: 54%, mean age of patients: 63.

| | |
|---|---|
| Intestinal infectious diseases (A00-A09) | ✗ |
| Malignant neoplasms (C00-C97) | ✗ |
| Mental and behavioural disorders due to psychoactive substance use (F10-F19) | ✗ |
| Hypertensive diseases (I10-I15) | ✗ |
| Diseases of veins, lymphatic vessels and lymph nodes, not elsewhere classified (I80-I89) | ✗ |
| Acute upper respiratory infections (J00-J06) | ✓ |
| Other acute lower respiratory infections (J20-J22) | ✓ |
| Other diseases of upper respiratory tract (J30-J39) | ✗ |
| Hernia (K40-K46) | ✗ |
| Other diseases of intestines (K55-K63) | ✗ |
| Arthropathies (M00-M25) | ✗ |
| Dorsopathies (M40-M54) | ✗ |
| Diseases of male genital organs (N40-N51) | ✗ |
| Noninflammatory disorders of female genital tract (N80-N98) | ✗ |

Table 109: Inclusion and exclusion criteria for cluster 109. Female ratio: 45%, mean age of patients: 10.

| | |
|---|---|
| Episodic and paroxysmal disorders (G40-G47) | ✗ |
| Diseases of middle ear and mastoid (H65-H75) | ✗ |
| Hypertensive diseases (I10-I15) | ✗ |
| Acute upper respiratory infections (J00-J06) | ✗ |
| Other diseases of upper respiratory tract (J30-J39) | ✓ |
| Chronic lower respiratory diseases (J40-J47) | ✓ |
| Other disorders of the musculoskeletal system and connective tissue (M95-M99) | ✗ |

Table 110: Inclusion and exclusion criteria for cluster 110. Female ratio: 41%, mean age of patients: 34.

| | |
|---|---|
| Diseases of middle ear and mastoid (H65-H75) | ✓ |
| Hypertensive diseases (I10-I15) | ✗ |
| Acute upper respiratory infections (J00-J06) | ✓ |
| Other diseases of upper respiratory tract (J30-J39) | ✓ |

Table 111: Inclusion and exclusion criteria for cluster 111. Female ratio: 37%, mean age of patients: 9.

| | |
|---|---|
| Intestinal infectious diseases (A00-A09) | ✓ |
| Episodic and paroxysmal disorders (G40-G47) | ✗ |
| Diseases of middle ear and mastoid (H65-H75) | ✗ |
| Hypertensive diseases (I10-I15) | ✗ |
| Acute upper respiratory infections (J00-J06) | ✗ |
| Other diseases of upper respiratory tract (J30-J39) | ✓ |
| Chronic lower respiratory diseases (J40-J47) | ✗ |
| Other disorders of the musculoskeletal system and connective tissue (M95-M99) | ✗ |

Table 112: Inclusion and exclusion criteria for cluster 112. Female ratio: 48%, mean age of patients: 16.

| | |
|---|---|
| Intestinal infectious diseases (A00-A09) | ✗ |
| Mental and behavioural disorders due to psychoactive substance use (F10-F19) | ✓ |
| Mood [affective] disorders (F30-F39) | ✗ |
| Neurotic, stress-related and somatoform disorders (F40-F48) | ✗ |
| Hypertensive diseases (I10-I15) | ✗ |
| Other diseases of upper respiratory tract (J30-J39) | ✗ |
| Chronic lower respiratory diseases (J40-J47) | ✓ |
| Other diseases of intestines (K55-K63) | ✗ |
| Diseases of liver (K70-K77) | ✗ |
| Arthropathies (M00-M25) | ✗ |
| Noninflammatory disorders of female genital tract (N80-N98) | ✗ |

Table 113: Inclusion and exclusion criteria for cluster 113. Female ratio: 28%, mean age of patients: 55.

| | |
|---|---|
| Intestinal infectious diseases (A00-A09) | ✗ |
| Other bacterial diseases (A30-A49) | ✗ |
| Infections with a predominantly sexual mode of transmission (A50-A64) | ✗ |
| Viral infections characterized by skin and mucous membrane lesions (B00-B09) | ✗ |
| Other viral diseases (B25-B34) | ✗ |
| Malignant neoplasms (C00-C97) | ✗ |
| Benign neoplasms (D10-D36) | ✗ |
| Neoplasms of uncertain or unknown behaviour (D37-D48) | ✗ |
| Disorders of thyroid gland (E00-E07) | ✗ |
| Metabolic disorders (E70-E90) | ✗ |
| Mental and behavioural disorders due to psychoactive substance use (F10-F19) | ✗ |
| Schizophrenia, schizotypal and delusional disorders (F20-F29) | ✗ |
| Mood [affective] disorders (F30-F39) | ✗ |
| Neurotic, stress-related and somatoform disorders (F40-F48) | ✗ |
| Episodic and paroxysmal disorders (G40-G47) | ✗ |
| Nerve, nerve root and plexus disorders (G50-G59) | ✗ |
| Disorders of eyelid, lacrimal system and orbit (H00-H06) | ✗ |
| Disorders of lens (H25-H28) | ✗ |
| Disorders of choroid and retina (H30-H36) | ✗ |
| Disorders of ocular muscles, binocular movement, accommodation and refraction (H49-H52) | ✗ |
| Diseases of middle ear and mastoid (H65-H75) | ✗ |
| Diseases of inner ear (H80-H83) | ✗ |
| Other disorders of ear (H90-H95) | ✗ |
| Hypertensive diseases (I10-I15) | ✗ |
| Other forms of heart disease (I30-I52) | ✗ |
| Cerebrovascular diseases (I60-I69) | ✗ |
| Diseases of veins, lymphatic vessels and lymph nodes, not elsewhere classified (I80-I89) | ✗ |
| Acute upper respiratory infections (J00-J06) | ✗ |
| Influenza and pneumonia (J09-J18) | ✗ |
| Other acute lower respiratory infections (J20-J22) | ✗ |
| Other diseases of upper respiratory tract (J30-J39) | ✗ |
| Chronic lower respiratory diseases (J40-J47) | ✗ |
| Diseases of oral cavity, salivary glands and jaws (K00-K14) | ✗ |
| Diseases of oesophagus, stomach and duodenum (K20-K31) | ✗ |
| Diseases of appendix (K35-K38) | ✗ |
| Hernia (K40-K46) | ✗ |
| Noninfective enteritis and colitis (K50-K52) | ✗ |
| Other diseases of intestines (K55-K63) | ✗ |
| Disorders of gallbladder, biliary tract and pancreas (K80-K87) | ✗ |
| Infections of the skin and subcutaneous tissue (L00-L08) | ✗ |
| Disorders of skin appendages (L60-L75) | ✗ |
| Other disorders of the skin and subcutaneous tissue (L80-L99) | ✓ |
| Arthropathies (M00-M25) | ✗ |
| Dorsopathies (M40-M54) | ✗ |
| Soft tissue disorders (M60-M79) | ✗ |
| Osteopathies and chondropathies (M80-M94) | ✗ |
| Renal tubulo-interstitial diseases (N10-N16) | ✗ |
| Urolithiasis (N20-N23) | ✗ |
| Other diseases of urinary system (N30-N39) | ✗ |
| Diseases of male genital organs (N40-N51) | ✗ |
| Disorders of breast (N60-N64) | ✗ |
| Inflammatory diseases of female pelvic organs (N70-N77) | ✗ |
| Noninflammatory disorders of female genital tract (N80-N98) | ✗ |

Table 114: Inclusion and exclusion criteria for cluster 114. Female ratio: 55%, mean age of patients: 39.

| | |
|---|---|
| Intestinal infectious diseases (A00-A09) | ✗ |
| Mental and behavioural disorders due to psychoactive substance use (F10-F19) | ✗ |
| Hypertensive diseases (I10-I15) | ✗ |
| Diseases of veins, lymphatic vessels and lymph nodes, not elsewhere classified (I80-I89) | ✓ |
| Other diseases of upper respiratory tract (J30-J39) | ✗ |
| Diseases of oesophagus, stomach and duodenum (K20-K31) | ✓ |
| Other diseases of intestines (K55-K63) | ✗ |
| Arthropathies (M00-M25) | ✗ |
| Dorsopathies (M40-M54) | ✗ |
| Noninflammatory disorders of female genital tract (N80-N98) | ✗ |

Table 115: Inclusion and exclusion criteria for cluster 115. Female ratio: 45%, mean age of patients: 57.

| | |
|---|---|
| Intestinal infectious diseases (A00-A09) | ✗ |
| Malignant neoplasms (C00-C97) | ✗ |
| Hypertensive diseases (I10-I15) | ✗ |
| Diseases of veins, lymphatic vessels and lymph nodes, not elsewhere classified (I80-I89) | ✗ |
| Other diseases of upper respiratory tract (J30-J39) | ✗ |
| Diseases of oesophagus, stomach and duodenum (K20-K31) | ✗ |
| Hernia (K40-K46) | ✓ |
| Other diseases of intestines (K55-K63) | ✓ |
| Arthropathies (M00-M25) | ✗ |
| Noninflammatory disorders of female genital tract (N80-N98) | ✗ |

Table 116: Inclusion and exclusion criteria for cluster 116. Female ratio: 29%, mean age of patients: 58.

| | |
|---|---|
| Hypertensive diseases (I10-I15) | ✗ |
| Other diseases of upper respiratory tract (J30-J39) | ✗ |
| Arthropathies (M00-M25) | ✓ |
| Soft tissue disorders (M60-M79) | ✓ |
| Osteopathies and chondropathies (M80-M94) | ✓ |

Table 117: Inclusion and exclusion criteria for cluster 117. Female ratio: 52%, mean age of patients: 56.

| | |
|---|---|
| Organic, including symptomatic, mental disorders (F00-F09) | ✓ |
| Hypertensive diseases (I10-I15) | ✓ |
| Other forms of heart disease (I30-I52) | ✓ |
| Diseases of arteries, arterioles and capillaries (I70-I79) | ✗ |
| Renal failure (N17-N19) | ✓ |

Table 118: Inclusion and exclusion criteria for cluster 118. Female ratio: 65%, mean age of patients: 86.

| | |
|---|---|
| Intestinal infectious diseases (A00-A09) | ✗ |
| Mental and behavioural disorders due to psychoactive substance use (F10-F19) | ✗ |
| Hypertensive diseases (I10-I15) | ✗ |
| Diseases of veins, lymphatic vessels and lymph nodes, not elsewhere classified (I80-I89) | ✗ |
| Other diseases of upper respiratory tract (J30-J39) | ✗ |
| Other diseases of intestines (K55-K63) | ✗ |
| Arthropathies (M00-M25) | ✗ |
| Dorsopathies (M40-M54) | ✗ |
| Other diseases of urinary system (N30-N39) | ✓ |
| Diseases of male genital organs (N40-N51) | ✓ |
| Noninflammatory disorders of female genital tract (N80-N98) | ✗ |

Table 119: Inclusion and exclusion criteria for cluster 119. Female ratio: 0%, mean age of patients: 54.

| | |
|---|---|
| Intestinal infectious diseases (A00-A09) | ✗ |
| Other bacterial diseases (A30-A49) | ✗ |
| Infections with a predominantly sexual mode of transmission (A50-A64) | ✗ |
| Viral infections characterized by skin and mucous membrane lesions (B00-B09) | ✗ |
| Other viral diseases (B25-B34) | ✓ |
| Malignant neoplasms (C00-C97) | ✗ |
| Benign neoplasms (D10-D36) | ✗ |
| Neoplasms of uncertain or unknown behaviour (D37-D48) | ✗ |
| Disorders of thyroid gland (E00-E07) | ✗ |
| Metabolic disorders (E70-E90) | ✗ |
| Mental and behavioural disorders due to psychoactive substance use (F10-F19) | ✗ |
| Schizophrenia, schizotypal and delusional disorders (F20-F29) | ✗ |
| Mood [affective] disorders (F30-F39) | ✗ |
| Neurotic, stress-related and somatoform disorders (F40-F48) | ✗ |
| Episodic and paroxysmal disorders (G40-G47) | ✗ |
| Nerve, nerve root and plexus disorders (G50-G59) | ✗ |
| Disorders of eyelid, lacrimal system and orbit (H00-H06) | ✗ |
| Disorders of lens (H25-H28) | ✗ |
| Disorders of choroid and retina (H30-H36) | ✗ |
| Disorders of ocular muscles, binocular movement, accommodation and refraction (H49-H52) | ✗ |
| Diseases of middle ear and mastoid (H65-H75) | ✗ |
| Diseases of inner ear (H80-H83) | ✗ |
| Other disorders of ear (H90-H95) | ✗ |
| Hypertensive diseases (I10-I15) | ✗ |
| Other forms of heart disease (I30-I52) | ✗ |
| Diseases of veins, lymphatic vessels and lymph nodes, not elsewhere classified (I80-I89) | ✗ |
| Acute upper respiratory infections (J00-J06) | ✗ |
| Influenza and pneumonia (J09-J18) | ✗ |
| Other acute lower respiratory infections (J20-J22) | ✗ |
| Other diseases of upper respiratory tract (J30-J39) | ✗ |
| Chronic lower respiratory diseases (J40-J47) | ✗ |
| Diseases of oral cavity, salivary glands and jaws (K00-K14) | ✗ |
| Diseases of oesophagus, stomach and duodenum (K20-K31) | ✗ |
| Diseases of appendix (K35-K38) | ✗ |
| Hernia (K40-K46) | ✗ |
| Noninfective enteritis and colitis (K50-K52) | ✗ |
| Other diseases of intestines (K55-K63) | ✗ |
| Disorders of gallbladder, biliary tract and pancreas (K80-K87) | ✗ |
| Infections of the skin and subcutaneous tissue (L00-L08) | ✗ |
| Disorders of skin appendages (L60-L75) | ✗ |
| Arthropathies (M00-M25) | ✗ |
| Dorsopathies (M40-M54) | ✗ |
| Soft tissue disorders (M60-M79) | ✗ |
| Osteopathies and chondropathies (M80-M94) | ✗ |
| Renal tubulo-interstitial diseases (N10-N16) | ✗ |
| Urolithiasis (N20-N23) | ✗ |
| Other diseases of urinary system (N30-N39) | ✗ |
| Diseases of male genital organs (N40-N51) | ✗ |
| Disorders of breast (N60-N64) | ✗ |
| Inflammatory diseases of female pelvic organs (N70-N77) | ✗ |
| Noninflammatory disorders of female genital tract (N80-N98) | ✗ |

Table 120: Inclusion and exclusion criteria for cluster 120. Female ratio: 47%, mean age of patients: 18.

| | |
|---|---|
| Intestinal infectious diseases (A00-A09) | ✗ |
| Other bacterial diseases (A30-A49) | ✗ |
| Infections with a predominantly sexual mode of transmission (A50-A64) | ✗ |
| Viral infections characterized by skin and mucous membrane lesions (B00-B09) | ✗ |
| Other viral diseases (B25-B34) | ✗ |
| Malignant neoplasms (C00-C97) | ✗ |
| In situ neoplasms (D00-D09) | ✗ |
| Benign neoplasms (D10-D36) | ✗ |
| Neoplasms of uncertain or unknown behaviour (D37-D48) | ✗ |
| Disorders of thyroid gland (E00-E07) | ✗ |
| Diabetes mellitus (E10-E14) | ✗ |
| Obesity and other hyperalimentation (E65-E68) | ✓ |
| Metabolic disorders (E70-E90) | ✗ |
| Mental and behavioural disorders due to psychoactive substance use (F10-F19) | ✗ |
| Schizophrenia, schizotypal and delusional disorders (F20-F29) | ✗ |
| Mood [affective] disorders (F30-F39) | ✗ |
| Neurotic, stress-related and somatoform disorders (F40-F48) | ✗ |
| Behavioural and emotional disorders with onset usually occurring in childhood and adolescence (F90-F98) | ✗ |
| Episodic and paroxysmal disorders (G40-G47) | ✗ |
| Nerve, nerve root and plexus disorders (G50-G59) | ✗ |
| Disorders of eyelid, lacrimal system and orbit (H00-H06) | ✗ |
| Disorders of lens (H25-H28) | ✗ |
| Disorders of choroid and retina (H30-H36) | ✗ |
| Disorders of ocular muscles, binocular movement, accommodation and refraction (H49-H52) | ✗ |
| Diseases of middle ear and mastoid (H65-H75) | ✗ |
| Diseases of inner ear (H80-H83) | ✗ |
| Other disorders of ear (H90-H95) | ✗ |
| Hypertensive diseases (I10-I15) | ✗ |
| Ischaemic heart diseases (I20-I25) | ✗ |
| Other forms of heart disease (I30-I52) | ✗ |
| Cerebrovascular diseases (I60-I69) | ✗ |
| Diseases of veins, lymphatic vessels and lymph nodes, not elsewhere classified (I80-I89) | ✗ |
| Acute upper respiratory infections (J00-J06) | ✗ |
| Influenza and pneumonia (J09-J18) | ✗ |
| Other acute lower respiratory infections (J20-J22) | ✗ |
| Other diseases of upper respiratory tract (J30-J39) | ✗ |
| Chronic lower respiratory diseases (J40-J47) | ✗ |
| Diseases of oral cavity, salivary glands and jaws (K00-K14) | ✗ |
| Diseases of oesophagus, stomach and duodenum (K20-K31) | ✗ |
| Diseases of appendix (K35-K38) | ✗ |
| Hernia (K40-K46) | ✗ |
| Noninfective enteritis and colitis (K50-K52) | ✗ |
| Other diseases of intestines (K55-K63) | ✗ |
| Disorders of gallbladder, biliary tract and pancreas (K80-K87) | ✗ |
| Infections of the skin and subcutaneous tissue (L00-L08) | ✗ |
| Dermatitis and eczema (L20-L30) | ✗ |
| Urticaria and erythema (L50-L54) | ✗ |
| Disorders of skin appendages (L60-L75) | ✗ |
| Other disorders of the skin and subcutaneous tissue (L80-L99) | ✗ |
| Arthropathies (M00-M25) | ✗ |
| Dorsopathies (M40-M54) | ✗ |
| Soft tissue disorders (M60-M79) | ✗ |
| Osteopathies and chondropathies (M80-M94) | ✗ |
| Renal tubulo-interstitial diseases (N10-N16) | ✗ |
| Urolithiasis (N20-N23) | ✗ |
| Other diseases of urinary system (N30-N39) | ✗ |
| Diseases of male genital organs (N40-N51) | ✗ |
| Disorders of breast (N60-N64) | ✗ |
| Inflammatory diseases of female pelvic organs (N70-N77) | ✗ |
| Noninflammatory disorders of female genital tract (N80-N98) | ✗ |

Table 121: Inclusion and exclusion criteria for cluster 121. Female ratio: 71%, mean age of patients: 33.

| | |
|---|---|
| Intestinal infectious diseases (A00-A09) | ✗ |
| Metabolic disorders (E70-E90) | ✗ |
| Mental and behavioural disorders due to psychoactive substance use (F10-F19) | ✗ |
| Mood [affective] disorders (F30-F39) | ✓ |
| Hypertensive diseases (I10-I15) | ✗ |
| Other diseases of upper respiratory tract (J30-J39) | ✗ |
| Other diseases of intestines (K55-K63) | ✗ |
| Arthropathies (M00-M25) | ✗ |
| Dorsopathies (M40-M54) | ✓ |
| Noninflammatory disorders of female genital tract (N80-N98) | ✗ |

Table 122: Inclusion and exclusion criteria for cluster 122. Female ratio: 70%, mean age of patients: 52.

| | |
|---|---|
| Intestinal infectious diseases (A00-A09) | ✗ |
| Other bacterial diseases (A30-A49) | ✗ |
| Infections with a predominantly sexual mode of transmission (A50-A64) | ✗ |
| Viral infections characterized by skin and mucous membrane lesions (B00-B09) | ✗ |
| Other viral diseases (B25-B34) | ✗ |
| Malignant neoplasms (C00-C97) | ✗ |
| Benign neoplasms (D10-D36) | ✗ |
| Neoplasms of uncertain or unknown behaviour (D37-D48) | ✗ |
| Disorders of thyroid gland (E00-E07) | ✗ |
| Metabolic disorders (E70-E90) | ✗ |
| Mental and behavioural disorders due to psychoactive substance use (F10-F19) | ✗ |
| Schizophrenia, schizotypal and delusional disorders (F20-F29) | ✗ |
| Mood [affective] disorders (F30-F39) | ✗ |
| Neurotic, stress-related and somatoform disorders (F40-F48) | ✗ |
| Episodic and paroxysmal disorders (G40-G47) | ✗ |
| Nerve, nerve root and plexus disorders (G50-G59) | ✗ |
| Disorders of eyelid, lacrimal system and orbit (H00-H06) | ✗ |
| Disorders of lens (H25-H28) | ✗ |
| Disorders of choroid and retina (H30-H36) | ✗ |
| Disorders of ocular muscles, binocular movement, accommodation and refraction (H49-H52) | ✗ |
| Diseases of middle ear and mastoid (H65-H75) | ✗ |
| Diseases of inner ear (H80-H83) | ✗ |
| Other disorders of ear (H90-H95) | ✗ |
| Hypertensive diseases (I10-I15) | ✗ |
| Ischaemic heart diseases (I20-I25) | ✗ |
| Other forms of heart disease (I30-I52) | ✗ |
| Cerebrovascular diseases (I60-I69) | ✗ |
| Diseases of veins, lymphatic vessels and lymph nodes, not elsewhere classified (I80-I89) | ✗ |
| Acute upper respiratory infections (J00-J06) | ✗ |
| Influenza and pneumonia (J09-J18) | ✗ |
| Other acute lower respiratory infections (J20-J22) | ✗ |
| Other diseases of upper respiratory tract (J30-J39) | ✗ |
| Chronic lower respiratory diseases (J40-J47) | ✗ |
| Diseases of oral cavity, salivary glands and jaws (K00-K14) | ✗ |
| Diseases of oesophagus, stomach and duodenum (K20-K31) | ✗ |
| Diseases of appendix (K35-K38) | ✗ |
| Hernia (K40-K46) | ✗ |
| Noninfective enteritis and colitis (K50-K52) | ✗ |
| Other diseases of intestines (K55-K63) | ✗ |
| Disorders of gallbladder, biliary tract and pancreas (K80-K87) | ✗ |
| Infections of the skin and subcutaneous tissue (L00-L08) | ✗ |
| Dermatitis and eczema (L20-L30) | ✓ |
| Disorders of skin appendages (L60-L75) | ✗ |
| Other disorders of the skin and subcutaneous tissue (L80-L99) | ✗ |
| Arthropathies (M00-M25) | ✗ |
| Dorsopathies (M40-M54) | ✗ |
| Soft tissue disorders (M60-M79) | ✗ |
| Osteopathies and chondropathies (M80-M94) | ✗ |
| Renal tubulo-interstitial diseases (N10-N16) | ✗ |
| Urolithiasis (N20-N23) | ✗ |
| Other diseases of urinary system (N30-N39) | ✗ |
| Diseases of male genital organs (N40-N51) | ✗ |
| Disorders of breast (N60-N64) | ✗ |
| Inflammatory diseases of female pelvic organs (N70-N77) | ✗ |
| Noninflammatory disorders of female genital tract (N80-N98) | ✗ |

Table 123: Inclusion and exclusion criteria for cluster 123. Female ratio: 57%, mean age of patients: 27.

| | |
|---|---|
| Intestinal infectious diseases (A00-A09) | ✗ |
| Benign neoplasms (D10-D36) | ✓ |
| Episodic and paroxysmal disorders (G40-G47) | ✗ |
| Diseases of middle ear and mastoid (H65-H75) | ✗ |
| Hypertensive diseases (I10-I15) | ✗ |
| Acute upper respiratory infections (J00-J06) | ✗ |
| Other diseases of upper respiratory tract (J30-J39) | ✓ |
| Chronic lower respiratory diseases (J40-J47) | ✗ |
| Diseases of oral cavity, salivary glands and jaws (K00-K14) | ✗ |
| Other disorders of the musculoskeletal system and connective tissue (M95-M99) | ✗ |

Table 124: Inclusion and exclusion criteria for cluster 124. Female ratio: 57%, mean age of patients: 40.

| | |
|---|---|
| Episodic and paroxysmal disorders (G40-G47) | ✗ |
| Diseases of middle ear and mastoid (H65-H75) | ✗ |
| Hypertensive diseases (I10-I15) | ✗ |
| Acute upper respiratory infections (J00-J06) | ✗ |
| Other diseases of upper respiratory tract (J30-J39) | ✓ |
| Other disorders of the musculoskeletal system and connective tissue (M95-M99) | ✓ |

Table 125: Inclusion and exclusion criteria for cluster 125. Female ratio: 43%, mean age of patients: 34.

| | |
|---|---|
| Intestinal infectious diseases (A00-A09) | ✗ |
| Episodic and paroxysmal disorders (G40-G47) | ✗ |
| Diseases of middle ear and mastoid (H65-H75) | ✗ |
| Hypertensive diseases (I10-I15) | ✗ |
| Acute upper respiratory infections (J00-J06) | ✗ |
| Other diseases of upper respiratory tract (J30-J39) | ✓ |
| Chronic lower respiratory diseases (J40-J47) | ✗ |
| Diseases of oral cavity, salivary glands and jaws (K00-K14) | ✓ |
| Other disorders of the musculoskeletal system and connective tissue (M95-M99) | ✗ |

Table 126: Inclusion and exclusion criteria for cluster 126. Female ratio: 47%, mean age of patients: 28.

| | |
|---|---|
| Intestinal infectious diseases (A00-A09) | ✓ |
| Metabolic disorders (E70-E90) | ✗ |
| Hypertensive diseases (I10-I15) | ✗ |
| Acute upper respiratory infections (J00-J06) | ✗ |
| Other acute lower respiratory infections (J20-J22) | ✓ |
| Other diseases of upper respiratory tract (J30-J39) | ✗ |
| Arthropathies (M00-M25) | ✗ |
| Noninflammatory disorders of female genital tract (N80-N98) | ✗ |

Table 127: Inclusion and exclusion criteria for cluster 127. Female ratio: 42%, mean age of patients: 10.

| | |
|---|---|
| Intestinal infectious diseases (A00-A09) | ✗ |
| Other bacterial diseases (A30-A49) | ✗ |
| Infections with a predominantly sexual mode of transmission (A50-A64) | ✗ |
| Viral infections characterized by skin and mucous membrane lesions (B00-B09) | ✗ |
| Other viral diseases (B25-B34) | ✗ |
| Malignant neoplasms (C00-C97) | ✗ |
| In situ neoplasms (D00-D09) | ✓ |
| Benign neoplasms (D10-D36) | ✗ |
| Neoplasms of uncertain or unknown behaviour (D37-D48) | ✗ |
| Disorders of thyroid gland (E00-E07) | ✗ |
| Metabolic disorders (E70-E90) | ✗ |
| Mental and behavioural disorders due to psychoactive substance use (F10-F19) | ✗ |
| Schizophrenia, schizotypal and delusional disorders (F20-F29) | ✗ |
| Mood [affective] disorders (F30-F39) | ✗ |
| Neurotic, stress-related and somatoform disorders (F40-F48) | ✗ |
| Episodic and paroxysmal disorders (G40-G47) | ✗ |
| Nerve, nerve root and plexus disorders (G50-G59) | ✗ |
| Disorders of eyelid, lacrimal system and orbit (H00-H06) | ✗ |
| Disorders of lens (H25-H28) | ✗ |
| Disorders of choroid and retina (H30-H36) | ✗ |
| Disorders of ocular muscles, binocular movement, accommodation and refraction (H49-H52) | ✗ |
| Diseases of middle ear and mastoid (H65-H75) | ✗ |
| Diseases of inner ear (H80-H83) | ✗ |
| Other disorders of ear (H90-H95) | ✗ |
| Hypertensive diseases (I10-I15) | ✗ |
| Ischaemic heart diseases (I20-I25) | ✗ |
| Other forms of heart disease (I30-I52) | ✗ |
| Cerebrovascular diseases (I60-I69) | ✗ |
| Diseases of veins, lymphatic vessels and lymph nodes, not elsewhere classified (I80-I89) | ✗ |
| Acute upper respiratory infections (J00-J06) | ✗ |
| Influenza and pneumonia (J09-J18) | ✗ |
| Other acute lower respiratory infections (J20-J22) | ✗ |
| Other diseases of upper respiratory tract (J30-J39) | ✗ |
| Chronic lower respiratory diseases (J40-J47) | ✗ |
| Diseases of oral cavity, salivary glands and jaws (K00-K14) | ✗ |
| Diseases of oesophagus, stomach and duodenum (K20-K31) | ✗ |
| Diseases of appendix (K35-K38) | ✗ |
| Hernia (K40-K46) | ✗ |
| Noninfective enteritis and colitis (K50-K52) | ✗ |
| Other diseases of intestines (K55-K63) | ✗ |
| Disorders of gallbladder, biliary tract and pancreas (K80-K87) | ✗ |
| Infections of the skin and subcutaneous tissue (L00-L08) | ✗ |
| Dermatitis and eczema (L20-L30) | ✗ |
| Disorders of skin appendages (L60-L75) | ✗ |
| Other disorders of the skin and subcutaneous tissue (L80-L99) | ✗ |
| Arthropathies (M00-M25) | ✗ |
| Dorsopathies (M40-M54) | ✗ |
| Soft tissue disorders (M60-M79) | ✗ |
| Osteopathies and chondropathies (M80-M94) | ✗ |
| Renal tubulo-interstitial diseases (N10-N16) | ✗ |
| Urolithiasis (N20-N23) | ✗ |
| Other diseases of urinary system (N30-N39) | ✗ |
| Diseases of male genital organs (N40-N51) | ✗ |
| Disorders of breast (N60-N64) | ✗ |
| Inflammatory diseases of female pelvic organs (N70-N77) | ✗ |
| Noninflammatory disorders of female genital tract (N80-N98) | ✗ |

Table 128: Inclusion and exclusion criteria for cluster 128. Female ratio: 97%, mean age of patients: 38.

| | |
|---|---|
| Intestinal infectious diseases (A00-A09) | ✗ |
| Other bacterial diseases (A30-A49) | ✗ |
| Infections with a predominantly sexual mode of transmission (A50-A64) | ✗ |
| Viral infections characterized by skin and mucous membrane lesions (B00-B09) | ✗ |
| Other viral diseases (B25-B34) | ✗ |
| Malignant neoplasms (C00-C97) | ✗ |
| In situ neoplasms (D00-D09) | ✗ |
| Benign neoplasms (D10-D36) | ✗ |
| Neoplasms of uncertain or unknown behaviour (D37-D48) | ✗ |
| Disorders of thyroid gland (E00-E07) | ✗ |
| Diabetes mellitus (E10-E14) | ✗ |
| Obesity and other hyperalimentation (E65-E68) | ✗ |
| Metabolic disorders (E70-E90) | ✗ |
| Mental and behavioural disorders due to psychoactive substance use (F10-F19) | ✗ |
| Schizophrenia, schizotypal and delusional disorders (F20-F29) | ✗ |
| Mood [affective] disorders (F30-F39) | ✗ |
| Neurotic, stress-related and somatoform disorders (F40-F48) | ✗ |
| Behavioural and emotional disorders with onset usually occurring in childhood and adolescence (F90-F98) | ✗ |
| Episodic and paroxysmal disorders (G40-G47) | ✗ |
| Nerve, nerve root and plexus disorders (G50-G59) | ✗ |
| Disorders of eyelid, lacrimal system and orbit (H00-H06) | ✗ |
| Disorders of lens (H25-H28) | ✗ |
| Disorders of choroid and retina (H30-H36) | ✗ |
| Disorders of ocular muscles, binocular movement, accommodation and refraction (H49-H52) | ✗ |
| Diseases of middle ear and mastoid (H65-H75) | ✗ |
| Diseases of inner ear (H80-H83) | ✗ |
| Other disorders of ear (H90-H95) | ✗ |
| Hypertensive diseases (I10-I15) | ✗ |
| Ischaemic heart diseases (I20-I25) | ✗ |
| Other forms of heart disease (I30-I52) | ✗ |
| Cerebrovascular diseases (I60-I69) | ✗ |
| Diseases of veins, lymphatic vessels and lymph nodes, not elsewhere classified (I80-I89) | ✗ |
| Other and unspecified disorders of the circulatory system (I95-I99) | ✓ |
| Acute upper respiratory infections (J00-J06) | ✗ |
| Influenza and pneumonia (J09-J18) | ✗ |
| Other acute lower respiratory infections (J20-J22) | ✗ |
| Other diseases of upper respiratory tract (J30-J39) | ✗ |
| Chronic lower respiratory diseases (J40-J47) | ✗ |
| Diseases of oral cavity, salivary glands and jaws (K00-K14) | ✗ |
| Diseases of oesophagus, stomach and duodenum (K20-K31) | ✗ |
| Diseases of appendix (K35-K38) | ✗ |
| Hernia (K40-K46) | ✗ |
| Noninfective enteritis and colitis (K50-K52) | ✗ |
| Other diseases of intestines (K55-K63) | ✗ |
| Disorders of gallbladder, biliary tract and pancreas (K80-K87) | ✗ |
| Infections of the skin and subcutaneous tissue (L00-L08) | ✗ |
| Dermatitis and eczema (L20-L30) | ✗ |
| Urticaria and erythema (L50-L54) | ✗ |
| Disorders of skin appendages (L60-L75) | ✗ |
| Other disorders of the skin and subcutaneous tissue (L80-L99) | ✗ |
| Arthropathies (M00-M25) | ✗ |
| Dorsopathies (M40-M54) | ✗ |
| Soft tissue disorders (M60-M79) | ✗ |
| Osteopathies and chondropathies (M80-M94) | ✗ |
| Renal tubulo-interstitial diseases (N10-N16) | ✗ |
| Urolithiasis (N20-N23) | ✗ |
| Other diseases of urinary system (N30-N39) | ✗ |
| Diseases of male genital organs (N40-N51) | ✗ |
| Disorders of breast (N60-N64) | ✗ |
| Inflammatory diseases of female pelvic organs (N70-N77) | ✗ |
| Noninflammatory disorders of female genital tract (N80-N98) | ✗ |

Table 129: Inclusion and exclusion criteria for cluster 129. Female ratio: 62%, mean age of patients: 31.

| | |
|---|---|
| Intestinal infectious diseases (A00-A09) | ✗ |
| Other bacterial diseases (A30-A49) | ✗ |
| Infections with a predominantly sexual mode of transmission (A50-A64) | ✗ |
| Viral infections characterized by skin and mucous membrane lesions (B00-B09) | ✗ |
| Other viral diseases (B25-B34) | ✗ |
| Malignant neoplasms (C00-C97) | ✗ |
| In situ neoplasms (D00-D09) | ✗ |
| Benign neoplasms (D10-D36) | ✗ |
| Neoplasms of uncertain or unknown behaviour (D37-D48) | ✗ |
| Disorders of thyroid gland (E00-E07) | ✗ |
| Diabetes mellitus (E10-E14) | ✗ |
| Metabolic disorders (E70-E90) | ✗ |
| Mental and behavioural disorders due to psychoactive substance use (F10-F19) | ✗ |
| Schizophrenia, schizotypal and delusional disorders (F20-F29) | ✗ |
| Mood [affective] disorders (F30-F39) | ✗ |
| Neurotic, stress-related and somatoform disorders (F40-F48) | ✗ |
| Episodic and paroxysmal disorders (G40-G47) | ✗ |
| Nerve, nerve root and plexus disorders (G50-G59) | ✗ |
| Disorders of eyelid, lacrimal system and orbit (H00-H06) | ✗ |
| Disorders of lens (H25-H28) | ✗ |
| Disorders of choroid and retina (H30-H36) | ✗ |
| Disorders of ocular muscles, binocular movement, accommodation and refraction (H49-H52) | ✗ |
| Diseases of middle ear and mastoid (H65-H75) | ✗ |
| Diseases of inner ear (H80-H83) | ✗ |
| Other disorders of ear (H90-H95) | ✗ |
| Hypertensive diseases (I10-I15) | ✗ |
| Ischaemic heart diseases (I20-I25) | ✗ |
| Other forms of heart disease (I30-I52) | ✗ |
| Cerebrovascular diseases (I60-I69) | ✗ |
| Diseases of veins, lymphatic vessels and lymph nodes, not elsewhere classified (I80-I89) | ✗ |
| Acute upper respiratory infections (J00-J06) | ✗ |
| Influenza and pneumonia (J09-J18) | ✗ |
| Other acute lower respiratory infections (J20-J22) | ✗ |
| Other diseases of upper respiratory tract (J30-J39) | ✗ |
| Chronic lower respiratory diseases (J40-J47) | ✗ |
| Diseases of oral cavity, salivary glands and jaws (K00-K14) | ✗ |
| Diseases of oesophagus, stomach and duodenum (K20-K31) | ✗ |
| Diseases of appendix (K35-K38) | ✗ |
| Hernia (K40-K46) | ✗ |
| Noninfective enteritis and colitis (K50-K52) | ✗ |
| Other diseases of intestines (K55-K63) | ✗ |
| Disorders of gallbladder, biliary tract and pancreas (K80-K87) | ✗ |
| Infections of the skin and subcutaneous tissue (L00-L08) | ✗ |
| Dermatitis and eczema (L20-L30) | ✗ |
| Urticaria and erythema (L50-L54) | ✓ |
| Disorders of skin appendages (L60-L75) | ✗ |
| Other disorders of the skin and subcutaneous tissue (L80-L99) | ✗ |
| Arthropathies (M00-M25) | ✗ |
| Dorsopathies (M40-M54) | ✗ |
| Soft tissue disorders (M60-M79) | ✗ |
| Osteopathies and chondropathies (M80-M94) | ✗ |
| Renal tubulo-interstitial diseases (N10-N16) | ✗ |
| Urolithiasis (N20-N23) | ✗ |
| Other diseases of urinary system (N30-N39) | ✗ |
| Diseases of male genital organs (N40-N51) | ✗ |
| Disorders of breast (N60-N64) | ✗ |
| Inflammatory diseases of female pelvic organs (N70-N77) | ✗ |
| Noninflammatory disorders of female genital tract (N80-N98) | ✗ |

Table 130: Inclusion and exclusion criteria for cluster 130. Female ratio: 54%, mean age of patients: 27.

| | |
|---|---|
| Intestinal infectious diseases (A00-A09) | ✗ |
| Other bacterial diseases (A30-A49) | ✗ |
| Infections with a predominantly sexual mode of transmission (A50-A64) | ✗ |
| Viral infections characterized by skin and mucous membrane lesions (B00-B09) | ✗ |
| Other viral diseases (B25-B34) | ✗ |
| Malignant neoplasms (C00-C97) | ✗ |
| In situ neoplasms (D00-D09) | ✗ |
| Benign neoplasms (D10-D36) | ✗ |
| Neoplasms of uncertain or unknown behaviour (D37-D48) | ✗ |
| Disorders of thyroid gland (E00-E07) | ✗ |
| Diabetes mellitus (E10-E14) | ✗ |
| Metabolic disorders (E70-E90) | ✗ |
| Mental and behavioural disorders due to psychoactive substance use (F10-F19) | ✗ |
| Schizophrenia, schizotypal and delusional disorders (F20-F29) | ✗ |
| Mood [affective] disorders (F30-F39) | ✗ |
| Neurotic, stress-related and somatoform disorders (F40-F48) | ✗ |
| Behavioural and emotional disorders with onset usually occurring in childhood and adolescence (F90-F98) | ✓ |
| Episodic and paroxysmal disorders (G40-G47) | ✗ |
| Nerve, nerve root and plexus disorders (G50-G59) | ✗ |
| Disorders of eyelid, lacrimal system and orbit (H00-H06) | ✗ |
| Disorders of lens (H25-H28) | ✗ |
| Disorders of choroid and retina (H30-H36) | ✗ |
| Disorders of ocular muscles, binocular movement, accommodation and refraction (H49-H52) | ✗ |
| Diseases of middle ear and mastoid (H65-H75) | ✗ |
| Diseases of inner ear (H80-H83) | ✗ |
| Other disorders of ear (H90-H95) | ✗ |
| Hypertensive diseases (I10-I15) | ✗ |
| Ischaemic heart diseases (I20-I25) | ✗ |
| Other forms of heart disease (I30-I52) | ✗ |
| Cerebrovascular diseases (I60-I69) | ✗ |
| Diseases of veins, lymphatic vessels and lymph nodes, not elsewhere classified (I80-I89) | ✗ |
| Acute upper respiratory infections (J00-J06) | ✗ |
| Influenza and pneumonia (J09-J18) | ✗ |
| Other acute lower respiratory infections (J20-J22) | ✗ |
| Other diseases of upper respiratory tract (J30-J39) | ✗ |
| Chronic lower respiratory diseases (J40-J47) | ✗ |
| Diseases of oral cavity, salivary glands and jaws (K00-K14) | ✗ |
| Diseases of oesophagus, stomach and duodenum (K20-K31) | ✗ |
| Diseases of appendix (K35-K38) | ✗ |
| Hernia (K40-K46) | ✗ |
| Noninfective enteritis and colitis (K50-K52) | ✗ |
| Other diseases of intestines (K55-K63) | ✗ |
| Disorders of gallbladder, biliary tract and pancreas (K80-K87) | ✗ |
| Infections of the skin and subcutaneous tissue (L00-L08) | ✗ |
| Dermatitis and eczema (L20-L30) | ✗ |
| Urticaria and erythema (L50-L54) | ✗ |
| Disorders of skin appendages (L60-L75) | ✗ |
| Other disorders of the skin and subcutaneous tissue (L80-L99) | ✗ |
| Arthropathies (M00-M25) | ✗ |
| Dorsopathies (M40-M54) | ✗ |
| Soft tissue disorders (M60-M79) | ✗ |
| Osteopathies and chondropathies (M80-M94) | ✗ |
| Renal tubulo-interstitial diseases (N10-N16) | ✗ |
| Urolithiasis (N20-N23) | ✗ |
| Other diseases of urinary system (N30-N39) | ✗ |
| Diseases of male genital organs (N40-N51) | ✗ |
| Disorders of breast (N60-N64) | ✗ |
| Inflammatory diseases of female pelvic organs (N70-N77) | ✗ |
| Noninflammatory disorders of female genital tract (N80-N98) | ✗ |

Table 131: Inclusion and exclusion criteria for cluster 131. Female ratio: 29%, mean age of patients: 14.

# 9 List of diagnosis blocks

The following table lists all the diagnosis blocks considered in this work.

| ID | ICD-10 | Description |
| --- | --- | --- |
| 0 | A00-A09 | Intestinal infectious diseases |
| 1 | A15-A19 | Tuberculosis |
| 2 | A20-A28 | Certain zoonotic bacterial diseases |
| 3 | A30-A49 | Other bacterial diseases |
| 4 | A50-A64 | Infections with a predominantly sexual mode of transmission |
| 5 | A65-A69 | Other spirochaetal diseases |
| 6 | A70-A74 | Other diseases caused by chlamydiae |
| 7 | A75-A79 | Rickettsioses |
| 8 | A80-A91 | Viral infections of the central nervous system |
| 9 | A92-A99 | Arthropod-borne viral fevers and viral haemorrhagic fevers |
| 10 | B00-B09 | Viral infections characterized by skin and mucous membrane lesions |
| 11 | B15-B19 | Viral hepatitis |
| 12 | B20-B24 | Human immunodeficiency virus [HIV] disease |
| 13 | B25-B34 | Other viral diseases |
| 14 | B35-B49 | Mycoses |
| 15 | B50-B64 | Protozoal diseases |
| 16 | B65-B83 | Helminthiases |
| 17 | B85-B89 | Pediculosis, acariasis and other infestations |
| 18 | B90-B94 | Sequelae of infectious and parasitic diseases |
| 19 | B95-B97 | Bacterial, viral and other infectious agents |
| 20 | B99-B99 | Other infectious diseases |
| 21 | C00-C97 | Malignant neoplasms |
| 22 | D00-D09 | In situ neoplasms |
| 23 | D10-D36 | Benign neoplasms |
| 24 | D37-D48 | Neoplasms of uncertain or unknown behaviour |
| 25 | D50-D53 | Nutritional anaemias |
| 26 | D55-D59 | Haemolytic anaemias |
| 27 | D60-D64 | Aplastic and other anaemias |
| 28 | D65-D69 | Coagulation defects, purpura and other haemorrhagic conditions |
| 29 | D70-D77 | Other diseases of blood and blood-forming organs |
| 30 | D80-D89 | Certain disorders involving the immune mechanism |
| 31 | E00-E07 | Disorders of thyroid gland |
| 32 | E10-E14 | Diabetes mellitus |
| 33 | E15-E16 | Other disorders of glucose regulation and pancreatic internal secretion |
| 34 | E20-E35 | Disorders of other endocrine glands |
| 35 | E40-E46 | Malnutrition |
| 36 | E50-E64 | Other nutritional deficiencies |
| 37 | E65-E68 | Obesity and other hyperalimentation |
| 38 | E70-E90 | Metabolic disorders |
| 39 | F00-F09 | Organic, including symptomatic, mental disorders |
| 40 | F10-F19 | Mental and behavioral disorders due to psychoactive substance use |
| 41 | F20-F29 | Schizophrenia, schizotypal and delusional disorders |
| 42 | F30-F39 | Mood [affective] disorders |
| 43 | F40-F48 | Neurotic, stress-related and somatoform disorders |
| 44 | F50-F59 | Behavioral syndromes associated with physiological disturbances and physical factors |
| 45 | F60-F69 | Disorders of adult personality and behaviour |
| 46 | F70-F79 | Mental retardation |
| 47 | F80-F89 | Disorders of psychological development |
| 48 | F90-F98 | Behavioral and emotional disorders with onset usually occurring in childhood and adolescence |
| 49 | F99-F99 | Unspecified mental disorder |
| 50 | G00-G09 | Inflammatory diseases of the central nervous system |
| 51 | G10-G13 | Systemic atrophies primarily affecting the central nervous system |
| 52 | G20-G26 | Extrapyramidal and movement disorders |
| 53 | G30-G32 | Other degenerative diseases of the nervous system |
| 54 | G35-G37 | Demyelinating diseases of the central nervous system |
| 55 | G40-G47 | Episodic and paroxysmal disorders |
| 56 | G50-G59 | Nerve, nerve root and plexus disorders |
| 57 | G60-G64 | Polyneuropathies and other disorders of the peripheral nervous system |
| 58 | G70-G73 | Diseases of myoneural junction and muscle |
| 59 | G80-G83 | Cerebral palsy and other paralytic syndromes |
| 60 | G90-G99 | Other disorders of the nervous system |

| ID | ICD-10 | Description |
| --- | --- | --- |
| 61 | H00-H06 | Disorders of eyelid, lacrimal system and orbit |
| 62 | H10-H13 | Disorders of conjunctiva |
| 63 | H15-H22 | Disorders of sclera, cornea, iris and ciliary body |
| 64 | H25-H28 | Disorders of lens |
| 65 | H30-H36 | Disorders of choroid and retina |
| 66 | H40-H42 | Glaucoma |
| 67 | H43-H45 | Disorders of vitreous body and globe |
| 68 | H46-H48 | Disorders of optic nerve and visual pathways |
| 69 | H49-H52 | Disorders of ocular muscles, binocular movement, accommodation and refraction |
| 70 | H53-H54 | Visual disturbances and blindness |
| 71 | H55-H59 | Other disorders of eye and adnexa |
| 72 | H60-H62 | Diseases of external ear |
| 73 | H65-H75 | Diseases of middle ear and mastoid |
| 74 | H80-H83 | Diseases of inner ear |
| 75 | H90-H95 | Other disorders of ear |
| 76 | I00-I02 | Acute rheumatic fever |
| 77 | I05-I09 | Chronic rheumatic heart diseases |
| 78 | I10-I15 | Hypertensive diseases |
| 79 | I20-I25 | Ischaemic heart diseases |
| 80 | I26-I28 | Pulmonary heart disease and diseases of pulmonary circulation |
| 81 | I30-I52 | Other forms of heart disease |
| 82 | I60-I69 | Cerebrovascular diseases |
| 83 | I70-I79 | Diseases of arteries, arterioles and capillaries |
| 84 | I80-I89 | Diseases of veins, lymphatic vessels and lymph nodes, not elsewhere classified |
| 85 | I95-I99 | Other and unspecified disorders of the circulatory system |
| 86 | J00-J06 | Acute upper respiratory infections |
| 87 | J09-J18 | Influenza and pneumonia |
| 88 | J20-J22 | Other acute lower respiratory infections |
| 89 | J30-J39 | Other diseases of upper respiratory tract |
| 90 | J40-J47 | Chronic lower respiratory diseases |
| 91 | J60-J70 | Lung diseases due to external agents |
| 92 | J80-J84 | Other respiratory diseases principally affecting the interstitium |
| 93 | J85-J86 | Suppurative and necrotic conditions of lower respiratory tract |
| 94 | J90-J94 | Other diseases of pleura |
| 95 | J95-J99 | Other diseases of the respiratory system |
| 96 | K00-K14 | Diseases of oral cavity, salivary glands and jaws |
| 97 | K20-K31 | Diseases of oesophagus, stomach and duodenum |
| 98 | K35-K38 | Diseases of appendix |
| 99 | K40-K46 | Hernia |
| 100 | K50-K52 | Noninfective enteritis and colitis |
| 101 | K55-K63 | Other diseases of intestines |
| 102 | K65-K67 | Diseases of peritoneum |
| 103 | K70-K77 | Diseases of liver |
| 104 | K80-K87 | Disorders of gallbladder, biliary tract and pancreas |
| 105 | K90-K93 | Other diseases of the digestive system |
| 106 | L00-L08 | Infections of the skin and subcutaneous tissue |
| 107 | L10-L14 | Bullous disorders |
| 108 | L20-L30 | Dermatitis and eczema |
| 109 | L40-L45 | Papulosquamous disorders |
| 110 | L50-L54 | Urticaria and erythema |
| 111 | L55-L59 | Radiation-related disorders of the skin and subcutaneous tissue |
| 112 | L60-L75 | Disorders of skin appendages |
| 113 | L80-L99 | Other disorders of the skin and subcutaneous tissue |
| 114 | M00-M25 | Arthropathies |
| 115 | M30-M36 | Systemic connective tissue disorders |
| 116 | M40-M54 | Dorsopathies |
| 117 | M60-M79 | Soft tissue disorders |
| 118 | M80-M94 | Osteopathies and chondropathies |
| 119 | M95-M99 | Other disorders of the musculoskeletal system and connective tissue |
| 120 | N00-N08 | Glomerular diseases |
| 121 | N10-N16 | Renal tubulo-interstitial diseases |
| 122 | N17-N19 | Renal failure |
| 123 | N20-N23 | Urolithiasis |
| 124 | N25-N29 | Other disorders of kidney and ureter |
| 125 | N30-N39 | Other diseases of urinary system |
| 126 | N40-N51 | Diseases of male genital organs |
| 127 | N60-N64 | Disorders of breast |
| 128 | N70-N77 | Inflammatory diseases of female pelvic organs |
| 129 | N80-N98 | Noninflammatory disorders of female genital tract |
| 130 | N99-N99 | Other disorders of the genitourinary system |

Table 132: List of all diagnosis blocks considered in this work.



\end{document}